\documentclass[a4paper,11pt]{article}

\usepackage{jheppub}
\usepackage{amsmath,amssymb,amsthm}
\usepackage{mathrsfs}
\usepackage{hyperref}
\usepackage{stackrel}
\usepackage{arydshln}
\usepackage[dvipsnames]{xcolor}

\newcommand{\R}{\mathbb{R}}
\newcommand{\C}{\mathbb{C}}
\newcommand{\xUDarrow}[1]{%
 {\left\updownarrow\vbox to #1{}\right.\kern-\nulldelimiterspace}
}

\makeatletter
\newcommand\xLRarrow[2][]{%
  \ext@arrow 9999{\longleftrightarrowfill@}{#1}{#2}}
\newcommand\longleftrightarrowfill@{%
  \arrowfill@\leftarrow\relbar\rightarrow}
 \newcommand{\xdashleftrightarrow}[2][]{\ext@arrow 3359\leftrightarrowfill@@{#1}{#2}}
 \def\leftrightarrowfill@@{\arrowfill@@\leftarrow\relbar\rightarrow}
\def\arrowfill@@#1#2#3#4{%
  $\m@th\thickmuskip0mu\medmuskip\thickmuskip\thinmuskip\thickmuskip
   \relax#4#1
   \xleaders\hbox{$#4#2$}\hfill
   #3$%
}
\makeatother

\newtheorem{thm}{Theorem}[section]

\newtheorem{rmk}[thm]{Remark}
\newtheorem{dfn}[thm]{Definition}
\newtheorem{con}[thm]{Conjecture}
\newtheorem{ass}[thm]{Assumption}

\def\be{\begin{equation}}
\def\ee{\end{equation}}
\def\IR{{\mathbb{R}}}
\def\IZ{{\mathbb{Z}}}
\def\IP{{\mathbb{P}}}
\def\IC{{\mathbb{C}}}

\def\IN{{\mathbb{N}}}

\def\CM{{\mathcal{M}}}

\def\CH{{\mathcal{H}}}
\def\CN{{\mathcal{N}}}
\def\CQ{{\mathcal{Q}}}
\def\CT{{\mathcal{T}}}

\def\CO{\mathcal{O}}

\def\sgn{{\rm{sgn}}}
\def\fq{\mathfrak{q}}
\def\ft{\mathfrak{t}}
\def\sl{\mathfrak{sl}}
\def\dd{\mathbf{d}}

\usepackage{graphicx}

\definecolor{purple}{rgb}{0.7,0,0.7}

\title{Knot homologies and generalized quiver partition functions}
\author[a,b]{Tobias Ekholm}
\author[c,d]{, Piotr Kucharski}
\author[e]{, and Pietro Longhi}

\affiliation[a]{Department of Mathematics, Uppsala University, Box 480, 751 06 Uppsala, Sweden}
\affiliation[b]{Institut Mittag-Leffler, Aurav. 17, 182 60 Djursholm, Sweden}
\affiliation[c]{Walter Burke Institute for Theoretical Physics, California Institute of Technology, \\ Pasadena, CA 91125, USA}
\affiliation[d]{Faculty of Physics, University of Warsaw, ul. Pasteura 5, 02-093 Warsaw, Poland}
\affiliation[e]{Institute for Theoretical Physics, ETH Zurich, CH - 8093, Zurich, Switzerland}

\emailAdd{tobias.ekholm@math.uu.se, piotrek@caltech.edu, longhip@phys.ethz.ch}

\abstract{
We introduce generalized quiver partition functions of a~knot $K$ and conjecture a~relation to generating functions of symmetrically colored HOMFLY-PT polynomials and corresponding HOMFLY-PT homology Poincar\'e polynomials. 
We interpret quiver nodes as certain basic holomorphic disks in the~resolved conifold, with boundary on the~knot conormal~$L_K$, a~positive multiple of a~unique closed geodesic, and with their (infinitesimal) boundary linking density measured by the~adjacency matrix of the~generalized quiver. The~basic holomorphic disks that are quiver nodes appear in a~certain $U(1)$-symmetric configuration. We propose an extension of the~quiver partition function to arbitrary, not $U(1)$-symmetric, configurations as a~function with values in chain complexes. The~chain complex differential is trivial at the~$U(1)$-symmetric configuration, under deformations the~complex changes, but its homology remains invariant.      
We also study recursion relations for the~partition functions connected to knot homologies. We show that, after a~suitable change of variables, any (generalized) quiver partition function satisfies the~recursion relation of a~single toric brane in $\mathbb{C}^3$. 
}

\begin{document}

\maketitle

\section{Introduction}
Polynomial knot invariants such as the~Jones and HOMFLY-PT polynomials, originally defined combinatorially, have been interpreted and further explained from physical and geometric points of view. In physics the invariants appear through quantum field theory (Chern-Simons theory) \cite{witten1989}, topological strings and M-theory in combination with the~conifold transition \cite{Ooguri:1999bv}, and in geometry through Gromov-Witten counts of bare curves \cite{ES,ES2}. Many polynomial knot invariants admit categorifications, where the~polynomial is expressed as the~(graded) Euler characteristic of a~chain complex associated to the~knot, with homology which is a~knot invariant. The~first example is Khovanov's categorification of the~Jones polynomial~\cite{Khovanov}. Connections between categorified knot invariants and BPS states in the~physical theories underlying the~original knot polynomials have been proposed, see~e.g.~\cite{GSV0412,GS1112}. 

Following \cite{Ekholm:2018eee,Ekholm:2019lmb}, we study connections between knot invariants and quivers as in \cite{Kucharski:2017poe,Kucharski:2017ogk}, where the~motivic generating function of a~quiver gives the~generating series for HOMFLY-PT polynomials from the~perspective of BPS counts. Here quiver nodes are basic BPS states, their interactions are governed by the~quiver arrows, and the~interacting nodes generate the~whole spectrum. Geometrically, basic BPS states correspond to M2-branes wrapping basic holomorphic disks in the~resolved conifold with boundary on an~M5-brane wrapping the~knot conormal, and the~quiver adjacency matrix encodes boundary linking data. 

The main theme of this paper is a~conjectural extension of the~correspondence between knot invariants and quivers to the~categorified level, see \eqref{eq:generalized-quiver-P}. The~extension involves new types of quiver nodes that geometrically correspond to certain `stretched' (non-embedded near the boundary) holomorphic disks and a~generalization of the~quiver motivic generating function that we conjecture gives the~generating series of the~Poincar\'e polynomial of symmetrically colored HOMFLY-PT homology, see section \ref{ssec:motivations}.
This conjecture updates \cite[Conjecture 1.1]{Ekholm:2018eee}, see remark \ref{rmk:correction} for details about the~relation and sections \ref{eq:geometric-t-degree-conjectural} and \ref{ssec:motivations} for motivations. 
The basic disks that appear as quiver nodes arise in a~(possibly degenerate) $U(1)$-symmetric situation, 
in line with earlier proposals originating from string dualities~\cite{Aganagic:2011sg}.

We then consider deformations that break the~$U(1)$ symmetric configuration, and 
propose in section \ref{sec:geom+HOMFLY} an extension of the~generalized quiver partition function. 
We first show that the~moduli spaces of stretched basic disks can be carried along generic 1-parameter families of complex structures that remain stretched near the~Lagrangian, so that these moduli spaces together with linking information suffices to compute the~refined partition function. We also perform a~bifurcation analysis of such deformations. We generalize the~quiver partition function to a~function with values in chain complexes, and show that at any generic parameter in a~generic path of deformations there are differentials on these chain complexes with homology that remains invariant. Here, for higher symmetric colorings, the~chain complexes correspond to certain subspaces of the~homology which nevertheless contain sufficient information to recover all of the~homology, see section \ref{sec:pathcategorification} for details.

In the~light of the~conjectured connection between HOMFLY-PT homology and generalized quivers, one might expect that structures in one of the~theories are reflected in the~other. We first consider structures in HOMFLY-PT homology that we expect originate from the~geometry of the~basic disks in the~$U(1)$-symmetric configuration. There are $\mathfrak{sl}_{N}$ differentials $d_{N}$, which act on the~HOMFLY-PT homology with resulting homology being $\mathfrak{sl}_N$ Kohovanov-Rozansky homology. The~$\mathfrak{sl}_1$ homology is very simple, it has rank one for every knot (in reduced normalization). We conjecture that this structure is reflected on the~level of generalized quivers and that the~quiver of a~knot can be obtained from one `spectator' node and a~quiver of half the~size of the~original quiver together with a~`universal disk' that comes from the~closed sector and carries the~quantum numbers of the~$\mathfrak{sl}_1$-differential $d_1$. The~application of the~multi-cover skein unlinking operation of \cite{Ekholm:2019lmb} to a~basic once-around disk and the~universal disk gives a~pair of generators connected by~$d_1$. This pair corresponds to two holomorphic curves which differ by a~unit of winding around the~base $\C \IP^1$ of the~resolved conifold, where the~heavier disk can be viewed as a~bound state of the~lighter and the~universal disk. The~connection between reduced and unreduced homologies seems to be reflected in the~structure of the~generalized quiver in a~similar way.

We also make predictions for the~general structure of HOMFLY-PT homology colored by symmetric representations with $r$ boxes.
We view the~generators of colored HOMFLY-PT homology in the~$r$-th symmetric representation as equivariant vortices of vorticity $r$ in the~theory $T[L_{K}]$, see section \ref{sec:generalized-quiver}. The~connection between quiver nodes and equivariant vortices is somewhat involved:
a~basic curve on a~knot conormal with boundary that wraps $k$ times around the~circle in the~conormal gives $2^{k-1}$ generators in the~$k$-th symmetrically colored homology. 
At level $r=1$ every generator corresponds to a~node of the~generalized quiver. 
The quiver partition function describes a~tower of contributions from $r=1$ generators to all higher levels with $r>1$, which is completely determined by homology data of these nodes and their mutual linking.
Knowing the~structure of these contributions from level $1$ to level $2$ allows to separate them from the~genuinely new generators that appear on level $2$. We continue in this way inductively: taking out contributions from all nodes of levels $< r$ allows us to identify the~genuinely new generators on level $r$.  We verify the~claim about homology contributions of level $r$ generators against conjectural expressions for homologies of knots $9_{42}$ and $10_{132}$ for $r=2$, see section \ref{sec:Examples}.

Further structures appear when we deform away from the~$U(1)$-symmetric configuration. Here the~pure level $r$ generators generate a~chain complex of `Bott-equivariant' vortices, see \eqref{eq:Bottvortex}, which are transformed by chain homotopy under deformation. The~homology of the~level $r$ generators is non-vanishing only for finitely many $r$ and, together with linking data, it recovers the~usual $U(1)$-symmetric HOMFLY-PT homology.

Generating series for knot polynomials satisfy polynomial recursion relations. Geometrically, such relations originate from counts of punctured holomorphic curves at infinity with boundary on the~knot conormal and asymptotic to Reeb chords at punctures, see \cite{Ekholm:2018iso, ES2}. From the~viewpoint of the~quiver or the~basic disks without punctures, the~recursion relation is generated by similar relations localized around individual basic disks. This leads to an~algebraic description of the~ensemble of basic disks in terms of noncommutative variables, see \cite{Ekholm:2019lmb}. Here we take this further and show that after a~suitable change of noncommutative variables the~recursion relation of \emph{any} quiver \emph{is} the~recursion relation of the~basic toric brane in $\C^{3}$. We generalize this simple recursion relation to include also the~new basic curves discussed above and look at the~implications of such a~relation for both knot polynomials and their categorifications. 
As the toric brane has a unique embedded holomorphic disk ending on it, this universal relation to the toric brane is in line with $U(1)$-symmetric configurations of knot conormals, where all holomorphic curves have boundaries which are multiples of a unique simple closed geodesic.

As already mentioned, the form of generalized quiver partition functions that we introduce for multiply wrapped basic disks is conjectural. We check it against the few available (also conjectural) results for knot homologies of $9_{42}$ and $10_{132}$ and it goes without saying that further tests are important. In particular, calculations of colored knot homologies of knots with more than eight crossings would provide support or indicate possible changes to our conjecture. In this context we point out that our conjecture has both a structural and a technical aspect. The structural aspect says that the basic generalized holomorphic disks (generalized quiver nodes) and their boundary linking densities (weighted quiver arrows) contain all information about symmetrically colored HOMOFLY-PT homology. The technical aspect gives the `change of variables' for extracting this information. It is thus possible that the structural part is the correct, even if the technical part needs modification. As discussed above, there are knots for which standard quiver partition functions cannot reproduce HOMFLY-PT homologies and, therefore, a generalization is needed. Our proposal is in a sense a minimal extension, compatible with available data and with the geometry of holomorphic curves under SFT-stretching.

\section{Background}\label{sec:background}
In this section we review earlier results on knot polynomials and knot homologies and their connections to quivers and (refined) curve counting. This is the starting point for our study in later sections. 

\subsection{Knot polynomials and homologies}
\begin{figure}[htb]
    \centering
    \includegraphics[width=.45\textwidth]{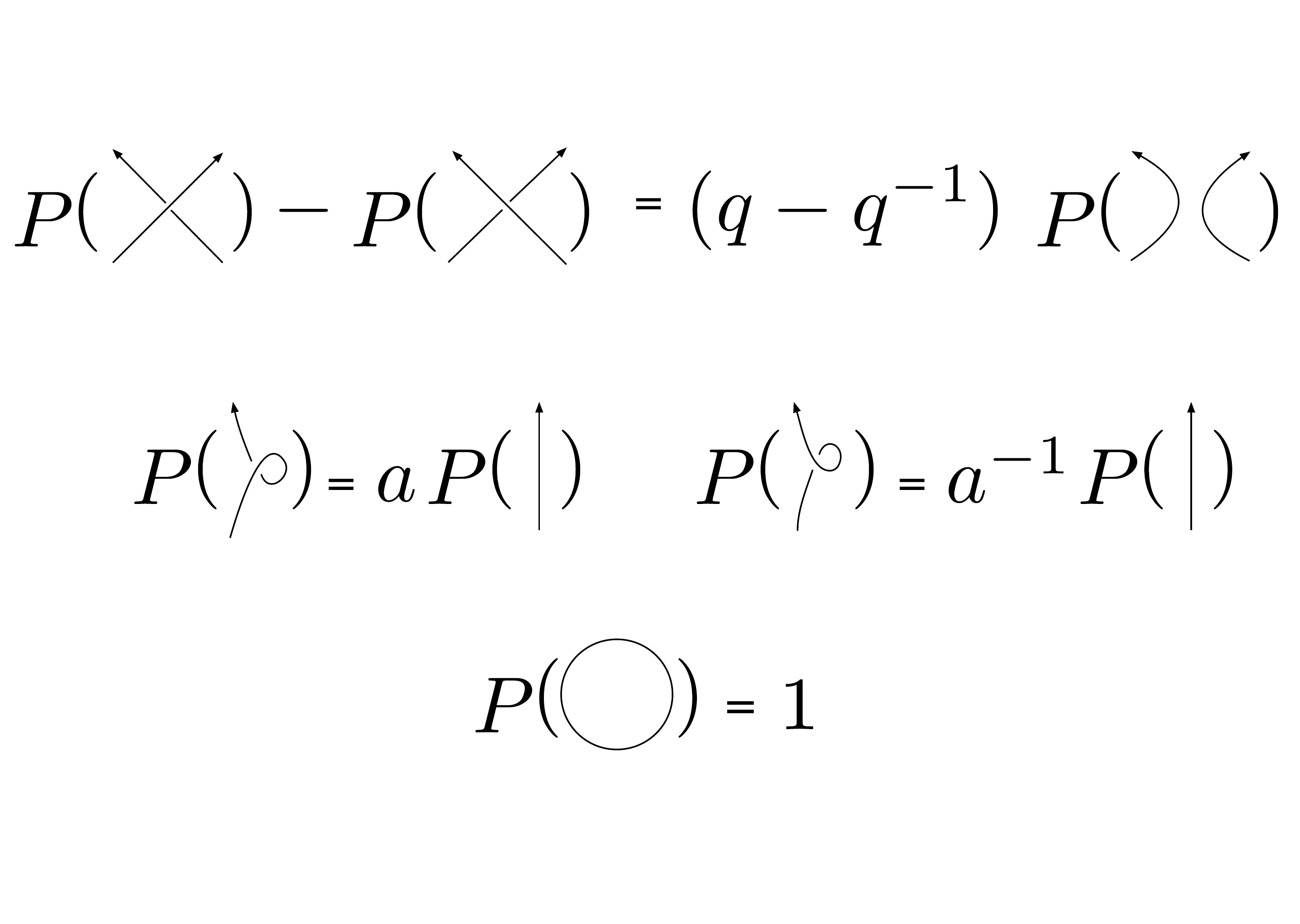}
    \caption{The framed skein relation}
    \label{fig:framedskein}
\end{figure}

If $K\subset S^{3}$ is a~framed knot, then its framed \emph{HOMFLY-PT polynomial} $P(K;a,q)$ \cite{freyd1985,PT} is a~two-variable polynomial that can be calculated from a~knot diagram (a projection of $K$ with over/under information at crossings and framing given by the~projection direction) via the~framed skein relation, see figure \ref{fig:framedskein}. 
The polynomial is an~invariant of framed knots up to framed isotopy. The~standard HOMFLY-PT polynomial of a~knot is the~framed HOMFLY-PT of that knot when equipped with the~framing for which its self-linking number vanishes. For $a=q^2$, the~HOMFLY-PT reduces to the~Jones polynomial $J(K;q)$ \cite{Jon85}, and for $a=q^N$ to the~$\mathfrak{sl}_N$ Jones polynomial $J_{N}(K;q)$ \cite{RT90}.

More generally, the~\emph{colored} HOMFLY-PT polynomials $P_{R}(K;a,q)$ are similar polynomial knot invariants depending also on a~representation $R$ of the~Lie algebra ${\mathfrak{u}}(N)$. In this setting, the~original HOMFLY-PT corresponds to the~standard representation. Also the~colored version admits a~diagrammatic description: it is given by a~linear combination of the~standard polynomial of certain satellite links of $K$. From the~physical point of view, the~colored HOMFLY-PT polynomial with $a=q^N$ is the~expectation value of the~knot viewed as a~Wilson line in $U(N)$ Chern-Simons gauge theory on $S^{3}$ \cite{witten1989}. 
In order to simplify the~notation, we will write the~HOMFLY-PT polynomial also when we refer to the~more general colored version.

HOMFLY-PT polynomials also have an~interpretation in terms of open topological string or holomorphic curve counts. Physically, $P(K;a,q)$ can be interpreted as the~partition function of open topological string of the~Lagrangian conormal of the~knot after transition from $T^{\ast} S^{3}$ to the~resolved conifold, where $q=e^{g_{s}}$ and $a$ is the~K\"ahler parameter in the~resolved conifold \cite{Ooguri:1999bv}. From the~mathematical point of view, this results from invariant counts of open holomorphic curves by the~values of their boundaries in the~skein module of the~Lagrangian, see \cite{ES,ES2}.   

We next consider knot homologies. In \cite{Khovanov}, Khovanov introduced a~knot invariant which is a~far-reaching generalization of the~Jones polynomial. To any knot $K$ he associated a~doubly-graded chain complex, the~homology of which is a~knot invariant and such that the~Jones polynomial arises as its (graded) Euler characteristic:
\begin{equation}
    J(K;q)=\sum_{i,j}(-1)^{j}q^{i}\dim\mathcal{H}^{\mathfrak{sl}_2}_{i,j}(K)\,.
    \label{eq:Jones as Euler}
\end{equation}
In this sense $\mathcal{H}^{\mathfrak{sl}_2}_{i,j}(K)$ is a~categorification of $J(K;q)$. Khovanov-Rozansky homology, which categorifies the~$\mathfrak{sl}_N$ polynomial, was defined in \cite{KhR1}:
\begin{equation}
    J_{N}(K;q)=\sum_{i,j}(-1)^{j}q^{i}\dim\mathcal{H}^{\mathfrak{sl}_N}_{i,j}(K)\,.
    \label{eq:slN Jones as Euler}
\end{equation}
In this paper we focus mostly on HOMFLY-PT homology \cite{KhR2}, which is a~categorification of the~(original) HOMFLY-PT polynomial:
\begin{equation}
    P(K;a,q)=\sum_{i,j,k}(-1)^{k}a^{i}q^j\dim\mathcal{H}_{i,j,k}(K)\,.
    \label{eq:HOMFLY as Euler}
\end{equation}
The corresponing Poincar\'{e} polynomial then 
provides a~$t$-refinement of HOMFLY-PT polynomial called the~\emph{superpolynomial} \cite{DGR0505}:
\begin{equation}
    P(K;a,q,t)=\sum_{i,j,k}a^{i}q^{j}t^{k}\dim\mathcal{H}_{i,j,k}(K)\,,
    \label{eq:superpolynomial as Poincare}
\end{equation}
with $P(K;a,q,-1)=P(K,a,q)$.
Categorifications  of colored HOMFLY-PT polynomials were considered in \cite{GS1112}, where it was conjectured that there exists a~colored HOMFLY-PT homology, which is invariant under isotopy and such that
\begin{equation}
\begin{split}
    P_{R}(K;a,q)&=\sum_{i,j,k}(-1)^{k}a^{i}q^j\dim\mathcal{H}_{R;i,j,k}(K)\,,
    \\
    P_{R}(K;a,q,t)&=\sum_{i,j,k}a^{i}q^{j}t^{k}\dim\mathcal{H}_{R;i,j,k}(K)\,.
    \label{eq:colored polynoamials and homologies}
\end{split}
\end{equation}
There are also colored Khovanov ($\mathcal{H}^{\mathfrak{sl}_2}_{R}$) and Khovanov-Rozansky ($\mathcal{H}^{\mathfrak{sl}_N}_{R}$)  homologies, which categorify $J_R(K;q)$ and $J_{N,R}(K;q)$ respectively. We point out that there are many constructions of colored HOMFLY-PT and $\mathfrak{sl}_N$ homologies. They do not always give the~same results and some work only in special cases (e.g., for $\Lambda^r$ representations), see \cite{Cau1611,RW1702,ETW1703,QRS1802,GW1904,OR2010,GHM2103} and references therein. Here we focus on symmetric representations $R=S^{r}$, corresponding to Young diagrams with a~single row of $r$ boxes, and we will write $P_r$ and $\mathcal{H}_r$ instead of $P_{S^{r}}$ and $\mathcal{H}_{S^{r}}$ to simplify the~notation. (For example, $\mathcal{H}_1$~will denote the~original HOMFLY-PT homology corresponding to the~standard representation.)

\subsection{Observed relations between knots and quivers}

In \cite{Kucharski:2017ogk,Kucharski:2017poe}, knot polynomials were related with representations of quivers. A~quiver $Q$ is an~oriented graph with a~finite number of vertices connected by finitely many signed arrows. We denote the~set of vertices by $Q_0$ and the~set of arrows by $Q_1$. A~\emph{dimension vector} for $Q$ is a~vector in the~integral lattice with basis $Q_{0}$, $\boldsymbol{d}\in \IN Q_0$. We number the~vertices of $Q$ by $1,2,\dots,m=|Q_{0}|$. A~quiver representation with dimension vector $\boldsymbol{d}=(d_{1},\dots,d_{m})$ is the~assignment of a~vector space of dimension $d_i$ to the~node $i\in Q_0$ and of a~linear map $\gamma_{ij}\colon\IC^{d_i} \to \IC^{d_j}$ to each arrow from vertex $i$ to vertex $j$. The~\emph{adjacency matrix} of $Q$ is the~$m\times m$ integer matrix with entries $C_{ij}$ equal to the~algberaic number of arrows from $i$ to $j$. A~quiver is symmetric if its adjacency matrix is. 

Quiver representation theory studies moduli spaces of stable quiver representations (see e.g., \cite{kirillov2016quiver} for an~introduction to this subject).
While explicit expressions for invariants describing those spaces are hard to find in general, they are quite well understood in
the~case of symmetric quivers \cite{Kontsevich:2008fj,KS1006,2011arXiv1103.2736E}. Important information (such as the~intersection homology Betti numbers of the~moduli space of all semi-simple representations of $Q$ of dimension vector $\boldsymbol{d}$, see \cite{MR1411,FR1512}) about the~moduli space of representations of a~symmetric quiver with trivial potential is encoded in the~\emph{motivic generating series} defined as 
\be\label{eq:Efimov}
	P_{Q}(\boldsymbol{x};q)=\sum_{d_{1},\ldots,d_{m}\geq0}(-q)^{\sum_{1\leq i,j\leq m}C_{ij}d_{i}d_{j}}\prod_{i=1}^{m}\frac{x_{i}^{d_{i}}}{(q^{2};q^{2})_{d_{i}}} = \sum_{\boldsymbol{d}}(-q)^{\boldsymbol{d}\cdot C\cdot\boldsymbol{d}}\frac{\boldsymbol{x}^{\boldsymbol{d}}}{(q^{2};q^{2})_{\boldsymbol{d}}}\,, 
\ee
where the~denominator is the~so called $q$-Pochhammer symbol:
\be
	(z;q^2)_r  = \prod_{s=0}^{r-1} (1-z q^{2s}) \,.
\ee
We will often refer to $P_{Q}(\boldsymbol{x};q)=P_{Q}(x_1,\ldots,x_m;q)$ as the~quiver partition function.
We point out that the~quiver representation theory involves the~choice of an~element, the~potential, in the~path algebra of the~quiver and that the~trivial potential is the~zero element.

A correspondence between knots and quivers proposed in \cite{Kucharski:2017ogk,Kucharski:2017poe} associates quivers to knots by equating the~motivic generating series with the~generating series of superpolynomials or  HOMFLY-PT generating series in the~variable $x$:
\be\label{eq:HOMFLY-PT series}
	P_{K}(x,a,q,t)=\sum_{r=0}^{\infty}\frac{P_{r}(K;a,q,t)}{(q^{2};q^{2})_{r}}x^{r}\,, \quad  P_{K}(x,a,q)=\sum_{r=0}^{\infty}\frac{P_{r}(K;a,q)}{(q^{2};q^{2})_{r}}x^{r}\,.
\ee
More precisely, the~most basic version of the~correspondence states that for each knot $K$ there exist a~symmetric quiver $Q$ and integers $\{a_{i},q_{i}\}_{i\in {Q}_{0}}$, such that
\be\label{eq:KQ-corr-basic}
	P_{K}(x,a,q)=P_{Q}(\boldsymbol{x};q) \quad \textrm{for} \quad x_{i}=x a^{a_{i}}q^{q_{i}-C_{ii}}\,.
\ee
The refined version conjectures the~bijection between $Q_0$ and the~set of generators of $\mathcal{H}_1$, which fixes $(a_i,q_i,C_{ii})$ to be $(a,q,t)$-degrees of respective generators:
\be\label{eq:KQ-corr-refined}
	P_{K}(x,a,q,t)=P_{Q}(\boldsymbol{x};q) \quad \textrm{for} \quad x_{i}=x a^{a_{i}}q^{q_{i}-C_{ii}}(-t)^{C_{ii}}\,.
\ee
For $t=-1$, this equation reduces to \eqref{eq:KQ-corr-basic}, following the~relation between the~superpolynomials and HOMFLY-PT polynomials.

The correspondence between knot invariants and quivers stated above was proved for all 2-bridge knots in~\cite{Stosic:2017wno} and for all arborescent knots in~\cite{SW2004} but \emph{it does not hold} for all knots, e.g., the~knot $9_{42}$ gives a~counter example, see section \ref{sec:Examples}. We will discuss a~modification of the~correspondence below that includes new types of nodes. 


\subsection{Physics and geometry of knots and quivers}

\subsubsection{3d \texorpdfstring{$\mathcal{N}=2$}{N=2} gauge theories}

The physics around the~knots-quivers correspondence is a~duality between two 3d $\mathcal{N}=2$ theories: one associated to the~knot and denoted $T[L_K]$, and the~other associated to the~quiver and denoted $T[Q]$, see \cite{Ekholm:2018eee}.

The theory associated to the~knot $K$, which we denote $T[L_{K}]$, arises from M-theory over the~resolved conifold~$X$ with a~single M5-brane wrapping the~conormal Lagrangian of the~knot $L_K$ \cite{Ooguri:1999bv}: 
\be\label{eq:M-theory-setup}
\begin{split}
	\text{space-time}: \quad& \IR^4 \times S^1 \times X \\
			& \cup \phantom{ \ \times S^1 \times \ \ } \cup\\
	\text{M5}: \quad & \IR^2\times S^1 \times L_K.
\end{split}
\ee
In general $T[L_K]$ does not admit a~simple Lagrangian description, but its 
vortex partition function is known to count M2-branes wrapping embedded holomorphic curves ending on~$L_K$.

In contrast, $T[Q]$ is easier to describe: the~gauge group is $U(1)^{\times Q_0}$ and there is one charged chiral for each $U(1)$ factor.
Interactions among the~different sectors are mediated by (mixed) Chern-Simons couplings, encoded by the~quiver adjacency matrix.
The quiver variables $x_i$ encode exponentiated Fayet-Ilioupoulos couplings.
The~partition function of vortices again counts M2-branes wrapping embedded holomorphic disks.
But in this case there are only $|Q_0|$ of them that are linked according to the~adjacency matrix.
Here, all other holomorphic curves are branched covers of the~basic embedded disks which after perturbation are counted in the~$U(1)$-skein of the~Lagrangian projected to homology and linking \cite{Ekholm:2019lmb, ES}.
The duality with $T[L_K]$ is encoded by the~change of variables (\ref{eq:KQ-corr-basic}). 


An important distinction to keep in mind is that while $T[Q]$ has a~simple Lagrangian description, $T[L_K]$ is closer to knot homology, see appendix \ref{app:TLK-vs-TQ}.
Indeed, its vortex partition function equals the~generating series of colored HOMFLY-PT polynomials: 
\be
	P_{K}(x,a,q)= \sum_{r=0}^{\infty}\frac{P_{r}(K;a,q)}{(q^2;q^2)_r}x^{r}
	= \sum_{r\geq 0} Z_r^{\text{vortex}}(a,q) \, x^{r} = Z^{\text{vortex}}(x,a,q)\,.
\ee
The coefficient $Z_r^{\text{vortex}}(a,q)$ is a~character for the~moduli space $\CM_{r}$ of $r$ vortices. 
It takes the~form ${Z_r^{\text{internal}}(a,q)}/{(q^2;q^2)_r}$ since 
schematically $\CM_{r} \simeq \CM_r^{\text{internal}}\times (\IC^r / S_r)$, where the~second factor parametrizes positions of $r$ vortices in the~plane, and generates the~denominator, see (\ref{eq:Morse}) below.
Therefore, $Z_r^{\text{internal}}(a,q)$ is a~polynomial whose coefficients correspond to net counts (with signs, including cancellations) of cohomology generators of  $\CM_r^{\text{internal}}$.
Passing to categorification, this polynomial corresponds to an~index for the~Hilbert space arising from quantization of $\CM_r^{\text{internal}}$, which serves as a~model for $\oplus_{i,j,k}\CH_{r;i,j,k}(K)$.

\subsubsection{Basic holomorphic curves}

In the~previous subsection we saw that $T[L_{K}]$ arises from M-theory as the~effective theory on the~surface of the~M5-brane, and that its BPS particles originate from M2-branes ending on the~M5. From the~symplectic geometric point of view, BPS states correspond to generalized holomorphic curves with boundary on the~Lagrangian submanifold $L_{K}$. 

We recall the~definition of generalized holomorphic curves in the~resolved conifold $X$ with boundary on a~knot conormal $L_{K}\subset X$ (as defined in \cite{Ekholm:2018iso,Ekholm:2018eee}) from the~skeins on branes approach to open curve counts in \cite{ES}. The~key observation in \cite{ES} is that the~count of bare curves (i.e., curves without constant components) counted by the~values of their boundaries in the~skein module remains invariant under deformations. The~count of such curves also requires the~choice of a~4-chain $C_{K}$ with $\partial C_K=2L_K$. Intersections of the~interior of a~holomorphic curve and the~4-chain contribute to the~framing variable $a$ in the~skein module. For generalized curves there is a~single brane on $L_{K}$, which leads to $a=q$. Then, the~map from the~skein module to `homology class and linking' is well-defined and thus counting curves this way, less refined than the~$U(1)$-skein, also remains invariant. In $L_{K}\simeq S^{1}\times\R^{2}$ one can define such a~map that depends on the~choice of a~framing of the~torus at infinity. More precisely, one fixes bounding chains for the~holomorphic curve boundaries that agree with multiples of the~longitude at infinity and replace linking with intersections between curve boundaries and bounding chains. In \cite{Ekholm:2018iso} an~explicit construction of such bounding chains and compatible 4-chain $C_{K}$ from a~certain Morse function of $L_{K}$ was described and shown to give invariant curve counts in 1-parameter families.   

Consider now holomorphic disks with boundary in the~basic homology class. Such disks are generically embedded and can never be further decomposed under deformations. Assuming, in line with~\cite{Gopakumar:1998ii, Gopakumar:1998jq}, that all actual holomorphic curves with boundary on $L_{K}$ lie in neighborhoods of such holomorphic disks attached to the~conormal, it would then follow that all generalized holomorphic curves are combinations of branched covers of the~basic disks. Using the~multiple cover formula  the~count of generalized curves then agrees with the~quiver partition function with nodes at the~basic disks and with arrows according to linking and additional contributions to the~vertices given by $4$-chain intersections.

From this point of view, the~theory $T[Q]$ can be thought of as changing the~perspective, starting from a~neighborhood of the~Lagrangian in its cotangent bundle and attaching small neighborhoods of the~basic holomorphic disks along curves near the~central $S^1$. The~resulting neighborhood is then determined up to symplectomorphism by the~framed link of the~boundaries of the~disks attached. In this paper we will extend the~collection of basic holomorphic curves to include certain non-standard disks that are not embedded. Such curves have boundary that wraps the~homology generator of the~conormal several times, but are not combinations of curves going once around that generator. As we will discuss below, the~most basic such curves are not embedded disks: an~embedded disk that goes $d$ times around is expressed as $d$ such basic holomorphic curves. We give a~conjectural picture of the~contribution from holomorphic curves in a~neighborhood of such basic curves to the~partition function, also on the~refined level. From the~quiver point of view, these curves could perhaps be considered as a~new type of `orbifold' quiver nodes, see section \ref{sec:orbifolds} for the~underlying geometry.

\subsection{Omega-background and refined indices}\label{eq:Omega-background and refined indices}

Having reviewed the~geometric and physical features of the~quiver-like description of knot invariants in the~context of topological strings, we now turn to the~question of refinement.
We begin with a~review of well-known facts about refinement in the~context of closed topological string theory. We then give a~brief description of the~counterpart for the~open sector for a~knot conormal $L_K$ from the~point of view of the~theory $T[L_K]$. In section~\ref{sec:Homological degree of M2-branes} we discuss the~extension of refinement to the~open string sector further, reviewing and clarifying the~role of a~certain $U(1)$ symmetry and its interpretation in the~context of branes wrapping knot conormal Lagrangians.

\subsubsection{Closed string sector}

One of the~motivations leading to refined topological strings was the~explicit evaluation of partition functions of 4d $\CN=2$ gauge theories via localization \cite{Nekrasov:2002qd, Nekrasov:2003rj}.
These gauge theories may be engineered by type IIA string theory on a~Calabi-Yau threefold $X$ times $\IR^4$, denoted $T_{4d}[X]$. 
This construction may be further viewed as a~circle compactification of M-theory on $X\times S^1_\beta\times \IR^4$, which engineers a~5d theory $T_{5d}[X]$.
The eight supercharges of $\CN=2$ SUSY in 4d transform as spinors under $SO(4)\simeq SU(2)_L\oplus SU(2)_R$, and are further charged under a~$U(1)\times SU(2)_I$ R-symmetry. This $U(1)$ is often anomalous and we will ignore it.
Supercharges transform under $SU(2)_L\times SU(2)_R\times SU(2)_I$ as $(\bar Q^I_{\dot\alpha}, {Q_{\alpha}^I})= {({\bf{2}},{\bf{1}};{\bf{2}})} \oplus {({\bf{1},\bf{2};\bf{2}})} $.

If $S^1\times \IR^4$ is replaced by an~$\IR^4$-bundle over $S^1$ with holonomy $g\in SO(4)$, in order to preserve some supersymmetry one can turn on a~nontrivial R-symmetry background as follows \cite{witten1988}. 
Via the~identification $SO(4)\simeq SU(2)_L\times SU(2)_R$, we may split $g = (g_L, g_R)$. 
One can turn on a~holonomy $g_I$ of $SU(2)_I$ for the~rank-2 R-symmetry bundle when going around $S^1$, in particular one can set  $g_I = g_R$.
It is then natural to consider the~diagonal subgroup $SU(2)_d\subset SU(2)_R\times SU(2)_I$. Supercharges transform under $SU(2)_L\times SU(2)_d$ as
\be
	({\bf{2},\bf{1};\bf{2}}) \to \underbrace{({\bf{2},\bf{2}})}_{\tilde Q_\mu}\,,
	\qquad
	({\bf{1},\bf{2};\bf{2}}) \to \underbrace{({\bf{1},\bf{1}})}_{\tilde Q} \oplus \underbrace{({\bf{1},\bf{3}})}_{\tilde Q_{\mu\nu}}\,.
\ee
There is a~distinguished supercharge $\tilde Q$, which is invariant under the~$S^1$ holonomy for any choice of $(g_L, g_R)$. 

Next we restrict to a~specific choice of bundle, the~so-called Omega-background, see~\cite{Nekrasov:2003rj}
for an~in-depth description.
In this case the~holonomies $g_{L}, g_{R}$ are restricted to $U(1)_L\times U(1)_R$ subgroups.
Notably, this implies that one only needs a~$U(1)_I$ subgroup of $SU(2)_I$ to perform a~topological twist. This fact will be important later when introducing branes.
After twisting, the~holonomy group is then $U(1)_L\times U(1)_d$, and acts by rotations $(g_L,g_d) = (e^{\epsilon_1 + \epsilon_2}, e^{\epsilon_1 - \epsilon_2})$ of $\IR^4\sim \IR^2_{\epsilon_1}\oplus \IR^2_{\epsilon_2}$.
Thanks to this restriction there are actually two conserved supercharges in the~Omega-background.

To understand this, let us choose a~basis for spinors in which $J^3_{L,R}$ are diagonal. 
Doing the~same for $SU(2)_I$, the~eight supercharges of the~4d theory have charges 
\be
	\underbrace{\left(\pm \frac{1}{2},0;\pm \frac{1}{2}\right)}_{\bar Q^I_{\dot\alpha}} \oplus \underbrace{\left(0,\pm \frac{1}{2};\pm \frac{1}{2}\right)}_{Q_{\alpha}^I}
	\quad\text{under} \quad 
	U(1)_L \times U(1)_R\times U(1)_I\,,
\ee
where signs are chosen independently.
With the~topological twist, one may preserve the~following two linear combinations:
\be
	\tilde Q_\pm \sim \left(0,\pm \frac{1}{2};\mp \frac{1}{2}\right)\,,
\ee
where now the~\emph{signs are correlated}. In fact, under $U(1)_L\times U(1)_d$ both of these transform as~$(0,0)$.\footnote{From these, one recovers $\tilde Q \sim \tilde Q_+ - \tilde Q_-$
being the~singlet ${\bf 1\subset 2\otimes 2}$.. 
With the~Omega-background, as opposed to more general $SU(2)$ holonomy, one may in addition preserve 
$\tilde Q'  \sim \tilde Q_+ + \tilde Q_-$.}
As already hinted by our parametrization of $(g_L, g_d)$ in terms of $\epsilon_1,\epsilon_2$, $J^3_{L,R}$ are diagonal/anti-diagonal combinations of $J_{1,2}$, the~latter being generators of rotations in $\IR_{\epsilon_1}^2\oplus\IR_{\epsilon_2}^2$. 
The~surviving supercharges therefore transform as
\be\label{eq:Q-tilde-pm}
	\tilde Q_{+} : \left( \frac{1}{2}, \frac{1}{2}; - \frac{1}{2}\right)\,,
	\qquad
	\tilde Q_{-} : \left( -\frac{1}{2}, -\frac{1}{2};  \frac{1}{2}\right)
	\quad\text{under} \quad 
	U(1)_1 \times U(1)_2\times U(1)_I\,.
\ee
Note that $\tilde Q_- = \tilde Q_+^\dagger$.
Taking into account that our conventions for R-charges are opposite to those of \cite{Aganagic:2011sg}, we can readily identify these supercharges with $Q_r,\, \bar Q_r$ there, respectively. 

Note that $[F, \tilde Q_{\pm}]=\pm \tilde Q_\pm$ with $F=2(J_L^3+J_R^3)$ ensures that $\{\tilde Q_{\pm},(-1)^F\}=0$.
Reducing the~theory along $\IR^2_{\epsilon_1}\oplus \IR^2_{\epsilon_2}$ gives a~$\CN=2$ quantum mechanics on $S^1$, where $\{\tilde Q_+,\tilde Q_-\}\sim H$ and $[\tilde Q_\pm, H]=0$. 
It is well-known that $\tilde Q_+$ cohomology only gets contributions from groundstates with $H=0$, due to cancellations among bosons and fermions for all excited states.
Deformations of the~theory may lead some excited states to become groundstates, or vice versa, however the~count of bosonic ($F\in 2\IZ$) minus fermionic ($F\in 2\IZ+1$) groundstates remains invariant. 
Witten index ${\rm Tr} (-1)^F e^{-\beta H}$ coincides with the~invariant difference of dimensions of spaces of groundstates ${\rm dim} \CH^B_{(0)}- {\rm dim} \CH^F_{(0)}$ \cite{Witten:1982im}.
By construction $\tilde Q_\pm$ are invariant under $J^3_L$, and $J^3_R+J^3_I$, therefore one may introduce additional grading on the~whole Hilbert space (both groundstates and excited states), preserving cancellations induced by the~twisting $(-1)^F$  in the~trace.
This leads to the~following for the~5d theory~\cite{Nekrasov:2002qd}:
\be
	Z(\epsilon_1,\epsilon_2) = {\rm Tr}_{{\cal H}[T_{5d}]} (-1)^{F} e^{ - (\epsilon_1-\epsilon_2) J^3_{L}  - (\epsilon_1+\epsilon_2) (J^3_{R} +J^3_I) }\,.
\ee

Although this index will not be an~object of primary interest for us, we will use it as a~point of contact to match with conventions of \cite{Aganagic:2011sg}.
If we turn off R-symmetry and only rotate $\IR^4$, one readily sees that $J_1 = J^3_L+J^3_R$ and $J_2 = -J^3_L+J^3_R$ are the~generators of rotations of $\IR^2_{\epsilon_1}$ and $\IR^2_{\epsilon_2}$ respectively.
We then identify $S_1, S_2, S_R,e^{-\epsilon_1}, e^{-\epsilon_2}$ from \cite{Aganagic:2011sg} with $J_1, J_2,-J^3_I,  \fq, \ft^{-1} $ in the~present  paper.
With these identifications, the~index takes the~form
\be\label{eq:5d-index}
	Z(\fq,\ft) = {\rm Tr}_{{\cal H}[T_{5d}]} (-1)^{F} \fq^{S_1'} \ft^{-S_2'}\,,
\ee
where $S_i' = S_i-S_R$ for $i=1,2$, as claimed in \cite{{Aganagic:2011sg}}.
Adapting to conventions from our earlier work \cite{Ekholm:2019lmb, Ekholm:2018eee}, we will henceforth switch to
\be\label{eq:qt-change}
	q = \fq^{1/2},\qquad t = -(\ft/\fq)^{-1/2}\,.
\ee

\subsubsection{Open string sector}

Let $L$ be a~special Lagrangian submanifold of the~resolved conifold $X=\CO(-1)\oplus \CO(-1)\to \C \IP^1$.
The low-energy dynamics of an~M5-brane on $L\times S^1\times \IR^2$ is described by a~3d $\CN=2$ theory $T[L]$ on $S^1\times \IR^2$ \cite{Dimofte:2011ju, Terashima:2011qi}. 
From this viewpoint, $J_1$ is a~space-time rotation,  however $J_2$ is now an~R-symmetry.
As stressed above, it only makes sense to turn on fugacities $\fq,\ft$ in the~index if the~corresponding generators commute with the~surviving supercharges. 
The presence of the~M5-brane on $\IR^2$ inside $\IR^4$ breaks Lorentz invariance, therefore we must re-investigate whether supercharges are preserved or not.

Recall our discussion of the~5d theory in a~general background, which led to the~existence of a~single  preserved supercharge $\tilde Q$. 
Since this transforms as a~scalar in 4d, clearly it is also preserved by the~presence of the~M5 that breaks Lorentz invariance.
However to define an index and its deformation by $q,t$, we need two supercharges. Without the~brane this was made possible by the~choice of a~non-generic background, the~Omega-background which breaks $SO(4)$ to $U(1)_1\times U(1)_2$. 
As long as the~defect lies in a~plane $\IR^2$, that is either $\IR^2_{\epsilon_1}$ or $\IR^2_{\epsilon_2}$, it preserves the~group of rotations of the~Omega-background.
To preserve two supercharges it was also crucial to perform a~topological twist using a~$U(1)_I$ subgroup of the~$SU(2)_I$ R-symmetry.
In general there is no reason to expect that neither $SU(2)_I$ nor $U(1)_I$ should be preserved by the~presence of M5 on $L\times S^1\times \IR^2$. 
However, one may hope that there exist specific geometric configurations of $L$ that allow to preserve at least the~$U(1)_I$ subgroup required by the~topological twist.

We will assume that $U(1)_I\subset SU(2)_I$ is preserved by the~presence of M5 on $L$. This is a~nontrivial and crucial assumption made in \cite{Aganagic:2011sg}, whose geometric significance in our setting will be clarified in section \ref{sec:U1-symmetry-susy}.
With the~topological twist, one may preserve the~same two supercharges (\ref{eq:Q-tilde-pm}) discussed before introducing the~M5.
As already stressed, these coincide precisely with $Q_r,\, \bar Q_r$ from \cite{Aganagic:2011sg}.

Finally, we come to the~object of main interest for us: the~partition function of 3d BPS states and its refinement.
These consist of open M2-branes ending on the~M5-brane, which give rise to BPS vortices of the~3d $\CN=2$ theory $T[L]$. 
The 3d index counting such BPS states is a~Witten index, again deformed by fugacities coupled to symmetries of the~theory that commute with $\tilde Q_\pm$ and therefore with $H$:
\be\label{eq:3d-index}
	Z(\fq,\ft) = {\rm Tr}_{{\cal H}[T_{3d}]} (-1)^{F} \fq^{S_1'} \ft^{-S_2'} a^\CQ x^\CT\,.
\ee
Here $\CQ$ is the~charge of a~global $U(1)$ symmetry of $T[L]$ associated to rotations of the~base $\C\IP^1$ of $X$ \cite{Fuji:2012nx}, and $\CT$ is the~charge of the~topological $U(1)$ symmetry associated to the~gauge $U(1)$ of $T[L]$ \cite{Aharony:1997bx}.\footnote{The fact that $T[L]$ is a~$U(1)$ gauge theory follows from the~fact that $b_1(L\approx S^1\times \IR^2)=1$. The~abelian 2-form on M5 reduced along the~1-cycle in $L$ gives rise to an abelian gauge field in $T[L]$.}
Contributions to the~index by states with $a^{2 k} x^{n}$ correspond to~M2-branes wrapping a~relative homology cycle labeled by $(k,n) \in H_2(X,L)$.
It is natural to switch to a~different basis for $\fq,\ft$ to distinguish between the~spin of a~BPS state (along $\IR^2_{\epsilon_1}$) and its R-charge (along transverse directions). Adopting (\ref{eq:qt-change}) yields 
\be\label{eq:qt-3d-index}
\begin{split}
	Z(q,t) 
	& = {\rm Tr}_{{\cal H}[T_{3d}]} (-1)^{2(J_1+J_2+J_I)} q^{2 (J_1-J_2)} t^{2(J_2+J_I)}\,  a^\CQ x^\CT\,,
\end{split}
\ee
where we lightened notation $J_I^3\to J_I$.
Spin $J_1$ along the~direction tangent to M5 is detected only by $q$, not~$t$. Conversely, the~R-symmetry charge $J_I$ is only detected by $t$, not~$q$.
When $L$ is a~knot conormal, this 3d index is expected to correspond to the~refined HOMFLY-PT generating series of $K$ (with symmetric colors) \cite{Ooguri:1999bv,Aganagic:2011sg}.

\section{Generalized quivers and HOMFLY-PT homology}\label{sec:generalized-quiver}

In this section we motivate and state our main conjectures on the~relation between HOMFLY-PT homology and generalized quivers. 

\begin{rmk}
One of the~implicit assumptions in the~original formulation of the~knots-quivers correspondence \cite{Kucharski:2017ogk,Kucharski:2017poe} is that each node of the~quiver has an~associated  change of variables where $x_i$ is directly proportional to $x$.
From a~geometric perspective, this means that each one of the~basic disks represented by nodes of $Q$ has a~boundary that wraps around the~longitude of $L_K$ exactly once.
It was noticed in \cite{Ekholm:2018eee} that this assumption is not always satisfied (see section \ref{sec:Examples} for concrete counterexamples), and an~extension of the~knots quivers correspondence was formulated \cite[Conjecture 1.1]{Ekholm:2018eee}. This conjecture was erroneously stated on the~level of the~refined partition function -- the~problem is that it does not say what the~refined contribution for higher level nodes would be. In this section we discuss such contributions and present a~new and complete version of the~conjecture. 

In more physical terms, the~basic disks corresponding to the~nodes of $Q$ are mutually linked and interact according to the~quiver adjacency matrix.  Together they generate the~whole  spectrum of BPS states corresponding to M2-branes wrapping embedded holomorphic curves  \cite{Ekholm:2018eee, Ekholm:2019lmb}. 
The quiver description expresses a~BPS state winding $d$ times around as a~bound state of $d$ copies of once-around basic disks, each corresponding to one of the~nodes of $Q$. As mentioned above, for knots that do not admit such descriptions there are BPS states corresponding to curves whose boundary wraps $d$ times around the~longitude of $L_K$, which cannot be realized as bound states of once-around curves. 
In these cases the~original formulation of the~knots-quivers correspondence fails. 
\end{rmk}

\subsection{Geometry and combinatorics of multiply-wrapped basic disks}
\label{sec:geom-combin}

In this section we study the~geometry of the~basic holomorphic disks that are the~nodes of our generalized quiver. We will first describe their geometry and then derive their multi-cover formulas. As mentioned above, quiver node curves appear for a~$U(1)$-symmetric configuration, when the~knot conormal lies on top of the~unknot conormal. Since the~unknot conormal supports no basic higher genus curves, we expect all quiver nodes to correspond to disks. The~simplest such curves are embedded disks and their multiple-cover formula is well-known.

Consider now an embedded holomorphic disk with connected boundary in class $x^{\mu}$ for $\mu>1$. We apply so-called Symplectic Field Theory (SFT) stretching around $L_K$, see \cite{EGH}. This is a~degeneration of the~complex structure under which holomorphic curves degenerate into several level holomorphic buildings, with parts near $L_K$ and parts far from $L_K$ joined at Reeb orbits in the~unit cotangent bundle of $L_K$. In the~case at hand, we pick a~metric on $L_K\approx S^1\times\R^2$ with only one simple closed geodesic, and correspondingly only two simple closed Reeb orbits which are the~unit cotangent lifts of this geodesic with its two orientations. Then, since the~unknot conormal has holomorphic curves that go in only one of the~two directions, all nearby curves are multiples of the~basic cylinder stretching from the~simple Reeb orbit to the~geodesic (with the~positive orientation) in the~zero section. In particular, our basic once-around disk becomes a~two level building consisting of an~outside sphere with puncture and an~inside disk with puncture. The~contribution to the~partition function comes from multiple covers of this curve together with constant curves attached along it. 

Since all curves near the~Lagrangian in the~stretched limit are multiples of the~basic punctured disk (stretching between the~Reeb orbit and the~unique geodesic in $S^{1}\times\R^2$), we find that the~lower part of the~holomorphic building in the~limit is a~$\mu$-fold cover of this basic curve. The~outside part of the~curve, the~upper level, is a~once punctured curve asymptotic to the~$\mu$-fold cover of the~basic Reeb orbit. The~curve before the~limit is assumed not to be a~multiple cover. Assuming that it is somewhere injective also in the~limit, this limit curve is generically embedded as well. The~simplest such two-level building has a~sphere with one puncture as its upper level. We define such buildings as our new $\mu$-times around basic curves, and note that for $\mu=1$ we simply get stretched versions of our previous basic disks.

\subsubsection{Generalized curves, the four-chain and deformations of M2-M5 configurations}

We next consider how these two-level buildings and their multiple covers glue and how they contribute to the~partition function counting generalized curves. Consider $d$-fold covers of the~levels. The~inside piece looks like a~disk with an~$U(1)$-action with a~fixed point of order $\mu d$ at the~Reeb orbit. The~outside piece looks like a~$d$-fold cover of a~sphere with a~single puncture. In Gromov-Witten counts of connected curves, a~$\mu d$-fold cover of a~disk with a~single fixed point contributes
\be 
\frac{1}{\mu d}\left(\frac{1}{1-q^{2\mu d}}\right).
\ee 
To see this, we use a~deformation argument. Consider first an embedded annulus stretching between two Lagrangians $L$ and $L'$, and with standard normal bundle. Assume that the~boundary of the~annulus in $L$ is homologically essential and lies in homology class $\log x\ne 0\in H_1(L)$, and that the~boundary in $L'$ is contractible. The~$d$-fold cover of $L$ then contributes
\[
\frac{1}{d}x^{d}.
\]
Assume now that the~boundary in $L'$ shrinks to a~point. This leaves a~disk with a~$d$-fold branch point. As in the~skein count \cite{ES}, this disk is an instance of a~1-parameter family of disks that crosses $L'$. 
The intersection of this disk with the~4-chain of $L'$ changes by $2d$ as they cross ($d$ positive intersections become negative). If the~disk contributes $\xi$, then invariance in the~$U(1)$-skein projected to homology and linking gives the~equation
\be\label{eq:M2-M5-skein-disk}
q^{2d}\xi - \xi = \frac{1}{d}x^{d},
\ee
and we find
\[
\xi =\frac{x^{d}}{d(1-q^{2d})},
\]
as claimed.

Consider now gluing on the~outside piece. This outside piece breaks the~symmetry, there are $\mu$ ways of gluing (here we glue first the~underlying curve and then take multiple covers) giving instead the~contribution
\be 
\mu\cdot \frac{1}{\mu d}\left(\frac{1}{1-q^{2\mu d}}\right) = \frac{1}{d}\left(\frac{1}{1-(q^{2\mu})^{d}}\right).
\ee 
Therefore the~contribution to the~Gromov-Witten partition function from a~single basic two-level configuration with boundary going $\mu$ times around the~generator is
\be 
\exp\left(\sum_{d>0}\frac{1}{d}\frac{(x^{\mu})^d}{1-(q^{\mu})^{2d}}\right) = (x^{\mu}, q^{2\mu})_{\infty}^{-1}.
\ee

\begin{figure}[h!]
\begin{center}
\includegraphics[width=0.5\textwidth]{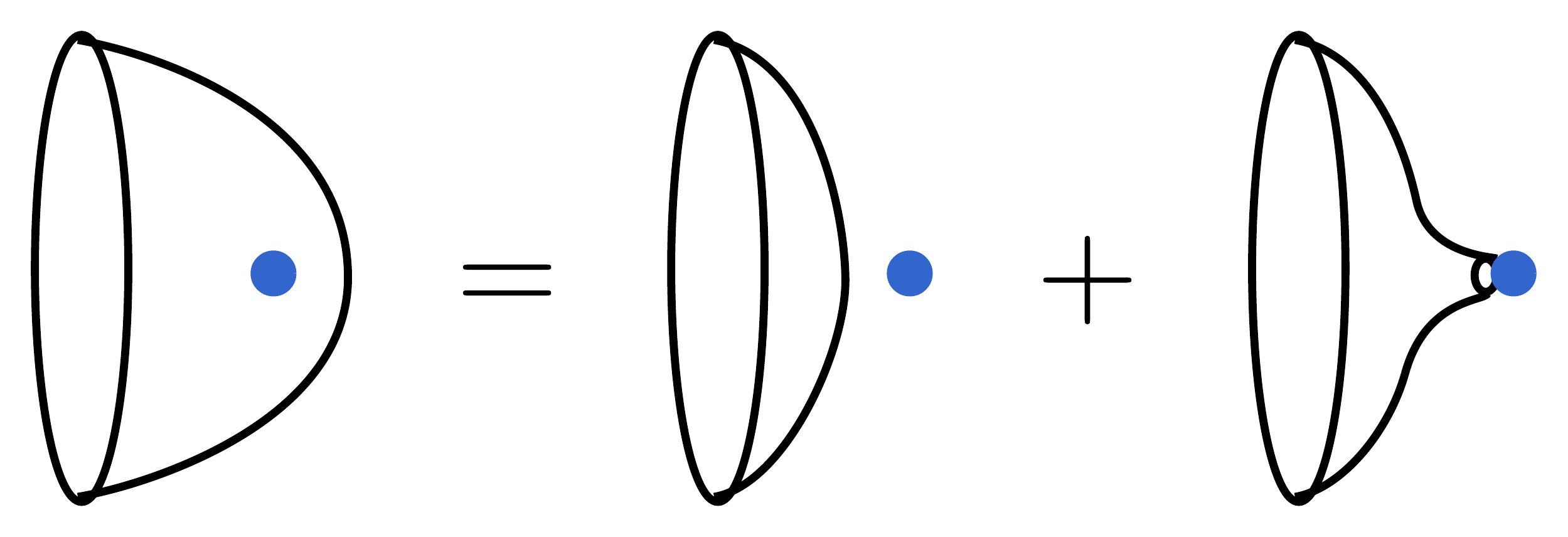}
\caption{Skein-like relation for an M2-brane wrapping an embedded holomorphic disk: when crossing an M5-brane, the~M2 breaks up and leaves behind a~new curve with a~new boundary on the~M5.}
\label{fig:M2-M5-skein}
\end{center}
\end{figure}

\begin{rmk}
Relation (\ref{eq:M2-M5-skein-disk}) has a~suggestive interpretation from a~physics viewpoint: the~left-hand side represents contributions of M2-branes wrapping holomorphic curves belonging to the~same family of deformations, and located on either side of an M5-brane on~$L'$. On the~right hand side, there is a~contribution of an annulus with a~boundary on $L'$.
This is a~kind of skein relation between M2 and M5-branes supported on higher-dimensional objects of total codimension one in the~Calabi-Yau, see figure \ref{fig:M2-M5-skein}. 
(The objects are linking from the~viewpoint of the~$A$-model on $X$, where M2 is represented by a~holomorphic curve and M5 by a~Lagrangian.
Lifting to M theory, the~M2 couples to the~three-form potential $A_3$ sourced by the~M5 by $j_{M5} \sim dF_4$ where $F_4=dA_3$ locally. 
Let $\tilde F_7=\star F_4$ be the~dual field in 11d; this is sourced by the~M2 current $j_{M2}$ with a~boundary term correction $d\tilde F_7 = j_{M2} + H_3 \wedge j_{M5}$, where $H_3$ is the~self-dual field strength on the~M5, which couples to the~M2 boundary \cite{Strominger:1995ac}.)
At the~level of (plethystic-)exponentiated partition functions, this relation is expressed by the~well-known property $(q^2x;q^2)_\infty = \frac{1}{1-x}(x;q^2)_\infty$. 
(This generalizes to $\mu$-times around curves, see (\ref{eq:M2-M5-skein-mu-times}))
Tracking 4-chain intersections is key for establishing this relation via Gromov-Witten counting.
Closing the~puncture corresponding to the boundary on $L$, yields a~similar relation between generating functions of M2-branes on spheres and disks $(q^2a^2;q^2;q^2)_\infty= (a^2;q^2)_\infty^{-1} (a^2;q^2;q^2)_\infty$, where $a^2$ is the~flux of M5-branes on $L$ linked by M2.
\end{rmk}

\subsubsection{Orbifold models of multiply wrapped disks}\label{sec:orbifolds}
As discussed above, two-level holomorphic buildings where the~inside (the part near the~Lagrangian) is a~punctured un-branched cover of a~punctured disk and where the~outside (the part far from the~Lagrangian) is an embedded punctured sphere are key objects in our generalized quivers.  
In order to understand the~behavior of such holomorphic objects, we can view them from inside and move the~embedded sphere part far away. This means shrinking the~outside punctured sphere together with the~Reeb orbit where it is attached to a~point and that gives an orbifold disk. In this section we study topological strings on orbifolds.

\begin{figure}[h!]
\begin{center}
\includegraphics[width=0.15\textwidth]{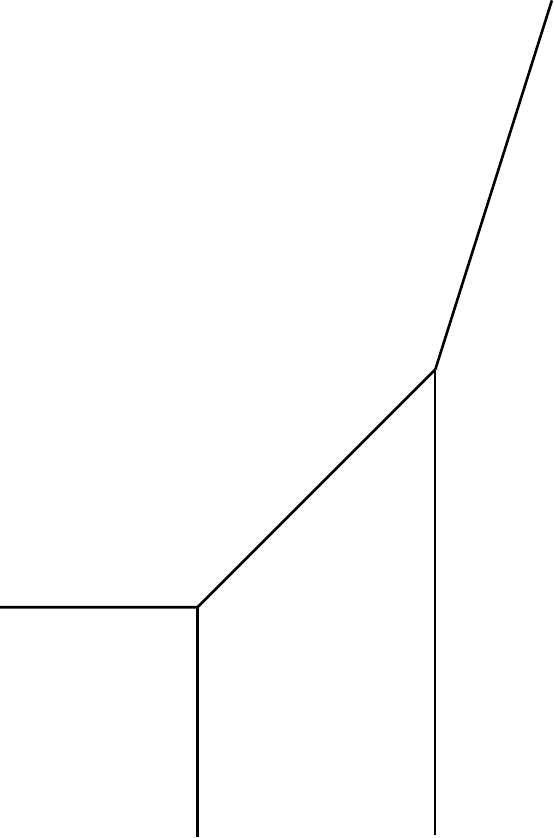}
\caption{Toric diagram of $\CO(0)\oplus  \CO(-2)\to \IC\IP^1$.}
\label{fig:C3-Z2-diagram-flipped}
\end{center}
\end{figure}

We start our discussion from the~following identity, relating the~partition function of a~$\mu$-times around basic disk to $\mu$ once-around basic disks, with specific $B$-field fluxes:
\be\label{eq:rootsofunity}
	(x^\mu;q^{2\mu})_\infty^{-1} = \prod_{n\geq 0}(1-x^\mu q^{2\mu n})^{-1} = \prod_{k=0}^{\mu-1}\prod_{n\geq 0}(1-\zeta^k xq^{2n})^{-1} = \prod_{k=0}^{\mu-1} (\zeta^{k} x;q^{2})_\infty^{-1}\,,
\ee
where $\zeta$ is a~primitive $\mu$-th root of unity. 
We will explain how such expressions appear from certain `fillings' of cylindrical versions of punctured orbifold singularities. We start in the~simplest non-trivial case, when $\mu=2$. Here we simply have
\be\label{eq:2-to-1-disks}
	(x^2;q^4)_\infty = (x;q^2)_\infty  (-x;q^2)_\infty \,.
\ee
This equation appears geometrically from the~toric brane in  $\CO(0)\oplus  \CO(-2)\to \IC\IP^1$ in a~certain limit. 
Consider the~toric diagram in figure \ref{fig:C3-Z2-diagram-flipped}. The~mirror curve is 
\be
	F(x,y) = 1 - y - (1 + Q) x + Q x^2  \,,
\ee
with a~single solution 
\be\label{eq:orbifold-classical-curve}
	y =(1 - x) (1 - Q x)\,.
\ee
In the~limit $Q\to e^{\pi i}=-1$ (the complexified K\"ahler modulus is purely $B$-field) this becomes 
\be
	1-y-x^2=0\,,
\ee
which is precisely the~classical limit of (\ref{eq:unrefined-quantum-curve}) with $C_{ii}=0$.

Switching from the~classical to the~quantum level, we consider a~toric brane on the~horizontal leg. 
Its partition function can be evaluated by topological vertex  techniques:\footnote{Compared to topological vertex conventions, we absorb inessential powers of $q$ in $x$ and $Q$ for cleaner expressions.}
\be
	Z(x) = (x;q^2)_\infty^{-1} (Qx;q^2)_\infty^{-1}\,.
\ee
This is annihilated by the~quantum curve 
\be
	\hat y = (1-\hat x)(1-Q \hat x)\,,
\ee
whose classical limit recovers (\ref{eq:orbifold-classical-curve}). Now an~orbifold limit  $Q\to -1$ turns the~curve into
\be\label{eq:C2-orbifold-curve}
	\hat y = (1-\hat x)(1+ \hat x) = 1-\hat x^2
\ee
and the~partition function collapses to
\be\label{eq:orbifold-partition-function}
	Z(x) \to (x;q^2)_\infty^{-1} (-x;q^2)_\infty^{-1} = (x^2;q^4)_\infty^{-1}\,.
\ee
This is precisely the~proposal for the~unrefined partition function  of a~twice-around  basic disk. Indeed,
\be
	(x^2;q^4)_\infty^{-1} = \sum_{k\geq 0}  \frac{x^{2k}}{(q^4;q^4)_k}
\ee
matches with the~form proposed in (\ref{eq:generalized-quiver-P}) with  $C_{11}=0$, $\mu_1=2$ and $x_1=x^2$.

In order to see the~geometry underlying these calculations, consider the~orbifold $\C\times (\C^2/(\mathbb{Z}/2\mathbb{Z}))$ and remove the~codimension four fixed point locus $\C\times [0]$, where $[0]$ denotes the~point in the~orbifold quotient which is the~image of $0\in\C^2$. We view the~resulting symplectic manifold as having a~negative end (the negative half of the~symplectization of standard contact $\R\IP^3$ times $\C$) near the~removed locus and note that there are two fillings: the~orbifold itself and $\mathcal{O}(-2)\oplus \mathcal{O}(0)$. In the~latter filling we view the~symplectic area of the~sphere as zero since the~filling is at negative infinity. We can now interpret the~limit $Q\to -1$ as this negative end splitting off from  $\mathcal{O}(-2)\oplus \mathcal{O}(0)$. When this happens, punctured versions of the~curves remain in the~punctured orbifold and these can be completed by curves in the~actual orbifold filling. As the~flux through the~sphere at negative infinity is set to $\pi i$, the~partition functions match. Geometrically, this can be interpreted as the~fact that upper level connected curves with odd asymptotics cancel, whereas those with even asymptotics add. This means that all non-zero curves can be filled with the~orbifold end or the~$\pi i$ flux $\mathcal{O}(-2)\oplus\mathcal{O}(0)$ and that the~curve counts agree. Thus, from the~viewpoint of curve counts in the~two-level symplectic manifold, the~two fillings give the~same result. Moreover, as the~filling $\mathcal{O}(-2)\oplus\mathcal{O}(0)$ moves toward negative infinity, the~moduli space of curves in the~upper level approaches the~corresponding space for the~orbifold. It is in this sense that taking the~limit $Q\to -1$ should be understood.   

\begin{figure}[h!]
\begin{center}
\includegraphics[width=0.35\textwidth]{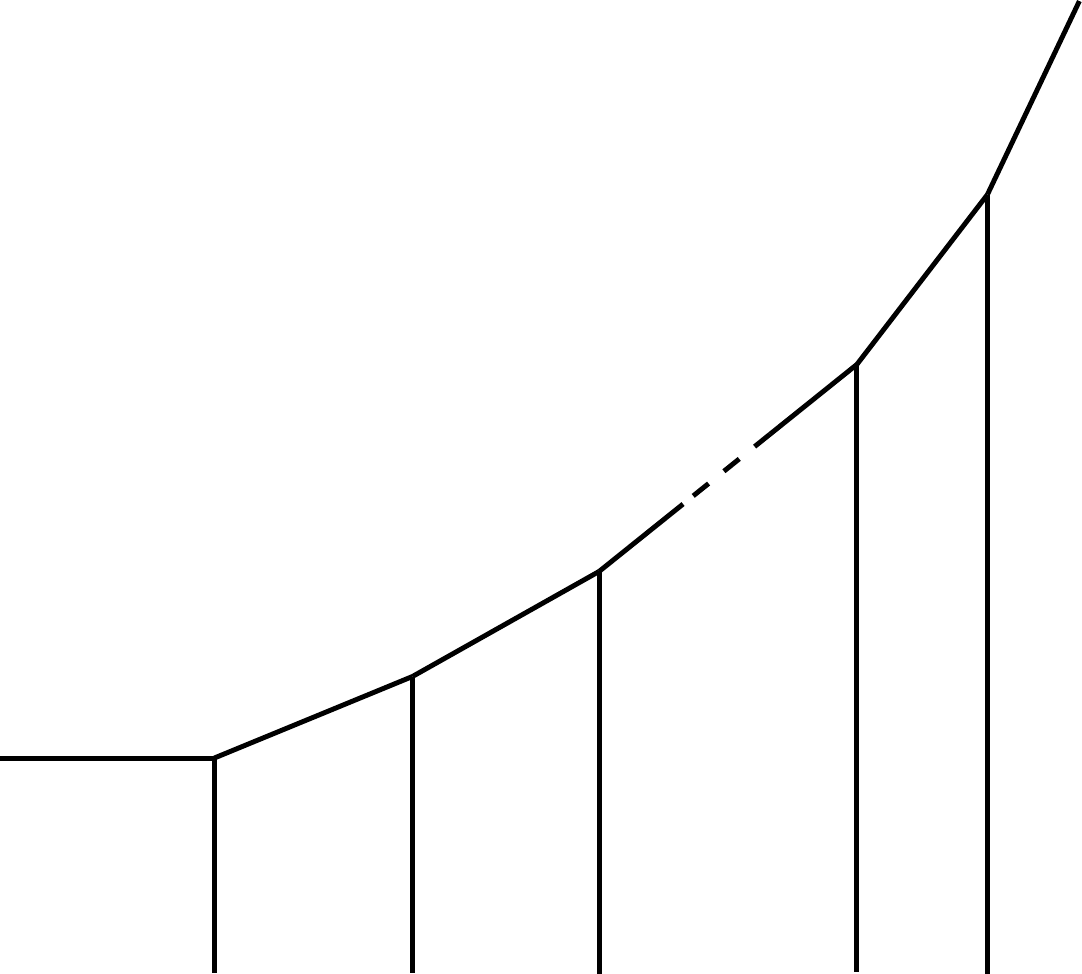}
\caption{Toric diagram of the~resolution of the~orbifold $\IC\times (\IC^2 / (\IZ/\mu \IZ))$.}
\label{fig:C3-ZN-diagram-un-flipped}
\end{center}
\end{figure}

Higher degree basic disks appear in more complicated orbifold quotients. The~corresponding toric diagram is shown in figure \ref{fig:C3-ZN-diagram-un-flipped} and gives the~mirror curve
\be
y=(1-x)(1-Q_1x)\dots(1-Q_{\mu-1}x).
\ee
In analogy with the~case $\mu=2$ above, we take the~limit $Q_k\to e^{\frac{2\pi i}{\mu} k}$ and get the~mirror curve of the~orbifold $\C\times(\C^2/(\mathbb{Z}/\mu\mathbb{Z}))$. 
\be\label{eq:muorbifold}
1-y-x^\mu =0.
\ee
On the~quantum level this reproduces the~identity~\eqref{eq:rootsofunity}. The~geometric interpretation is as above: we count two level curves in the~two level symplectic manifold. The~upper level of this manifold is the~orbifold with the~central $\C\times[0]$ removed, viewed as a~symplectic manifold with an end given by the~negative half of the~symplectization of the~standard contact lens space $L(\mu,1)$ times $\C$. The~filling of the~negative end is either the~orbifold itself or the~manifold given in figure \ref{fig:C3-ZN-diagram-un-flipped} with zero-area spheres, since they live at negative infinity, and with $B$-field fluxes $\frac{2\pi}{\mu} k$ so that curve counts are identical. The~convergence of moduli spaces works in direct analogy with the~$\mu=2$ case.  

\subsubsection{Semiclassical consequences of basic disk denominators}
It is worth noting that the~difference between the~standard denominators $(q^2;q^2)_{d_i}$ and $(q^{2\mu_i};q^{2\mu_i})_{d_i}$ we propose in (\ref{eq:generalized-quiver-P}) can be seen already at the~semiclassical level.
Consider a~single basic $\mu$-times around disk with no self-linking. Then the~partition function
\be
	P  =\sum_{d\geq 0}\frac{x^{d\mu}}{(q^{2\mu};q^{2\mu})_d} = (x^\mu;q^{2\mu})_\infty^{-1}
\ee
is annihilated by $1-\hat y - \hat x^\mu$, since
\be\label{eq:M2-M5-skein-mu-times}
	\hat y \cdot (x^{\mu};q^{2\mu})_\infty^{-1} = (q^{2\mu}x^{\mu};q^{2\mu})_\infty^{-1} = (1-x^\mu) (x^{\mu};q^{2\mu})_\infty^{-1}\,.
\ee
The classical curve is therefore
\be\label{eq:classical-curve-single}
	1-x^\mu - y=0\,.
\ee
On the~other hand, a~partition function like
\be
	\tilde P  =\sum_{d\geq 0}\frac{x^{d\mu}}{(q^{2};q^{2})_d} = (x^\mu;q^{2})_\infty^{-1}
\ee
would behave as follows:
\be
	\hat y \cdot (x^{\mu};q^{2})_\infty^{-1} = (q^{2\mu}x^{\mu};q^{2})_\infty^{-1} = (1-x^\mu)\dots(1-x^\mu q^{2\mu-2}) (x^{\mu};q^{2\mu})_\infty^{-1}\,.
\ee
The classical curve in this case is quite different:
\be
	(1-x^\mu)^\mu -y =0\,.
\ee
Therefore, in order to detect the~difference between the~two types of denominator, it is sufficient to compute the~unrefined contributions from genus-zero basic curves to the~augmentation polynomial.
In section \ref{sec:Examples} we will provide concrete examples where the~distinguished form of the~curve (\ref{eq:classical-curve-single}) appears.

\subsection{Holomorphic curves viewed as BPS generators}
\label{sec:Holomorphic  curves viewed as BPS generators}

The structure of HOMFLY-PT homology seems to indicate that it is more closely related to the~expansion of the~refined partition function of the~theory $T[L_{K}]$ in terms of equivariant vortices than to the~corresponding expansion in terms of basic holomorphic curves. This means that in order to extract homology information from basic holomorphic curves viewed as BPS generators, we must understand how to expand a~configuration of such objects as a~combination of vortices. In this section we discuss a~proposal for such an~expansion and its origins. We study the~expansions for single generators, first on the~unrefined and then on the~refined levels. The~discussion here is in a~sense localized near the~boundary of the~curve and independent of other charges, which on the~unrefined level comes from homology class, 4-chain intersections, self-linking, and -- on the~refined level -- also on a~certain framing density term that we discuss in section \ref{eq:geometric-t-degree-conjectural}. 

Consider a~basic holomorphic curve. As explained in section \ref{sec:geom-combin}, such a~curve consists of a~punctured embedded sphere on the~outside, asymptotic to a~multiple of the~unique Reeb orbit, and an unbranched multiple cover of the~basic cylinder stretching from the~simple Reeb orbit to the~geodesic in the~zero section on the~inside.

\begin{figure}[h!]
	\centering
	\includegraphics[width=0.7\linewidth]{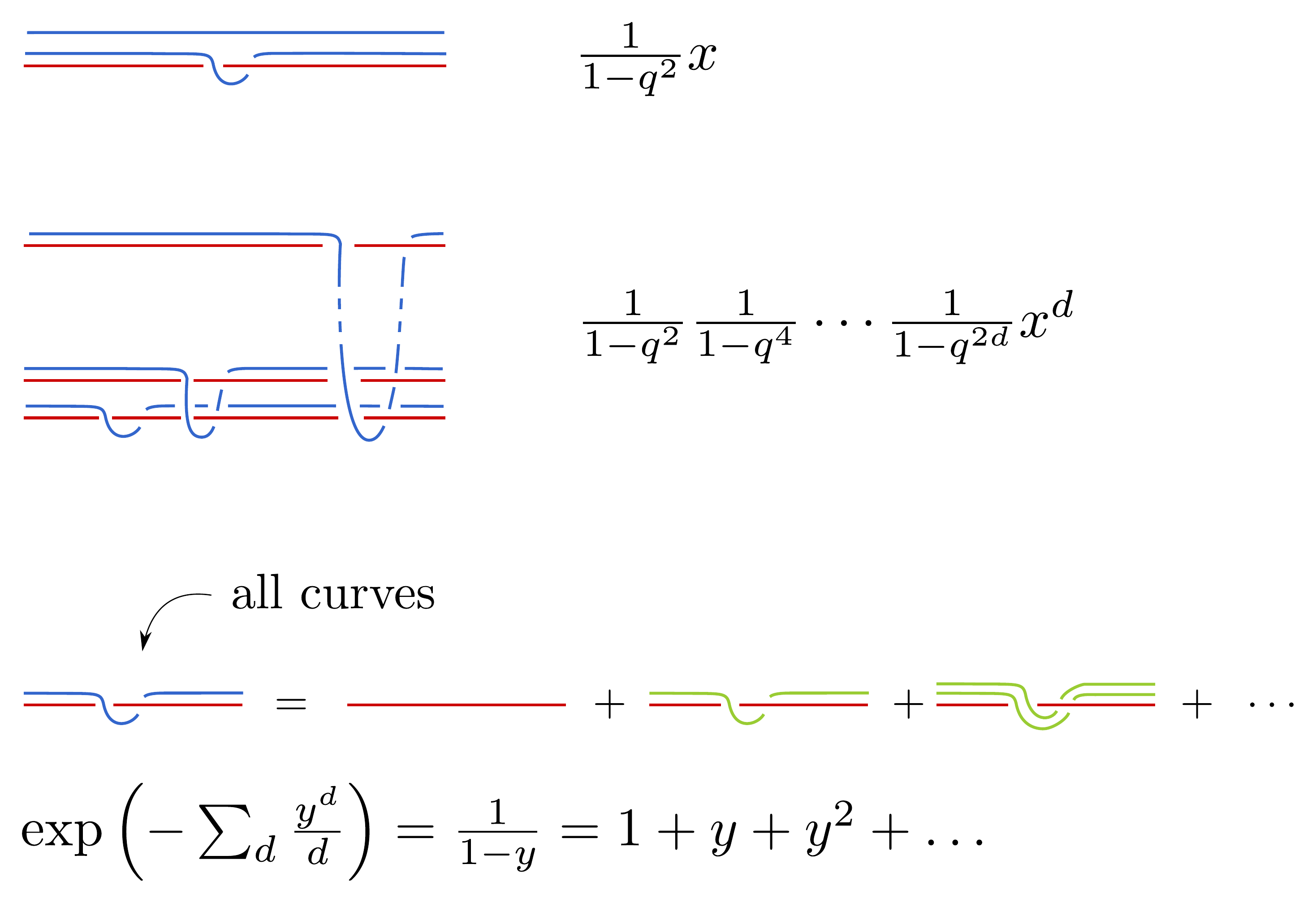}
	\caption{Contributions of pushed down constant annuli to generalized curve counts. Both boundaries of each annulus follow the~longitude, one of the~boundaries also wraps around a~small meridian, linking with the~boundary of the~$i$-th strand. The~overall holonomy is $y_i \cdot x\cdot x^{-1} = y_i$.}
	\label{fig:annuli}
\end{figure}

In particular, our basic once-around disk becomes a~two level building consisting of an~outside sphere with puncture and an~inside disk with puncture. The~contribution to the~partition function comes from multiple covers of this curve together with constant curves attached along it. Here we propose to push the~contributions from the~constants down to the~boundary. We observe that at the~level of generalized curves this can be done in a~specific way: the~contributions from constants correspond to very thin holomorphic annuli with one boundary component linking the~basic curve and the~other one not linking it. For a~$d$-fold multiple cover of a~basic disk, there is one such annulus on each level of magnification. The~first annulus links all strands, the~second all but one, etc., until the~last annulus which links only one strand, see figure \ref{fig:annuli}.

To support this picture, let us explain how it relates to the~more familiar count of multiple covers of a~single holomorphic disk with boundary on $L_K$.
Recall that $L_K$ carries a~$U(1)$ local system, with meridian and longitudinal holonomies at the~boundary denoted~$x,y$. These holonomies provide (exponentiated) Darboux coordinates for the~moduli space of abelian flat connections on $T^2\simeq \partial L_K$. We denote their deformation quantization by $\hat x, \hat y$, obeying the~relation $\hat y\hat x = q^2\hat x\hat y$.
Any disks ending on $L_K$ arise through the~large~$N$ geometric transition, descending from a~common holomorphic cylinder stretching between the~zero-section in $T^*S^3$ and $L_K$. As the~boundaries of this cylinder are the~original knot $K\subset S^3$ and the~longitude in $L_K$, 
its skein valued partition function will be
given by \eqref{eq:HOMFLY-PT series},
where $P_r(K;a,q)$ is the~HOMFLY-PT polynomial of $K$ in the~$r$-th symmetric representation, and
$x^r$ is the~$U(1)$ HOMFLY-PT skein element in $L_K$ projected to homology and linking.
For illustration, consider the~unknot in reduced normalization, for which $P_r(0_1;a,q) = 1$.
This gives the~following partition function of a~holomorphic disk
\be
	\psi(x)= P_{0_1}(x,a,q) = \sum_{r\geq 0} \frac{x^r}{(q^2;q^2)_r} = (x;q^2)_\infty^{-1}
\ee
From the~viewpoint of $U(1)$ Chern-Simons on $L_K$, boundaries of worldsheet instantons wrapping the~holomorphic disk give rise to infinite series of Wilson lines \cite{Witten:1992fb}. 
A Wilson line wrapping $r\geq 0$ times around $L_K$ with $k$ kinks on it contributes with $q^{2k} x^r$. The~number of such Wilson lines is $c_{k,r}$, defined by the~expansion
\be
	\psi(x) = \sum_{r\geq 0} \sum_{k\geq 0} c_{k,r} \, q^{2k} {x^r} \,.
\ee
The interpretation of $q^{2k}x^r$ as an $r$-times wrapped Wilson line with $k$ kinks can be seen by noting that kinks are counted by powers of $a$ as in figure~\ref{fig:framedskein}, and by noting that in $U(1)$ Chern-Simons $a=q$.
Incidentally, $c_{k,r}$ is also the~dimension of the~Hilbert space of BPS vortices of $T[L_K]$ with vorticity $r$ and spin $k$ \cite{Dimofte:2010tz}.

Kinks on the~boundary of a~Wilson line can be traded with linking with a~dual Wilson line using the~$U(1)$ skein algebra of variables $\hat x, \hat y$:
\be
	\frac{x^r}{(q^2;q^2)_r} = \, : \ \underbrace{\frac{1}{1-\hat y} \hat x \dots \frac{1}{1-\hat y} \hat x}_{r}\ :
\ee 
where $:\ :$ is the~normal ordering operation defined in \cite{Ekholm:2019lmb}.
Trading kinks with linking loops may be viewed as a~`half' of the~$U(1)$ skein relation on $L_K$: the~$\hat x$ Wilson line with a~small $\hat y$ loop around it can be viewed as the~over-crossing diagram in the~first line of figure \ref{fig:framedskein}, while the~kink corresponds to $q^2$ times the~smoothing also on the~first line.

Once again, this specific ensemble of Wilson lines can be given a~geometric interpretation in terms of curve counting.
The factor $\frac{1}{1-\hat y}$ corresponds to an annulus with a~boundary linking the~boundary of the~$x$-cylinder in $L_K$ once, and with the~other boundary linking zero times. 
Overall, the~partition function resembles an $x$-annulus ending on $L_K$, with a~linked $\hat y$-annulus on each strand of its multi-cover:
\be
	\psi(x) =\ :\ \sum_{r\geq 0} \left(\frac{1}{1-\hat{y}}\,\hat{x}\right)^{r} \ :  \ =\  : \ \frac{1}{1- \frac{1}{1-\hat{y}}\,\hat{x}} \ : 
\ee

The~corresponding recursion relation is 
\be 
\left(1 - \left(\frac{1}{1-\hat{y}}\,\hat{x}\right)\right)\psi  = 1\,,
\ee
equivalent, upon left-multiplication by $1-\hat{y}$, to the~more familiar
\be
    (1 - \hat{y} -\hat{x}) \psi = 0 \,.
\ee

As remarked above, a~single vortex would contribute $q^{2k}x^r$ corresponding to a~Wilson line.
On the~geometric side, we repackage Wilson lines into linked holomorphic annuli. On the~gauge theory side of $T[L_K]$, this repackaging corresponds to the~definition of equivariant vortices.

Given this, we now propose to think about an~equivariant vortex in the~theory $T[L_{K}]$ simply as a~configuration of the~form above: on level $d$, there are $d$ parallel copies of the~central curve linked in the~nested way by basic annuli, see figure \ref{fig:annuli}.

\begin{figure}
	\centering
	\includegraphics[width=0.6\linewidth]{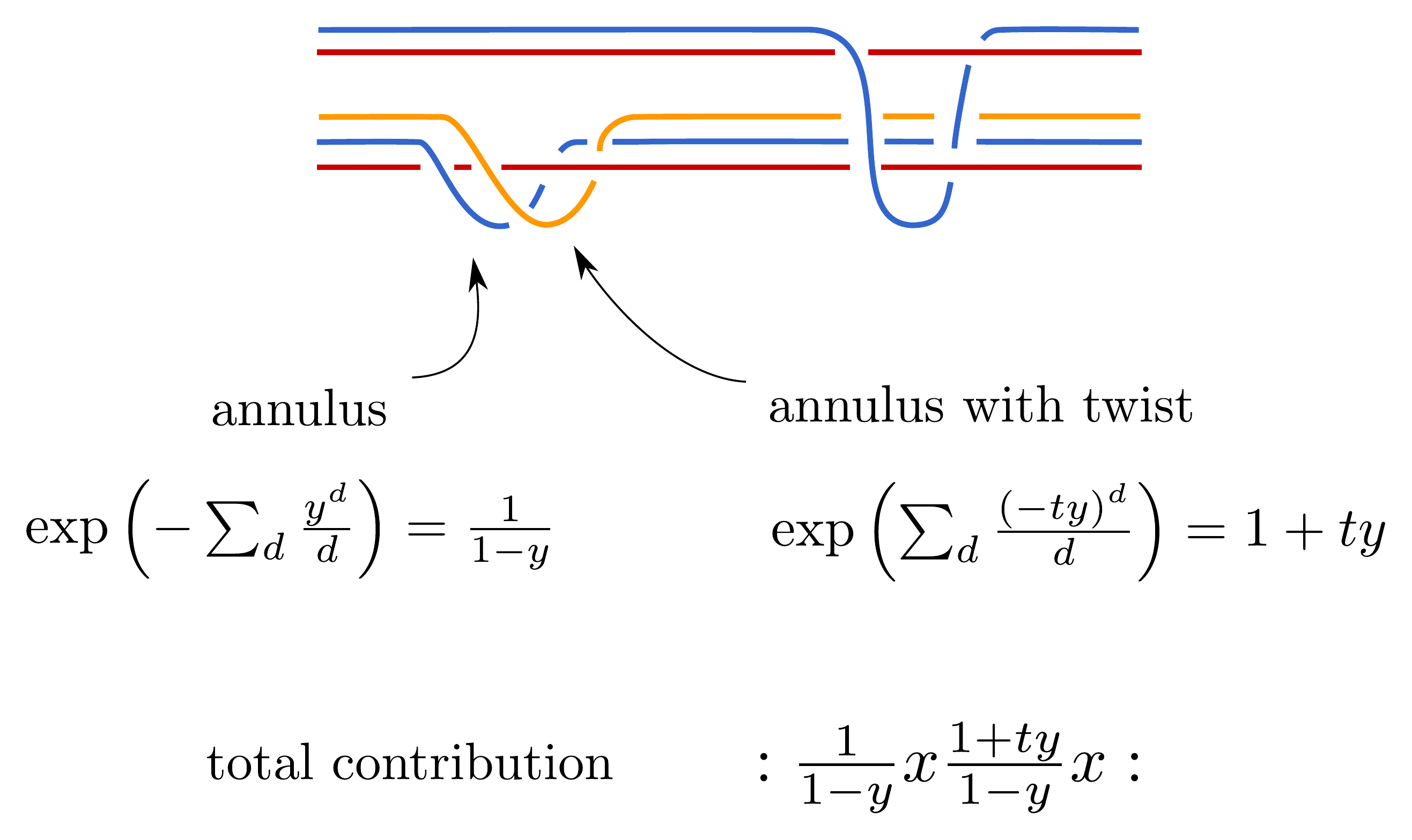}
	\caption{A degree 2 generator with constants pushed down.}
	\label{fig:twicearound}
\end{figure}

We next consider the~corresponding procedure applied to a~$\mu$ times around generator. We propose that such generators become multiples of $\mu$ with constants and anti-constants attached on all intermediate covers. The~constant and anti-constant push down to the~boundary as almost identical annuli, where the~second  annulus has a~twist in the~trivialization on the~boundary corresponding to $t$ or $t^{-1}$, depending on the~underlying half-framing of the~underlying punctured disk. Figure~\ref{fig:twicearound} shows the~configuration for a~twice-around generator.
The configuration for a~$\mu$ times around generator is clear and gives the~refined contribution
\be 
:\frac{1}{{1+t^{\pm1}\hat{y}}}\left(\frac{1+t^{\pm1}\hat{y}}{1-\hat{y}}\hat{x}\right)^{\mu}:.
\ee
The recursion relation for the~refined partition function of a~$\mu$-times around generator is then
\be 
\left(1-\frac{1}{{1+t^{\pm1}\hat{y}}}\left(\frac{1+t^{\pm1}\hat{y}}{1-\hat{y}}\hat{x}\right)^{\mu}\right)\psi = 1.
\ee    

\begin{figure}
	\centering
	\includegraphics[width=0.8\linewidth]{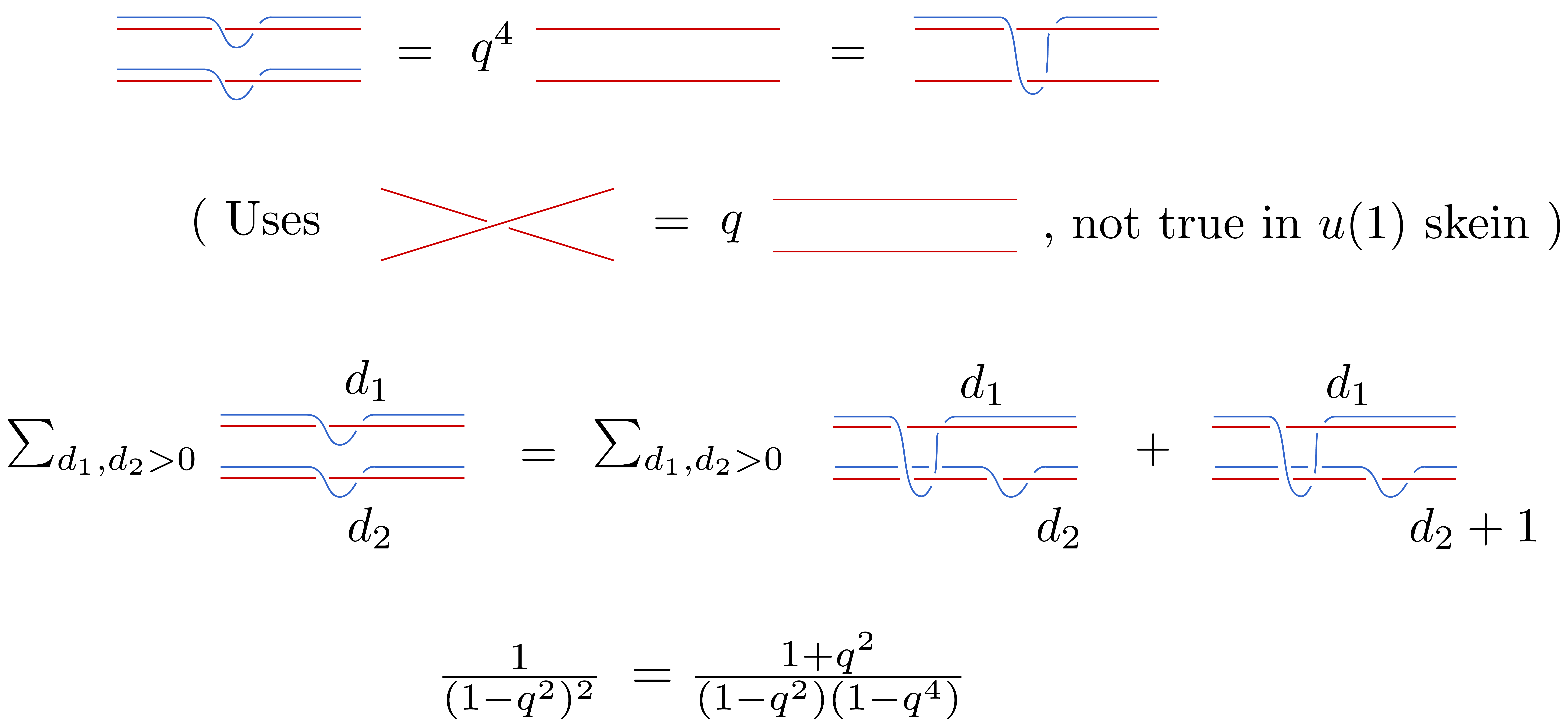}
	\caption{Contributions on second level from two unlinked basic disks.}
	\label{fig:quadratic}
\end{figure}

To illustrate the~procedure, we next consider the~second level of a~system of two unlinked basic disks. We must express the~configuration of two strands with an annulus linking each in terms of our standard basis, which leads to the~counting of intersections presented in figure \ref{fig:quadratic}.

We next look at this way of expanding states in terms of vortices and their Hilbert spaces. Consider an~equivariant vortex. Its partition function is given by
\be
\psi(x,q)=\sum_{d\ge 0}\frac{x^{d}}{(1-q^{2})\dots(1-q^{2d})}.
\ee
Expanding the~denominators using geometric series, it is convenient to think of the~Hilbert space at level $d$ as follows. Consider the~points in the~simplex $x_{1}\ge x_{2}\ge\dots x_{d}\ge 0$ with integer coordinates $(n_{1},\dots,n_{d})$. The~dimension of the~Hilbert space of states with $q$-charge $k$ is then the~number of points in the~intersection between the~simplex and the~plane $\sum_{j=1}^{d} x_{j}=k$. 

The Hilbert space of mixed states of two vortices on level two as above looks like the~integral points over the~first quadrant and can be expressed as two copies of the~simplex, one shifted by multiplication by $q^{2}$. Similarly, one generator going $\mu$ times around corresponds to the~points along the~diagonal $x_{1}=\dots=x_{\mu}$ and can be expressed as a~sum of shifted simplices. This indicates that refinement applies to the~Hilbert spaces associated to vortices individually.

We give one final perspective on vortices and homology. Viewing a~vortex as an~$U(1)$-invariant fixed point in $\mathbb{C}$, we may compute its contribution to the~partition function by Bott localization. Here we change coordinates and think of the~$d$-fold vortex as a~point in the~configuration space of $d$ points in $\mathbb{C}$ thought of as the~space of polynomials using the~relation between roots an~coeffcients. The~contribution
\be \label{eq:Morse}
\frac{1}{(1-q^{2})\dots(1-q^{2d})}
\ee    
now arises as the~equivariant Chern character. For a~$\mu$ times around generator, only the~top degree coefficients of the~polynomials are fixed, others are free to vary. This means we have a~corresponding $\C^{\mu-1}$ family of fixed points with similar action on normal bundles. The~contribution from the~corresponding Bott manifold is
\begin{equation}\label{eq:Bottvortex}
\frac{1}{(1-q^{2d})}\cdot [(\C\IP^1)^{d-1}],    
\end{equation}
where we have replaced $\mathbb{C}$ with $\C \IP^{1}$ and a~fixed point also at infinity and where when we pass from Bott to Morse we get the~contribution
\be \label{eq:Bott}
\frac{1+tq^2}{1-q^2}\dots\frac{1+tq^{2(\mu-1)}}{1-q^{2(\mu-1)}}\frac{1}{1-q^{2d}},
\ee 
corresponding to two fixed points in each $\C \IP^1$-factor.

\subsection{M2-branes wrapping holomorphic curves}
\label{sec:Homological degree of M2-branes}

In this section we discuss the~physical and geometric interpretations of refinement, connecting the~conjectures of \cite{Aganagic:2011sg} to our interpretation of the~knot-quiver correspondence \cite{Ekholm:2018eee, Ekholm:2019lmb}.
The 3d $\CN=2$  index (\ref{eq:3d-index}), proposed by \cite{Aganagic:2011sg}, corresponds to the~refined partition function of knot invariants.
However, this index is well-defined if not just one, but \emph{two} supercharges are preserved in the~Omega-background and with an~M5-brane inserted along $L\times S^1\times \IR^2$.
We will now provide motivation for the~existence of this additional supercharge, based on the~presence of a~certain geometric $U(1)$ symmetry, expanding on a~suggestion of \cite{Aganagic:2011sg}. 
Eventually, this will lead us to a~geometric interpretation of the~R-charges of BPS states in terms of `self-linking densities', consistent with our previous observations in \cite{Ekholm:2018eee}. Then, together with the~study in section \ref{sec:Holomorphic  curves viewed as BPS generators}, this gives our generalized quiver partition function.

The analysis of preserved supersymmetries that we develop here takes a~different perspective from the~one in section \ref{eq:Omega-background and refined indices}. 
There we started with supercharges of the~5d theory engineered by $X$ and examined which ones survive in presence of the~Omega-background, and eventually also in presence of the~M5-brane on $L\times S^1\times \IR^2$.
Here we shall start with the~worldvolume supersymmetry on the~M5 and analyze how it is broken by introducing the~Calabi-Yau background and by placing the~brane on $L$. An interesting novelty of this perspective is that it will clarify a~little-appreciated consequence of the~$U(1)$ symmetry postulated in \cite{Aganagic:2011sg}. As we will show, the~existence of a~$U(1)$ symmetry corresponds to a~point in the~moduli space of $T[L]$ with enhanced supersymmetry.

\subsubsection{Conserved supercharges in the~generic case}

We study M-theory on the~resolved conifold $X=\CO(-1)\oplus\CO(-1)\to \IC\IP^1$ times $S^1\times \IR^4$, with an~Omega-deformation turned on. This consists of rotations of the~planes $\IR^2\times\IR^2$ by independent  phases $\epsilon_{1,2}$ when going around the~$S^1$  \cite{Nekrasov:2002qd,Nekrasov:2003rj}.\footnote{Recall that these rotations arise from the~diagonal and anti-diagonal combination of Cartan generators of $U(1)_L\times U(1)_R \subset  SU(2)_L\times SU(2)_R \simeq SO(4)$, via the~identification $(g_L,g_R) = (e^{\epsilon_1 + \epsilon_2}, e^{\epsilon_1 - \epsilon_2})$.}

The Calabi-Yau background preserves eight supercharges.
This amount of supersymmetry is further reduced if we consider an~M5-brane wrapped on a~special Lagrangian submanifold $L\subset X$ times $S^1\times \IR^2_{\epsilon_1}$. Recall that a~special Lagrangian is a~Lagrangian calibrated by the~holomorphic top form on $X$, which must have a~constant phase along~$L$, the~phase determines which supercharges are preserved and which are broken by the~M5-brane. Under a~genericity assumption, such a~configuration preserves four supercharges. Let us  review how this works.

In a~flat space the~theory on an~M5-brane would be the~abelian 6d (2,0) SCFT, with 16 supercharges transforming as $({\bf 4, 4})$ spinors under $Spin(1,5)\times Spin(5)$, with the~second factor corresponding the~R-symmetry group.
With the~Calabi-Yau background, the~latter is broken to $Spin(3)\times Spin(2)$, where $Spin(3)$ is a~subgroup of the~local rotations on $X$ which leave $L$ fixed, and $Spin(2) \simeq U(1)_2$ is the~group of rotations of $\IR^2_{\epsilon_2}$.
The (2,0) 6d supercharges thus decompose as follows:
\be\label{eq:M5-Q-decomp}
	({\bf 4;4}) \to ({\bf 2,2;2})_{+1/2} \oplus ({\bf 2,2;2})_{-1/2} 
\ee
as representations of 
\be\label{eq:spin15xspin5-decomp}
	\underbrace{Spin(3)_{TL} \times Spin(1,2)_{S^1\times \IR^2_{\epsilon_1}}}_{\subset Spin(1,5)} \times \underbrace{Spin(3)_{NL} \times Spin(2)_{\IR^2_{\epsilon_2}}}_{{\subset Spin(5)}}\,.
\ee
The holonomy on $X$ belongs to an $SU(3)\subset SU(4)\simeq Spin(6)$ subgroup. This induces nontrivial holonomies for both $Spin(3)_{TL}$ and $Spin(3)_{NL}$, further reducing the~number of conserved supercharges.  
The special Lagrangian property of $L$ implies that holonomies of the~tangent and normal directions to $L$ are related.
This means that to any given loop $\gamma\in \pi_1(L)$ one may associate holonomies for both $NL$ and $TL$, denoted respectively  $g_{TL} \in Spin(3)_{TL}$ and $g_{NL} \in Spin(3)_{NL}$. 
Due to the~fact that $L$ is a~special Lagrangian of $X$, these group elements are in fact related to each other, allowing one to perform a~topological twist by considering the~diagonal subgroup $Spin(3)_{dL} \subset Spin(3)_{TL}\times Spin(3)_{NL}$. 
Then, spinors in the~$({\bf 2,2})$ transform as ${\bf 2\otimes 2 = 1\oplus 3}$,  implying that one out of four components is invariant under holonomy. This leaves four conserved supercharges, which transform as
\be\label{eq:4-generic-supercharges}
	({\bf 1,2})_{+1/2} \oplus ({\bf 1,2})_{-1/2} 
	\quad\text{under}\quad
	Spin(3)_{dL} \times Spin(1,2)_{S^1\times \IR^2_{\epsilon_1}} \times Spin(2)_{\IR^2_{\epsilon_2}}\,.
\ee
These are the~four supercharges of 3d $\CN=2$ theory $T[L]$ arising on the~M5-brane worldvolume, along directions $S^1\times \IR^2_{\epsilon_1}$. This amount of supersymmetry can be preserved under {generic} conditions under our assumptions, namely for an~M5-brane wrapping any special Lagrangian $L$ in any Calabi-Yau threefold $X$.

\subsubsection{Supersymmetry enhancement from a~geometric \texorpdfstring{$U(1)$}{U(1)}-symmetry}\label{sec:U1-symmetry-susy}
Following \cite{Aganagic:2011sg}, the~definition of a~refined index requires an~additional supercharge in addition to those in (\ref{eq:4-generic-supercharges}). In view of this we make the~following assumption. 

\begin{ass}{}\label{eq:assumption-U1-symmetry}
The Lagrangian $L$ is diffeomorphic to $S^1\times \IR^2$ and there exists a~$U(1)$-action on $L$ which rotates the~$S^1$-factor. The~corresponding vector field $V$ defines a~normal plane $\IR^2_V$ at each point. Then $\partial \IR^2_V$ is the~contractible meridian on the~torus $\partial L$.
\end{ass}

\medskip
The embedding $TL\subset TX$ defines a~dual vector field $JV$ normal to $L$ ($J$ is the~complex structure on $X$).
We view $JV$ as the~direction in which an~M2-brane wrapping a~holomorphic curve in $X$ attaches to $L$.
(If the~M2 boundary lies along a~flow line of $V$ in $TL$, and the~M2 wraps a~holomorphic curve in $X$, its normal is tangent to a~flow line of $JV$ in $J \cdot TL\simeq NL$, using the~local splitting $TX\simeq TL\oplus NL$.)
Let $\IR^2_{NL}$ be the~family of planes normal to $JV$ within $NL$.
Now we come to the~assumption: the~existence of a~$U(1)$-action implies that the~holonomy $g_{TL}\in Spin(3)_{TL}$ is actually \emph{not generic}, but belongs to a~subgroup $Spin(2)_{V}$ that rotates the~planes $\IR^2_{TL}$ tangent to $\IR^2_{V}$ as one goes around $S^1$ corresponding to the~boundary of a~holomorphic curve. 
Since $L$ is special Lagrangian, $g_{TL}=g_{NL}$ and therefore the~holonomy induces a~rotation of $\IR^2_{NL}$. 
The holonomy $g_V = g_{TL}=g_{NL} \in Spin(2)_{dL}\subset Spin(3)_{dL}$, so it rotates both summands in $\IR^2_{TL} \oplus \IR^2_{NL}$ by the~same amount, leaving fixed the~`origin' $S^1\times \IR \subset TL\times NL$, corresponding to the~tangent and normal directions of a~holomorphic curve wrapped by M2 attaching to $L$.

Our proposal is to identify $U(1)_I \equiv Spin(2)_{NL}$
as the~R-symmetry of the~5d theory $T_{5d}[X]$ that is necessary to perform a~topological twist to preserve additional supercharges leading to a~refined index, recall the~discussion from section \ref{eq:Omega-background and refined indices}. We will elaborate on the~details shortly.\footnote{This is a~subgroup $U(1)_I\subset SU(2)_I$ of the~R-symmetry group of the~4d $\CN=2$ theory on $\IR^4$ arising from reduction on the~M-theory circle. It corresponds to the~R-symmetry employed by \cite{Nekrasov:2002qd,Nekrasov:2003rj} in the~topological twist to define refined partition functions of 4d $\CN=2$ gauge theories.}

First, we wish to stress that $U(1)_I$ is an~artifact of the~string theory setup, and not a~property of a~garden-variety 3d $\CN=2$ theory. 
On the~one hand, while surviving 3d $\CN=2$ spinors are invariant under $Spin(2)_{dL}$, they transform nontrivially under $U(1)_I$.
On the~other hand, $U(1)_I$ is \emph{not} an~automorphism of the~3d $\CN=2$ super-Poincar\'e algebra, i.e., it is not an~R-symmetry of the~3d theory.
To establish these claims, recall that surviving spinors arise from the~branching rule ${\bf 2\otimes 2 = 1\oplus 3}$ for $Spin(3)_{dL}\subset Spin(3)_{TL}\times Spin(3)_{NL}$. Also recall that, following the~above assumption, we identified a~distinguished $Spin(2)_{TL}\times Spin(2)_{NL}\subset Spin(3)_{TL}\times Spin(3)_{NL}$, whose generators will be denoted $J^3_{TL}, J^3_{NL}$.
Working in a~basis where $J^3_{TL}, J^3_{NL}$ are diagonal, we may reclassify the~spinors in (\ref{eq:M5-Q-decomp}) 
as follows:
\be
	({\bf 4;4}) \to (\pm 1/2,{\bf 2};\pm 1/2)_{+1/2} \oplus (\pm 1/2,{\bf 2};\pm 1/2 )_{-1/2} 
\ee
as representations of of the~following subgroup of (\ref{eq:spin15xspin5-decomp})
\be
	Spin(2)_{TL} \times Spin(1,2)_{S^1\times \IR^2_{\epsilon_1}} \times Spin(2)_{NL} \times Spin(2)_{\IR^2_{\epsilon_2}}\,.
\ee
Topological twisting with reduced holonomy $Spin(2)_{TL}\times Spin(2)_{NL} $  implies retaining those supercharges that are invariant under the~diagonal subgroup. In this case, these are the~ones with opposite charges $j_V = -j_{JV}$ under $Spin(2)_{TL}\times Spin(2)_{NL}$:
\be
	(\pm 1/2,{\bf 2};\mp 1/2)_{+1/2} \oplus (\pm 1/2,{\bf 2};\mp 1/2 )_{-1/2} \,.
\ee
This leaves a~total of four plus four conserved supercharges, twice the~generic amount in  (\ref{eq:4-generic-supercharges}). 
To lighten notation,  we will sometimes omit the~representation labels under $Spin(1,2)_{S^1\times \IR^2_{\epsilon_1}} \times  Spin(2)_{\IR^2_{\epsilon_2}}$, and simply use $(j_V, j_{JV})$.

Among these, we can recognize the~singlet ${\bf 1}$ that survived in the~general case (\ref{eq:4-generic-supercharges}). This must be a~linear combination proportional to $(1/2,-1/2) - (-1/2,1/2)$, since it arose as the~second anti-symmetric power of ${\bf 2}$ for the~diagonal subgroup $Spin(3)_{dL}$.
Overall, there are four supercharges in the~singlet: a~${\bf 2}$ of  $Spin(1,2)$ with $j_2=+1/2$  and another with $j_2=-1/2$.
There is also a~second combination of supercharges that is conserved, namely $(1/2,-1/2) + (-1/2,1/2)$. While this is part of the~triplet ${\bf 3}$ and therefore not invariant under the~generic $Spin(3)_{dL}$, it is nonetheless invariant under the~reduced $Spin(2)_{dL}$.  Again, this corresponds to  four supercharges.
Therefore, our assumptions on the~geometry of $L$ and the~existence of a~$U(1)$-action imply doubling the~supersymmetry
from 3d $\CN=2$ in the~generic setting to 3d $ \CN=4$ in presence of $U(1)$.\footnote{This fact was observed also in \cite{Aganagic:2011sg} albeit somewhat implicitly. From a~physics viewpoint, this enhancement is also plausible from the~viewpoint of the~theory $T[L]$ of the~unknot conormal. This is $N_f=1$ SQED, which up to a~(neutral) adjoint chiral can indeed be viewed as a~3d $\CN=4$ theory in disguise. It is unclear if this enhancement is visible in the~theory $T[L]$ for other knot conormals.}

For Lagrangians $L=S^1\times\R^2$ as above this has the~following consequences. From the~point of view of the~theory $T[L]$ in the~generic case, without the~additional $U(1)$-symmetry, the~Hilbert space of states is generated by vortices  and contains states of fixed vorticity $x^r$ and fixed spin in the~$\R^2_{\epsilon_1}$-plane, $q^{2s}$. The~partition function is a~(super)trace which for fixed vorticity $x^r$ gives a~Laurent series in $q^{2}$. As the~state space is generated by vortices, the~coefficients stabilize as the~power of $q^2$ goes to infinity, and the~coefficient of $x^r$ can be expressed as $P_r(q^{\pm 2})(q^2;q^2)_r^{-1}$, where $P_r$ is a~polynomial with integer coefficients. With additional $U(1)$-symmetry, there is an additional $t$-charge that refines this theory and its states. The~Hilbert space is now a~sum of finite dimensional vector spaces of fixed charges $x^rq^{2s}t^l$. As before, after fixing the~vorticity $x^r$ the~Poincar\'e polynomial of the~vector spaces stabilizes and we can write it as $\mathcal P_r(q^{\pm 2},t^{\pm 1})(q^2;q^2)_r^{-1}$, where $\mathcal P_r$ is a~polynomial with positive integer coefficients and such that $\mathcal P_r(q^{\pm 2},-1)=P_r(q^{\pm 2})$. 

\subsubsection{Omega-background and 3d  index}

Now we come back to the~definition a~refined 3d index (\ref{eq:3d-index}). 
First we note that, while surviving supercharges are not neutral under $Spin(2)_{NL}$, neither do they transform in  a~representation of  $Spin(2)_{NL}$. In this sense $U(1)_I$ is not a~symmetry of the~3d $\CN=2$ superalgebra.
For this reason we consider linear combinations of the~two conserved supercharges. In particular, one may view $(1/2,-1/2)$ and $(-1/2,1/2)$ as being {separately} conserved. 
With this change of basis,  supercharges transform as representations of $U(1)_I$: the~first batch with charge $-1/2$, the~second batch with charge $+1/2$.
Overall, the~surviving supercharges are $({\bf 2})_{j_V, j_2}$, where $j_{V} = -j_{JV}$ and $j_{V},j_{2}\in \{-1/2,+1/2\}$ are chosen independently. Note that $j_I \equiv j_{JV} = - j_V $ is the~charge under $U(1)_I$.

The~point of having supercharges with a~well-defined charge under $U(1)_{I}$ is that one can turn on the~Omega-background and use the~$U(1)_I$ remnant of the~5d R-symmetry to perform a~topological twist. 
The Omega-background means that as we go around the~M-theory circle, we perform a~rotation of $U(1)_1\times U(1)_2\times U(1)_I$. In particular, this breaks $Spin(1,2)_{S^1 \times \IR^2}$ to $U(1)_1$ and spinors $({\bf 2})_{j_V, j_2}$ get labeled by their $U(1)$ charges $(j_1, j_2, j_I)$.
In this notation, the~surviving supercharges are those with $j_1=j_2=-j_I$, as derived in (\ref{eq:Q-tilde-pm}).
This can be shown to match precisely with the~supercharges considered in \cite{Aganagic:2011sg}.

\subsection{R-charge, homological degree, and self-linking}\label{eq:geometric-t-degree-conjectural}
We summarize the~geometry near the~Lagrangian considered above. The~Lagrangian itself is $S^1\times\R^2$ and a~neighborhood of it in the~Calabi-Yau looks like its cotangent bundle $T^\ast (S^1\times\IR^2)$. We furthermore have a~$U(1)$-action that rotates the~$S^1$-direction, and which splits the~normal bundle of the~Lagrangian as $NL=\R\times \IR^2_{NL}$. We can therefore write a~neighborhood of the~locus of the~M5-brane as
\be
(S^1\times\IR) \times (\IR^2_{TL}\times \R^2_{NL})\times \IR^2_q \times\IR^2_{q^{-1}t} \times S^1,
\ee
where normal directions lie in $\IR\times \IR^2_{NL}\times \IR^2_{q^{-1}t}$. Here $\IR^2_q$ and $\IR^2_{q^{-1}t}$ come from \eqref{eq:qt-3d-index}: they are the~planes that are rotated by $J_1$ and $J_2$, respectively. 
The topological twist reduces the~structure group of this bundle to a~diagonal $U(1)$ rotating both factors $\IR^2_{NL}\times \IR^2_{q^{-1}t}$ in the~same way.

\subsubsection{Once-wrapped disks}\label{sec:once-wrapped}
We next read off the~$t$-charge of an M2-brane wrapped on $(S^1\times\IR)\times S^1$. In the~Calabi-Yau space this is a~holomorphic curve along $S^1\times\IR$. By holomorphicity, its normal variation along the~boundary is given by a~vector field $(v,Jv)\in \IR^2_{TL}\times\IR^2_{NL}$. 
This implies that, going around a~loop in $S^1$, the~vector field $v$ rotates in $\IR^2_{TL}$ by an amount that equals that of $Jv$ in $\IR^2_{NL}$. 
The former furthermore equals the~self-linking $C_{ii}$ in $L$. By the~topological twist the~rotation of $Jv$ contributes to the~$t$-degree and it follows that the~corresponding $t$-charge of a~$d$-fold cover is $q^{-C_{ii}d}(-t)^{C_{ii}d}$, in line with the~conjectural identification of $C_{ii}$ and the~homological $t$-degree \cite{Kucharski:2017poe,Kucharski:2017ogk}.    

We point out that the~origin of the~contribution $q^{C_{ii}d^2}$ to the~partition function is different: it comes from counting $q$-power in the~$U(1)$-skein and then passing to homology and (self-)linking. 
In other words, the~quadratic growth of powers of $q$ corresponds to the~contribution to the~Gromov-Witten invariant from counting generalized holomorphic curves. 
In physics this corresponds to the~open topological string partition function with an $A$-brane on $L$, 
where $q$ arises through the~M-theory interpretation of the~string coupling constant \cite{Ooguri:1999bv}. 
There are $C_{ii}$ crossings between $d$-fold covers of an underlying once-around curve which gives $q^{C_{ii} d^2}$. Overall we have  contribution $q^{C_{ii}d^2-C_{ii}d}(-t)^{C_{ii}d}$ for the~$d$-fold cover of a~once-around disk. 
We stress that this reasoning is based on a~picture of M2-branes wrapping $d$ times around, viewed as tightly packed copies of an underlying once-around disk.   

Our prescription should recover the~contribution to the~unrefined partition function in the~limit $t=-1$. 
On this level we should perturb multi-covers of the~underlying curve and then count generalized curves. 
The geometry behind the~count is the~following. A~contribution to $C_{ii}$ can be thought of as an infinitesimal kink. 
After perturbation we find an actual kink with $d^2$ crossings; in order for this to be $d$ times a~curve in framing~$C_{ii}$, we deform to make all innermost kinks very small and then pull them out, making them infinitesimal again and trading them for framing.
This leaves $d(d-1)$ crossings, and summing contributions from all $C_{ii}$ infinitesimal twists we eventually get $q^{C_{ii}d(d-1)}$. This agrees with the~specialization to $t=-1$. 

\begin{rmk}

To round off our discussion on the~geometric interpretation of the~homological degree of BPS states, we offer yet another reason why such geometric information (the~`wiggling' of an~M2-brane boundary) should play a~role in the~definition of the~BPS partition function. 
M2-branes wrapping holomorphic curves produce Wilson lines for the~$U(1)$ Chern-Simons theory on $L$.
From the~viewpoint of the~full 6d $\CN=$(2,0) theory describing the~worldvolume dynamics of a~single M5-brane, such Wilson lines correspond to reduction of Wilson surfaces. 
By supersymmetry, these involve integration along the~boundary of the~M2 of the~abelian 2-form connection, as well as the~five scalars. In particular, the~latter transform as a~vector under the~$Spin$(5) R-symmetry of the~(2,0) theory. 
Recalling that R-symmetry corresponds to rotations transverse to the~brane, 
the~winding number of these scalars around the~brane must correspond to their charge under R-symmetry. 
At the~same time, the~scalars parametrize embeddings of M5 in the~transverse directions, and in particular they parametrize the~direction along which the~M2 ends on~M5.
Yet again, this suggests that the~wiggling of the~M2 boundary is captured by the~R-charge of a~BPS state, which in the~setup involving $L\subset X$ is eventually identified with the~homological degree. 

\end{rmk}

\subsubsection{Multiply-wrapped disks}
Our suggestion for refinement for multiply wrapped disks have two sources. The~first arises when we express the~contribution of the~curve in terms of standard vortices, or geometrically in terms of a~specific link as described in section \ref{sec:Holomorphic  curves viewed as BPS generators}. Here, internal $t$-charge of the~M2-brane is transferred to thin annuli and appears as a~difference in framing along their two boundary components. The~second comes from an external framing analogous to the~framing of the~standard basic disks discussed above. 

To describe this external $t$-degree, we refer to section \ref{sec:orbifolds} where multiply wrapped disks are related to disks in orbifolds and consider an M2-brane wrapped on a~disk in a~degree $\mu$ orbifold corresponding to \eqref{eq:muorbifold}. 
The $t$-charge arises from the~$U(1)$-action on~$L$, and the~M2-brane is invariant under this action (rotation along $S^1$) only if its boundary remains a~tightly wrapped $\mu$-fold cover of $S^1\times \text{point}$. 
This means that its boundary gives a~single Wilson line of charge $\mu$ rather than a~$\mu$-times around Wilson line of charge $1$. 
As in section \ref{sec:once-wrapped}, by holomorphicity and the~topological twist that identifies rotations in $\IR_{NL}$ and $\IR^2_{q^{-1}t}$, the~$t$-degree is the~framing of this charge-$\mu$ Wilson line, which in turn is determined by its self-linking. We define $C_{ii}$ to be the~framing density along the~$\mu_i$-fold boundary of the~M2-brane, thus $C_{ii}\in\frac{1}{\mu_i}\mathbb{Z}$. This means that a~single twist of the~underlying simple curve corresponds to $C_{ii}=\frac{1}{\mu_i}$. 
The total twist in the~normal bundle is then $C_{ii}\mu_i \in \IZ$, which leads to the~charge $x_i^d\sim q^{-C_{ii}\mu_id}(-t)^{C_{ii}\mu_id}$ for a~$d$-fold cover of the~$\mu$-wrapped M2-brane. 
Next we consider the~self-linking contribution to $q$-degree for generalized curves invariant under the~covering transformation. We think of the~framing $C_{ii}\mu_i$ of the~underlying curve as a~collection of infinitesimal kinks. 
These lift to $C_{ii}\mu_i^2$ kinks, giving the~total contribution to the~$q$-degree $q^{C_{ii}\mu_i^2 d^2}$ at the~$d$-fold level of the~brane. In summary we then have contributions proportional to $q^{C_{ii}\mu_i^2d^2-C_{ii}\mu_i d}(-t)^{C_{ii}\mu_id}$ at the~$d$-fold covering level.

Similarly to the~once-around case, we consider the~$t=-1$ specialization of this formula and the~contribution to generalized curves after perturbation. 
As above we consider the~lift to the~$d$-fold cover. 
A $\mu_i$-wrapped curve with $C_{ii}\mu_i^2$ infinitesimal kinks gives $C_{ii}\mu_i^2 d^2$ crossings at the~$d$-fold covering level, after perturbation.
In order to lift a~curve $d$ times around a~curve with framing $C_{ii}\mu_i$,
we trade the~innermost kinks at $C_{ii}\mu_i$ of the~crossings 
for framing, 
leaving a~net $C_{ii}\mu_i^2d^2-C_{ii}\mu_i d$ crossings and hence an overall contribution of $q^{C_{ii}\mu_id(\mu_id-1)}$.   
   
We next consider a~degree $\mu_i$ basic disk on a~more general Lagrangian of topology $S^1\times\R^2$. Here, as before, we get a~$U(1)$-symmetric boundary of the~disk by SFT-stretching. In the~limit where the~embedded disk is split off, the~lower piece close to the~Lagrangian can be identified with the~complement of the~orbifold point in an orbifold disk. 
The addition of the~orbifold point corresponds to adding the~embedded disk `at infinity'. 
Again, the~existence of a~$U(1)$-action on $L$ implies that all boundaries of holomorphic curves should sit very close to each other, within an~infinitesimal neighborhood of the~orbits $S^1\times \text{point}$.
The $t$-degree is then obtained from the~framing of the~once-around charge-$\mu$ boundary in the~plane field normal to the~orbits, and the~contribution to the~refined partition function at level $d$ is $(-t)^{C_{ii}\mu_i d} q^{C_{ii}\mu_i^2d^2-C_{ii}\mu_i d}$, exactly as above.

\subsection{Multiply-wrapped basic disks and generalized quiver partition functions}\label{ssec:motivations}

In this section we collect the~results of the~arguments in section \ref{eq:geometric-t-degree-conjectural} and define a~partition function for generalized quivers. 

A \emph{generalized quiver} $Q$ is an~oriented graph with a~finite number of vertices connected by finitely many signed arrows (signed oriented edges). We denote the~set of vertices by $Q_0$ and the~set of arrows by $Q_1$. A~\emph{dimension vector} for $Q$ is a~vector in the~integral lattice with basis $Q_{0}$, $\boldsymbol{d}\in \IZ Q_0$. We number the~vertices of $Q$ by $1,2,\dots,m=|Q_{0}|$. Each vertex~$i$ has a~positive integral \emph{multiplicity} $\mu_i$. The~\emph{adjacency matrix} of $Q$ is the~$m\times m$ rational matrix with entries $C_{ij}$ defined as follows. Let $N_{ij}\in\mathbb{Z}$ denote the~number of signed arrows from $i$ to $j$. Then 
\be
C_{ij} = 
\begin{cases}
N_{ij} &\text{ if } \mu_i\ne \mu_j,\\
\frac{1}{\mu}\cdot N_{ij} &\text{ if }\mu_i=\mu_j=\mu.
\end{cases}
\ee
A generalized quiver is symmetric if its adjacency matrix is. 

Let $Q$ be a~generalized quiver with nodes $1,2,\dots,m$ of multiplicities $\mu_1,\dots,\mu_m$ and adjacency matrix $C_{ij}$. We associate variables $x_i$ to the~generalized quiver nodes $i$ and define two partition functions as follows.
\begin{dfn}\label{def:genpartfunction}
The partition function of $Q$ is
\be\label{eq:generalized-quiver-P}
	{
	P_Q(\boldsymbol{x};q)= \sum_{\boldsymbol{d}} (-1)^{\sum_i C_{ii}\mu_id_i} q^{\sum_{i, j}C_{ij} (\mu_i d_i) (\mu_j d_j)}  
	\prod_{i=1}^{|Q_0|}\frac{x_i^{d_i}}{(q^{2\mu_i};q^{2\mu_i})_{d_i}}
	}\,,
\ee
and the~refined partition function of $Q$ is
\be\label{eq:generalized-quiver-P-refined}
	{
	P_{Q,t}(\boldsymbol{x};q)  = \sum_{\boldsymbol{d}} 
	(-1)^{\sum_i C_{ii}\mu_id_i}q^{\sum_{i,j} C_{ij} (\mu_i d_i)(\mu_j d_j)} 
	\prod_{i=1}^{|Q_0|}\frac{x_i^{d_i}}{(q^{2};q^{2})_{\mu_id_i}} \frac{(-q^2 t^{\sigma_i};q^2)_{d_i \mu_i}}{(-q^{2\mu_i} t^{\sigma_i};q^{2\mu_i})_{d_i}}
	}\,,
\ee
where $\boldsymbol{d}$ ranges over all dimension vectors and where $\sigma_i=\pm 1$.
\end{dfn}

Note that $P_{Q,t=-1}(\boldsymbol{x};q)=P_{Q}(\boldsymbol{x};q)$ and all exponents of $q$ and $t$ appearing in the~formulas are integers. 

Definition \ref{def:genpartfunction} is the~starting point for our main conjecture on HOMFLY-PT homology. Consider a~knot $K$ and its conormal $L_K$ in the~resolved conifold. We conjecture that it (after degeneration onto the~unknot conormal) admits a~$U(1)$-symmetry and that there is a~finite number of basic disks that we label $1,\dots,m$. We view these disks as generalized quiver nodes, where the~multiplicity of a~node is the~multiplicity of the~boundary of the~corresponding basic disk. Attaching data for the~normal bundles of the~basic disks give a~quiver adjacency matrix $C_{ij}$. Here, for $i\ne j$, $C_{ij}$ is the~infinitesimal mutual linking (a relative framing). In particular $C_{ij}$ is integral if $\mu_i\ne \mu_j$. 
For $\mu_i=\mu_j=\mu$, $C_{ij}$ is $\frac{1}{\mu}\cdot N_{ij}$, where $N_{ij}$ is the~self- or mutual-linking, depending on whether $i=j$ or $i\neq j$.
This defines a~symmetric generalized quiver $Q$ associated to $K$. Let $\log x$ be the~positive generator of $H_1(L_K)$ and let $\log a^2$ denote homology class of $\mathbb{CP}^1$ in the~resolved conifold. Let the~homology class of the~generalized disk $i$ be $\mu_i\log x +a_i\log a$ and its invariant self-linking (or 4-chain intersectiond far from the~boundary) be $q_i$. 
\begin{con}
If we substitute
\be
x_i = x^{\mu_i}q^{q_i-C_{ii}\mu_i}a^{a_i}(-t)^{C_{ii}\mu_i},
\ee
for $i=1,2,\dots,m$, then the~following hold.
\begin{itemize}
    \item The~generalized quiver partition function $P_{Q}(\boldsymbol{x};q)=P_{Q,t=-1}(\boldsymbol{x};q)$ equals the~generating function of the~Gromov-Witten invariants counting generalized holomorphic curves, or equivalently the~generating function of the~symmetrically colored HOMFLY-PT polynomials of $K$.
    \item The~refined generalized quiver partition function $P_{Q,t}(\boldsymbol{x};q)$ equals the~generating function of Poincar\'e polynomials of HOMFLY-PT homology in the~symmetric represtentions.
\end{itemize}
\end{con}

\begin{rmk}
The powers $\sigma_i=\pm1$ in the~refined partition function are interpreted as follows: view the~$\mu_i$ of the~boundary of the~generator $x_i$ as strands of a~braid over the~underlying once-around curve; we obtain a~connected $\mu_i$-times around loop by gluing the~endpoint of the~last strand to the~beginning of the~first one, and this can be done passing above $\sigma_i=+1$ or below $\sigma_i=-1$ all other intermediate strands, according to the~half framing of the~attaching data of the~neighborhood of the~basic disk. 
\end{rmk}

\begin{rmk}
The change of framing operation, counterpart to changing variables $x\to x(-y)^f$ with $f\in \IZ$ in the~augmentation curve, is given by the~usual shift of the~linking matrix $C_{ij}\to C_{ij}+f$, both for diagonal and off-diagonal elements, in the~partition function. 
\end{rmk}

\begin{rmk}\label{rmk:correction}
Equation \eqref{eq:generalized-quiver-P-refined} completes and corrects \cite[Conjecture 1.1]{Ekholm:2018eee} as follows. There, a~$d$~times around generator in framing zero was assumed to be an~embedded disk that contributes $(x^d;q^2)_\infty$ to the~partition function. Its contribution to the~refined partition function was not specified as can be seen already on level $d$. To do so, one would have to explain how to express the~current denominator $(q^2;q^2)_1$ in terms of $(q^2;q^2)_d$. Here the~basic object is different: it is an~embedded disk outside completed by a~multiply covered disk inside. It contributes $(x^d;q^{2d})_\infty$ to the~partition function and on the~refined level its denominators are corrected to $(q^2;q^2)_{kd}$ in \eqref{eq:generalized-quiver-P}.    
\end{rmk}

Equations \eqref{eq:generalized-quiver-P} and \eqref{eq:generalized-quiver-P-refined} differ from the~proposed knots-quivers correspondence of \cite{Kucharski:2017ogk,Kucharski:2017poe} in several ways. 
They are not quiver partition functions, nor are they characters of a~cohomological Hall algebra \cite{Kontsevich:2008fj, 2011arXiv1103.2736E}. Nevertheless, both expressions reflect  the~idea that the~whole spectrum of BPS states is generated by a~finite set of basic objects, interacting in a~certain way. The~basic objects  may be viewed as nodes of a~`generalized quiver', whose adjacency matrix encodes their interactions. The~main difference with the~knots-quivers correspondence is that we allow for nodes with $\mu>1$ and with new denominators. 

We list some arguments that show that some properties of our proposed partition function are necessary:
\begin{enumerate}
  \setcounter{enumi}{-1}
    \item In any quiver-like description, nodes with $\mu>1$ are necessary, see \cite{Ekholm:2018eee} and the~examples in section \ref{sec:Examples}.
    \item A~basic disk wrapping $\mu$-times around $L_K$ contributes to the~multi-cover formula with $q^2$ replaced by $q^{2\mu}$. 
    See section \ref{sec:geom-combin} for motivation and section \ref{sec:Examples} for examples.
    This leads to replacing denominators $(q^2;q^2)_{d_i}$ in (\ref{eq:Efimov}) by $(q^{2\mu_i},q^{2\mu_i})_{d_i}$ in (\ref{eq:generalized-quiver-P}).
    \item Poincar\'e polynomials of knot homologies seem to agree with the~refinement of vortex Hilbert spaces, as discussed in section \ref{sec:Holomorphic  curves viewed as BPS generators} where  denominators of $x^{\mu d}$ correspond to contributions from standard vortices, $(q^{2},q^{2})_{\mu d}$. The~change of variables then requires multiplication and division by the~finite polynomial: $(q^{2},q^{2})_{\mu d}/(q^{2\mu_i},q^{2\mu_i})_{d_i}$.
    \item In the~next step we consider refinement. The~correction introduced in the~numerator by $(q^{2},q^{2})_{\mu d}/(q^{2\mu_i},q^{2\mu_i})_{d_i}$ should arise from a~similar correction by polynomials in $(q,t)$ with positive coefficients, where negative signs in the~original expression come from (odd) powers of $t$. Algebraically, there are several possibilities. Our proposal is based on the~passage from Bott to Morse localization in \eqref{eq:Morse}--\eqref{eq:Bott}.
    We check it in concrete calculations of colored superpolynomials for knots $9_{42}$ and $10_{132}$ in section~\ref{sec:Examples}.
\end{enumerate}

\subsection{HOMFLY-PT homology and geometric deformations}\label{sec:geom+HOMFLY}
Sections \ref{sec:U1-symmetry-susy} and \ref{eq:geometric-t-degree-conjectural} explain how holomorphic disks in a~$U(1)$-symmetric setting, via branched covers with constant ghost bubbles attached, generate symmetrically colored HOMFLY-PT homology. In this section we isolate the~ingredients in this description and show how they give rise to a~deformation invariant version: holomorphic curves for generic almost complex structures (constraint near the~Lagrangian only) give a~chain complex with a~differential, the~homology of which remains invariant and equals certain filtered quotients of HOMFLY-PT homology that recover all of the~original HOMFLY-PT homology. Naturally, as a~chain complex is not determined by its homology, the~chain complexes we propose ar not uniquely determined. The~guiding principle for our definition is that it should reflect the~underlying geometry and that it should model the~initial $U(1)$ invariant situation as closely as possible.   

\subsubsection{Tracing the~refined partition function under deformation}\label{ssec:refpartfunctdef}
Recall the~data that gave the~generalized quiver partition function from a~$U(1)$-symmetric version of the~knot conormal: a~collection of numbered basic  disks $i\in\{1,\dots,n\}$ together with linking densities $C_{ij}$ measuring linking and self linking between their boundaries. Here the~holomorphic curves are in fact two-level holomorphic buildings in an SFT-stretched complex structure: the~outside parts are embedded once punctured spheres, while the~inside parts are unbranched covers of the~trivial cylinder over the~unique geodesic in $S^1\times\R^2$. In this limit all curve boundaries are multiples of the~unique geodesic, and the~linking densities $C_{ij}$ appear as the~(normalized, relative) framing of these boundary components. We point out that the~only curves near the~Lagrangian that are invariant under the~$U(1)$-action are these unbranched covers of the~punctured disk.

We next consider a~class of deformations for which the~picture above persists. To this end we express the~data we have as follows. A~collection of $\mathcal{M}_i$, moduli spaces of holomorphic punctured spheres asymptotic to Reeb orbits, and trivial cylinders near $L_K$ with linking densities $C_{ij}$, where $C_{ij}$ are integers if $\mu_i\ne \mu_j$ and $C_{i,j}\in \frac{1}{\mu}\mathbb{Z}$ if $\mu_i=\mu_j=\mu$. We will now keep the~trivial cylinders and linking densities $C_{ij}$, but allow for deformations of the~outside curves. More precisely, we consider the~moduli spaces $\mathcal{M}_i$ for varying almost complex structure $J_s$ on the~outside part, while we keep $J$ fixed and equal to the~stretched structure near the~negative end of the~symplectic manifold (i.e., near the Lagrangian). 

To organize this, we note that the~homology data of each punctured sphere in $\mathcal{M}_i$ is characterized by its Reeb multiplicity $\mu_i$, its homology class $a_i$, and its $4$-chain intersection~$q_i$. We write $\mathcal{M}(\mu,\alpha,\nu;\chi=1)$ for the~moduli space of all embedded punctured spheres $i$ with $\mu_i=\mu$, $a_i=\alpha$, and $q_i=\nu$. Then 
$$
\mathcal{M}(\mu,\alpha,\nu;\chi=1)\qquad =\quad \bigcup_{\{i\colon \mu_i=\mu, a_i=\alpha, q_i=\nu\}}\mathcal{M}_i
$$ 
is a~space with a~finite number of points, one for each disk $i$ that has these quantum numbers. Furthermore, if their orientations are twisted by their orientation data at the~boundary, all the~points come with a~positive sign.  

We next study the~moduli spaces $\mathcal{M}_s(\alpha,\nu,\mu;\chi=1)$ corresponding to a~1-parameter family of almost complex structures $J_s$. We argue that a~generic $1$-parameter family gives a~cobordism of moduli spaces with the~following properties. At generic times $s$ the~actual spheres in the~moduli space admit standard neighborhoods such that any other holomorphic curve inside such a~neighborhood must be a~branched cover of the~basic sphere. This is analogous to what happens in the~neighborhood of the~central sphere in the~resolved conifold: here any other holomorphic curve must lie completely over the~central sphere. There are finitely many instances where there are Morse modifications in our 1-dimensional cobordism. Such instances correspond to birth/death of curves. Also for this phenomenon, there is a~standard neighborhood of the~sphere at the~critical moment, that contains only the~central curve and the~two redundant curves for nearby times, while all other curves are multiple covers of these. We mention that also moduli spaces $\mathcal{M}(\alpha,\mu,\nu;\chi=\chi_0)$ of punctured embedded curves of Euler characteristic $\chi_0<1$ can be handled in the~the same way and that our conjecture says that these moduli spaces are empty in the~$U(1)$-symmetric situation. 

We next discuss how one might establish such a~structural result, and which objects the~corresponding moduli spaces actually contain. One starts with the~simplest punctured spheres with once-around boundary. Here the~result on moduli spaces is straightforward: by minimality there can be no splitting and the~moduli space gives a~cobordism. We next consider embedded punctured spheres with boundaries that go twice around. The~entire moduli space for these must also include certain two-level curves: two once-around spheres on the~outside, with a~branched covered trivial cylinder in the~symplectization level. We do not perturb the~cylinder, but equip it with both an obstruction bundle and with gluing data to the~spheres above it, which then allows us to count. Once we check that these configurations contain all possible limits, we would get the~required cobordism. The~only configuration not accounted for in the~SFT-compactification is then the~one given by a~branch-covered cylinder glued to two copies of the~same once-around sphere. However, since the~once-around sphere has a~standard neighborhood, the~whole curve would have to agree with this configuration at earlier times and then already from the~first step, which contradicts our sphere starting out as embedded. After establishing the~existence of neighborhoods for twice around disks, the~argument proceeds to the~next Reeb orbit covering level. It is clear how to continue: include all several level curves with branched covered cylinders at the~negative end, the~existence of the~standard neighborhoods at earlier steps shows that the~moduli space gives a~1-dimensional cobordism. Thus if we start from a~finite number of moduli spaces, the~cobordism class never changes as we deform~$J_s$. The~same argument works to show that the~algebraic count of embedded higher genus curves equals zero since there are no such curves at the~starting $U(1)$-symmetric configuration. We point out that an important property of the~knot conormal used in this argument is that it can be placed so that all outside curves go positively around the~basic geodesic. This is arranged by placing the~conormal close to the~toric Lagrangian on an external leg of the~conifold toric diagram.

\subsubsection{Partition functions with values in chain complexes}\label{sec:pathcategorification}
As explained in section \ref{ssec:refpartfunctdef}, we can find the~symmetrically colored HOMFLY-PT homology from the~moduli spaces of embedded curves $\mathcal{M}(\alpha,\nu,\mu;\chi=1)$ for any generic almost complex structure with standard negative end near the~Lagrangian, provided we know the~linking information. For an actual deformation this is easy to trace, but it may also be useful to have a~bifurcation analysis version of this. The~behavior of the~full chain complex is somewhat complicated, see section \ref{sec:unlinkingdiffl}, but certain filtered quotients, from which the~full homology can be recovered, are simpler to trace. In this section we introduce these quotients and study them under homotopy.

We start by considering the~filtration we will use for the~$U(1)$-symmetric configuration. Let $K$ be a~knot and $Q$ its associated generalized quiver with set of nodes $Q_0$. We subdivide the~nodes according to their multiplicity and write $Q_0=\bigcup_{\mu=1}^M Q_0(\mu)$, where $Q_0(\mu)$ is the~set of nodes corresponding to disks of multiplicity $\mu$. To each node $j$ we associate a~quiver variable $x_j$ and basic charges 
\be
	x_j=x^{\mu_j}q^{q_j}a^{a_j}q^{C_{jj}\mu_j}(-t)^{C_{jj}\mu_j}\,.
\ee
The refined partition then has the~form
\be
P_{Q}(x_1,\dots,x_m,q,t)=\sum_r h_r(a,q,t)\frac{x^r}{(q^2;q^2)_r},
\ee
where $h_r$ is a~Laurent polynomial with positive integer coefficients that -- we conjecture -- equals the~Poincar\'e polynomial of HOMFLY-PT homology in the~$r^{\rm th}$ symmetric coloring. Here we will categorify this in a~trivial way. We view the~monomials in $h_r(a,q,t)$ together with the~denominators as generators of a~vector space of equivariant vortices of corresponding charges. Explicitly, a~to summand $n \frac{a^kq^st^v}{(q^2;q^2)_r}$, $n>0$, we associate a~vector space $\C^n$ spanned by $n$ equivariant vortices of charges $a^kq^st^v x^r$. We further think of these vector spaces as a~chain complex with the~trivial differential. We write the~chain complex $\mathcal{C}_{0}=\bigoplus_r \mathcal{C}_{0,r}$ and its homology $\mathcal{H}=\bigoplus_r\mathcal{H}_r$,where the~extra subscript $0$ refers to the~starting point of the~deformation we will consider. As the~differential $\delta_0$ on $\mathcal{C}_0$ is trivial, we have $\mathcal{C}_0=\mathcal{H}$. 

We next consider a~filtration on $\mathcal{C}_0$. The~vector space $\mathcal{C}_{0,r}$ is spanned by equivariant vortices. Each such vortex is a~bound state of one of more basic disks $x_j$: it arises in the~refined partition function as a~coefficient of a~unique monomial of the~form  $x_{j_1}^{k_1}\dots x_{j_l}^{k_l}$ and we define its filtration degree as the~maximal multiplicity $\mu_{j_i}$ of the~quiver variables $x_{j_i}$ in this monomial. We then define $\mathcal{C}_{0,r}(k)\subset \mathcal{C}_{0,r}$ as the~subspace spanned by vortices of filtration degree $\le k$, and $\mathcal{C}_0(k)=\bigoplus_r\mathcal{C}_{0,r}(k)$. Then
\be
\mathcal{C}_{0}(1)\subseteq \mathcal{C}_{0}(2)\subseteq \dots \subseteq\mathcal{C}_0(M)=\mathcal{C}_0.
\ee
Consider now the~quotient $\mathcal{C}_0(r)/\mathcal{C}_0(r-1)$. It is rather complicated, spanned by all bound states of nodes of multiplicity $\le r$ that contain at least one node of multiplicity $r$. However, if we restrict to the~degree $r$ part of the~complex $\mathcal{C}_{0,r}(r)/\mathcal{C}_{0,r}(r-1)$, we find something simpler; here any monomial is a~single multiplicity $r$ generator and we can write the~contribution as a~sum of collections of vortices of the~following form:
\be
(1+t^{\pm1} q^2)(1+t^{\pm1} q^4)\dots(1+t^{\pm1} q^{2r-2}) \frac{x_j}{(q^2;q^2)_r}.
\ee

We define a~new chain complex of equivariant vortices generated only by the~$r$ times around nodes. We take $\mathcal{D}_0(r)$ to be the~vector space of equivariant vortices corresponding to $r$ times around disks. Recall that such vortices have partition functions
\be
\frac{x^{rd}}{(q^{2r};q^{2r})_d}.
\ee
We define $\mathcal{D}_0(r)$ to be the~chain complex with trivial differential associated to the~unrefined partition function, except for substituting the~variables $x_i$ by the~above powers of $x,a,q,t$:
\be
	\check P_{Q(r)}(x,q,t)= P_{Q(r)}(\boldsymbol{x};q^\mu)\,,
\ee
where $q^\mu$ denotes the~substitution of denominators $(q^2;q^2)_d$ by $(q^{2\mu};q^{2\mu})_d$ in the~quiver partition function (\ref{eq:Efimov}) and where $Q_K(r)$ denotes the~multiplicity $r$ part of the~quiver $Q_K$, i.e., all the~degree $r$ nodes and all weighted arrows connecting them. We let $\mathcal{G}(r)$ denote the~homology of $\mathcal{D}_0(r)$ with the~trivial differential. Geometrically, this is the~vector space generated by the~Bott equivariant vorticity $r$ vortices as in \eqref{eq:Bottvortex}.

We will show below that for a~generic path of complex structures $J_s$ we can define a~family of complexes $\mathcal{D}_s$ with differentials $\delta_s$ such that the~homology of $\delta_s$ is isomorphic to $\mathcal{G}$ for each generic $s$. 

\begin{rmk}
We have $\mathcal{D}_0(1)=\mathcal{C}_0(1)$. Furthermore, since the~quiver is finite, there is $R>0$ such that $\mathcal{G}_s(r)=0$ for $r>R$ and if we know $\mathcal{G}_{s,r}(r)$ for all $r\le R$, then together with the~framing density matrix $C_{ij}$ and half-framing information for each node of multiplicty $\ge 1$ we can recover $\mathcal{H}$. 
\end{rmk}

We next consider redundant nodes.  
Consider a~quiver node $i$ of multiplicity $\mu_i=\mu$ with associated quiver variable $x_i$ and linking density $C_{ii}=0\in\frac{1}{\mu}\mathbb{Z}$. Its $\mu$-refined partition function is given by
\be
	\check P_{(i)}(x,q,t)=\sum_{d_i} \frac{x_i^{d_i}}{(q^{2\mu};q^{2\mu})_{d_i}}.
\ee
We define the~\emph{redundant node} of $i$ as the~quiver node $i^*$ of multipicity $\mu_{i^*}=\mu$ and with linking density $C_{i^*i^*}=\frac{1}{\mu}$. Then the~$\mu$-refined partition function of $i^*$ in the~quiver variable $x_{i^*}$ reads 
\be
	\check P_{(i^*)}(x,q,t)=\sum_{d_{i^*}} (-1)^{d_{i^*}}q^{\mu^2d_{i^*}^2}\frac{x_{i^*}^{d_{i^*}\mu}}{(q^{2\mu};q^{2\mu})_{d_{i^*}}},
\ee
and setting $x_{i^*}=q^{-\mu}(-t)x_i$ we find that $P_{i,t=-1}(x_1)\cdot P_{i^*,t=-1}(x_{i^*})=1$. 
We call the~pair $(i,i^*)$ the~degree $\mu$ redundant pair of linking density zero. We think of the~redundant pair as a~two node generalized quiver with nodes $i$ and $i^*$ with linking density matrix 
\be
\left(
\begin{matrix}
0 & 0 \\
0 & \tfrac{1}{\mu}
\end{matrix}
\right).
\ee

We define the~degree $\mu$ redundant pair of linking density $f\in\frac{1}{\mu}{\IZ}$ by a~fractional framing change on the~linking density zero pair. Thus its linking density matrix is instead
\be
\left(
\begin{matrix}
f & f \\
f & f+ \tfrac{1}{\mu}
\end{matrix}
\right).
\ee
Since framing change alters the~partition function of a~quiver according to the~substitution $x_i\mapsto q^{\mu_i^2f} (-t)^{f\mu_i}x_i$, the~unrefined partition function of a~redundant pair is still $1$.

The~degree $1$ redundant pair is a~`standard' redundant pair noticed first in~\cite{Kucharski:2017ogk} and explained in terms of multi-cover skein relations in \cite{Ekholm:2019lmb}.

\begin{rmk}
We comment on the~geometry of the~redundant nodes. Think of the~node~$1_0$ as an orbifold disk. A~redundant disk should have underlying framing different by one twist. We can imagine undoing the~twist and picking up $\mu_1$ 4-chain intersection, the~shift in quiver variable $q^{-\mu_1}$ can be thought of as 4-chain intersections canceling these and returning the~original disk.

The fractional framing change of a~redundant pair corresponds to framing in the~$\mu_1$-fold cover where the~lift lives and indicates that a~framed redundant pair cannot be equivariantly separated. 
\end{rmk}

We now return to the~definition of our chain complex at generic complex structures. For each node $i$ of the~generalized quiver corresponding to an embedded disk $i$ pick a~redundant node $i^\ast$ with $C_{i^\ast i^\ast}=C_{ii}+\frac{1}{\mu_i}$, $C_{ii^\ast}=C_{ii}$, and $C_{ij}=C_{i^\ast j}$. We will define chain complexes along a~path $J_s$ of complex structures using a~level by level construction. 

Consider the~chain complexes $\mathcal{D}_0(r)$ for integers $r>0$. This complex is generated by the~punctured spheres in the~moduli spaces $\mathcal{M}(\mu,\alpha,\nu)$ where $\mu=r$. Let the~linking matrix be $C_{ij}$. Consider the~$1$-parameter family $J_s$ of almost complex structure and the~corresponding moduli spaces $\mathcal{M}_s(\mu,\alpha,\nu)$. When $J_s$ moves away from the~$U(1)$-symmetric configuration, the~moduli spaces undergo Morse modifications at birth/death instances and is otherwise unchanged. We will describe corresponding changes in the~chain complex. 

At a~birth moment $s_0$, the~chain complex at times slightly before $s_0$, $\mathcal{D}_{s_0-\epsilon}(r)$, changes by addition of
of a~redundant pair to a~new chain complex $\mathcal{D}_{s_0+\epsilon}(r)$ with differential $\delta_{s_0+\epsilon}$ such that 
the homology of $\delta_{s_0+\epsilon}$ is equal to that of $\delta_{s_0-\epsilon}$ (and hence still equal to $\mathcal{G}(r)$). 
Below we show that such a~differential exists. At death moments the~chain complex changes by removal of a~redundant pair, and similarly the~homology is unchanged.

\subsubsection{The differential on $\mathcal{D}_s$}

We show that the~required differential exists. Let $Q$ be a~generalized quiver with $m$ nodes of vorticity $\mu$.
The birth of a~redundant pair of nodes involves modifying $Q$ by adding a~pair of nodes which link with all previous $m$ nodes in the~same way, resulting in a~new quiver $Q'$. The~signs of the~two spheres in the~redundant pairs are opposite. This means that their framings differ by one unit when the~orientation data is pushed to the~boundary, and the~disks must form a~redundant pair as the~partition function stays unchanged at the~unrefined level. 

Therefore $P_{Q'}=P_Q\cdot 1$ where the~identity is represented as follows 
\be\label{eq:partition-of-unit-by-canceling-pair}
\begin{split}
	1 & = \sum_{r\geq 0}
	\sum_{d=0}^{r}
	(-1)^{(\rho+1) r + d}
	q^{\mu \rho r^2 + \mu (r-d)^2
	+ 2 \mu r (\alpha_1d_1+\dots+\alpha_m d_m)} 
	\frac{x_{m+1}^{d}x_{m+2}^{r-d}}{(q^{2\mu};q^{2\mu})_d(q^{2\mu};q^{2\mu})_{r-d}}\end{split}.
\ee
Here $\mu$ is the~multiplicity of the~redundant pair of disks $x_{m+1}\sim x_{m+2}\sim x^\mu$, while 
$\rho = \mu C_{m+1,m+1}=\mu C_{m+2,m+2}-1=\mu C_{m+1,m+2}$ is the~integer self and mutual linking of the~pair of new nodes, and $\alpha_i=C_{i,m+1}\mu_i=C_{i,m+2}\mu_i$ is the~linking between the~pair and the~$i^{\text{th}}$ pre-existing node.
The identity is true provided that $x_{m+2} = q^{-\mu} x_{m+1}$, and reduces to \cite[eq. (4.19)]{Ekholm:2019lmb} when $\mu=1$.

Recall that we allow $C_{m+1, m+2}$ to be valued in $\frac{1}{\mu}\IZ$ since $\mu_{m+1}=\mu_{m+2}=\mu$. 
In connection with this, note that at a~redundant pair of disks the~$U(1)$-symmetry has necessarily been broken, disk boundaries are no longer tightly packed, and single strands of a~$\mu$-times around disk can link with strands of other disks individually. For perturbations not too far from the~the $U(1)$-symmetric setting, we need to relax integrality of mutual linking $C_{ij}$ only for pairs of disks with the~same homology data (the same $\mu$- and $a$-degree). More general perturbations may however require further relaxing integrality of the~linking matrix. Understanding this would require a~further study of the~geometry of perturbations, which we leave to future work.

At the~refined level (\ref{eq:partition-of-unit-by-canceling-pair}) ceases to hold.
By our general rules on the~assignment of homological degree, we refine the~partition function by setting $x_{m+2} = -t q^{-\mu} x_{m+1}$, which reflects the~fact that $\mu C_{m+2,m+2} = \mu C_{m+1,m+1}+1$. This deforms (\ref{eq:partition-of-unit-by-canceling-pair}) as follows:
\be
\begin{split}
	& \sum_{r\geq 0}
	x_{m+1}^{r}
	\sum_{d=0}^{r}
	(-1)^{(\rho+1) r + d}
	q^{\mu \rho r^2 + \mu (r-d)^2
	+ 2 \mu r (\alpha_1d_1+\dots+\alpha_m d_m)} 
	\frac{(-t q^{-\mu})^{r-d}}{(q^{2\mu};q^{2\mu})_d(q^{2\mu};q^{2\mu})_{r-d}}
	\\
	= & \sum_{r\geq 0}
	x_{m+1}^{r} (-1)^{\rho r}
	q^{\mu (\rho+1) r^2 -\mu r + 2 \mu r (\alpha_1d_1+\dots+\alpha_m d_m)} 
	\frac{(1+t q^{2\mu(r-1)})\dots(1+q^{2\mu} t)(1+t)}{(q^{2\mu};q^{2\mu})_r} \,.
\end{split}
\ee
While this is not equal to $1$ anymore, it is an expression of the~form
\be
	1 +  (1+t) \sum_{r\geq 1} \frac{x^{\mu r}}{(q^{2\mu};q^{2\mu})_r} \cdot f_r(a,q,t)\,,
\ee
where the~Laurent polynomial $f_r(a,q,t)$ is a~sum of positive integer multiples of monomials. 
We define the~differential to map a~vortex corresponding to a~monomial $q^sa^rt^k$ in $1\cdot f_r(a,q,t)$ to $q^sa^rt^{k+1}$ in $t\cdot f_r(a,q,t)$ and to be $0$ on all other vortices. This gives a~differential of degree $(a,q,t)=(0,0,1)$ with desired properties.

\begin{rmk}
It is crucial to keep the~denominators $(q^{2\mu};q^{2\mu})_r$ of Bott-equivariant vortices here. Switching to the~usual denominator $(q^2;q^2)_{\mu r}$ of standard equivariant vortices spoils the~positivity of coefficient in the~numerator. The~positivity of the~coefficients is what allows us to define the~required differential, which then allows us to interpreted the~partition function as the~Poincar\'e polynomial of the~complex $\mathcal{D}$ of Bott-vortices defined above. In order to carry out the~same program for a~chain complex generated by standard vortices, one would have to introduce other types of generators, see section \ref{sec:unlinkingdiffl} for an analogous discussion.  
\end{rmk}

\begin{rmk}
The differential $\delta_s$ defined in this way depends on the~path of almost complex structures $J_s$ as well as choices of near which basic disks the~redundant pair should be located. It would be interesting to consider the~construction of chain homotopies of different differentials on $\mathcal{D}_s$ that came from different paths. (Note that any two such paths are homotopic since the~space of almost complex structures is contractible.)    
\end{rmk}

\subsection{Unlinking and multiple disks}\label{sec:unlinkingdiffl}
When defining the~differentials $\delta_s$ in the~complexes $\mathcal{D}_s$ in section \ref{sec:pathcategorification}, we worked with a~stretched almost complex structure that allowed us to keep holomorphic disk boundaries fixed. This was essential when demonstrating the~existence of chain complexes still generated by equivariant vortices and with invariant homology, along paths of almost complex structures. Similar constructions not keeping all symmetry require additional generators of new types. To illustrate this, we consider the~basic unlinking operation in the~presence of nodes with multiplicity $>1$.

Recall that in the~case of embedded disks there are, besides the~birth/death of redundant pairs of nodes, also other operations on quivers that do not affect the~partition function \cite{Ekholm:2019lmb}. The~most basic such move is unlinking, where two linked nodes are replaced by two nodes with mutual linking decreased by one unit, together with a~new node obtained by gluing the~other two. We discuss the~counterpart of this for multiply wrapped disks, and show that such an identity involving multiply wrapped disks cannot exist, unless we introduce certain additional holomorphic objects. This then indicates that $U(1)$-invariant configurations are essential for defining HOMFLY-PT homology, and that a~much more involved description is required if one allows to break that symmetry near the~boundary. To see how unlinking works for multiply covered disks, we consider crossing a~simple disk in class $x_1$ with a~$\mu$-times around disk in class $x_2$. Recall that we can write the~contribution to the~partition function by a~$\mu$-times around disk as a~product
\be
\Psi_{q^\mu}(x_2) = \Psi_q(\zeta)\Psi_q(e^{\frac{2\pi i}{\mu}}\zeta)\dots\Psi_q(e^{\frac{2\pi i(\mu-1)}{\mu}}\zeta),
\ee
where here $\zeta$ is a~root of $\zeta^\mu=x_2$, and $\Psi_q$ is defined in (\ref{eq:q-dilog-Psi}). We next assume that there is a~disk $x_1$ that links once equivariantly with $x_2$. To see the~partition function after unlinking, we average over all possible unlinkings, and get 
\be
\Psi_{q}(x_1)\Psi_{q}(x_2)\cdot \left(\frac{1}{\mu}\sum_{k=0}^{\mu-1} \Psi_{q}^{-1}(qe^{\frac{2\pi i k}{\mu}}\zeta x_1)\right)=
\Psi_{q}(x_1)\Psi_{q}(x_2)\phi(x_2 x_1^\mu),
\ee
where 
\be
\phi(x_2x_1^\mu) =\sum_{d=0}^{\infty}  (-q)^{\mu^2d^2} \frac{(q x_2x_1^\mu)^d}{(q^2;q^2)_{\mu d}} \,.
\ee
This latter function is a~new contribution from a~$\mu$-times around disk with a~simple disk attached. For unlinking of more complicated disks we would similarly get new objects, and more of them when we unlink further. To categorify these changes, we would need to find contributions on the~refined level as  well as necessary differentials.

\section{Recursion relations for generalized quiver partition functions}

A fundamental property of partition functions of knot invariants is the~fact that they obey certain types of recursion relations. In \cite{Ekholm:2018eee, Ekholm:2019lmb} we showed that quiver partition functions share this property, in fact leading to a~decomposition of (quantum) $A$-polynomials (and generalizations thereof) into universal building blocks. Here we review these properties, and extend them to include generalized partition functions with multiply wrapped disk nodes.

\subsection{Noncommutative variables}\label{eq:noncomm-variables}

In \cite{Ekholm:2019lmb} it was shown that the~partition function of a~quiver $Q$ can be conveniently expressed by introducing noncommutative variables
\be\label{eq:PQ-operator-form}
	\IP_{Q} = \Psi_{q}(X_m) \cdot \Psi_{q}(X_{m-1}) \cdots \Psi_{q}(X_1)\,,  
\ee
where 
\be 
\label{eq:q-dilog-Psi}
	\Psi_q(\xi) = \sum_{k\geq 0} \frac{q^k}{(q^2;q^2)_k}\xi^k, \qquad  \quad
	X_i = (-1)^{C_{ii}} q^{C_{ii}-1} \hat x_i \hat y_i^{C_{ii}} \prod_{j<i} \hat y_j^{C_{ij}}\,.
\ee
The noncommuting variables  obey quantum torus relations $X_{i} X_j = q^{2 A_{ij}} X_j X_i$, where $A$~is an~antisymmetrization of the~quiver matrix determined by a~choice of ordering of nodes. This is based on the~convention that $\hat y_i \hat x_j = q^{2\delta_{ij}} \hat x_j \hat y_i$.
The quiver partition function is obtained from $\IP_{Q}$ by normal ordering, defined as taking all the~$\hat y_i$ variables to the~right:
\be
	P_{Q} = \ :\,\IP_{Q}\, :
\ee

\subsubsection{Generalization to quivers with multiply wrapped basic disks}
We can apply this idea to generalized partition function and write (\ref{eq:generalized-quiver-P}) in operator form, as a~generalization of (\ref{eq:PQ-operator-form}).
We single out a~node, whose variable we denote $x_0$, and rewrite the~partition function as follows:
\be
\begin{split}
	P_Q  
	& = \sum_{d_0, \boldsymbol{d}} 
	\Bigg[
	(-1)^{C_{00} \mu_0 d_0 + \sum_i C_{ii}  \mu_i d_i}
	q^{C_{00} \mu_0^2 d_0^2 + 2 \sum_i C_{0i} \mu_0\mu_i d_0 d_i 
	+\sum_{i,j} C_{ij} (\mu_i d_i) (\mu_j d_j)}	 
	\\
	& \qquad\quad\times 
	\left( \frac{x_0^{d_0}}{(q^{2\mu_0};q^{2\mu_0})_{d_0}}\right)\left( \prod_{i>0}^{}\frac{x_i^{d_i}}{(q^{2\mu_i};q^{2\mu_i})_{d_i}}\right)\Bigg]
	\\
	& =\left[
	 \sum_{d_0} 
	(-1)^{C_{00} \mu_0 d_0} 
	q^{C_{00} \mu_0^2 d_0^2} 
	\left( \frac{x_0^{d_0}}{(q^{2\mu_0};q^{2\mu_0})_{d_0}}\right) 
	\left( \prod_{i>0}\hat y_i^{\mu_i C_{0i}} \right)^{\mu_0 d_0}
	\right]
	\\
	&\qquad\quad\times 
	\left[
	\sum_{\boldsymbol{d}} 
        (-1)^{\sum_{i} C_{ii}\mu_i  d_i }
        q^{\sum_{i,j} C_{ij} (\mu_i d_i) (\mu_j d_j)}
	\left( \prod_{i>0}^{}\frac{x_i^{d_i}}{(q^{2\mu_i};q^{2\mu_i})_{d_i}}\right)
	\right]\,,
\end{split}
\ee
where we employed quantum torus variables with the~usual $q$-commutator $\hat y_i \hat x_j = q^{2\delta_{ij}} \hat x_j \hat y_i$.
Next, we rewrite the~first sum as an~operator using 
$
	(x_0 \hat y_0^{k})^n = x_0^n \hat y_0^{k n } q^{(n^2-n)k}
$,
with $n=d_0$ and 
$k=C_{00}\mu_0^2$. 
This gives
\be\label{eq:one-disk-operator}
\begin{split}
	P_Q  
	& =\left[
	 \sum_{d_0} 
	(-1)^{C_{00} \mu_0 d_0} 
	q^{C_{00}  \mu_0^2d_0 }
	\left( \frac{(x_0 \hat y_0^{C_{00}\mu_0^2})^{d_0}}{(q^{2\mu_0};q^{2\mu_0})_{d_0}}\right) 
	\left( \prod_{i>0}\hat y_i^{\mu_i C_{0i}} \right)^{\mu_0 d_0}
	\right]
	\\
	&\qquad\qquad\times 
	\left[
	\sum_{\boldsymbol{d}} 
	(-1)^{\sum_i C_{ii} \mu_i d_i}
	q^{\sum_{i,j} C_{ij} (\mu_i d_i) (\mu_j d_j)}
	\left( \prod_{i>0}\frac{x_i^{d_i}}{(q^{2\mu_i};q^{2\mu_i})_{d_i}}\right)
	\right]
	\\
	& =\left[
	 \sum_{d_0} 
	\frac{( (-1)^{C_{00} \mu_0  } q^{C_{00}  \mu_0^2} \, x_0 \hat y_0^{C_{00}\mu_0^2})^{d_0}}{(q^{2\mu_0};q^{2\mu_0})_{d_0}}
	\left( \prod_{i>0}\hat y_i^{\mu_i C_{0i}} \right)^{\mu_0 d_0}
	\right]
	\\
	&\qquad\qquad\times 
	\left[
	\sum_{\boldsymbol{d}} 
	(-1)^{\sum_i C_{ii} \mu_i d_i}
	q^{\sum_{i,j} C_{ij} (\mu_i d_i) (\mu_j d_j)}
	\left( \prod_{i>0}^{}\frac{x_i^{d_i}}{(q^{2\mu_i};q^{2\mu_i})_{d_i}}\right)
	\right]
	\\
	& =
	\Psi_{q^{\mu_0}}\left( 
	(-1)^{C_{00}  \mu_0} q^{C_{00}  \mu_0^2 - \mu_0} \, x_0 \hat y_0^{C_{00}\mu_0^2}
	\prod_{i>0}\hat y_i^{\mu_0 \mu_i C_{0i}} 
	\right)
	\\
	&\qquad\qquad\times 
	\left[
	\sum_{\boldsymbol{d}} 
	(-1)^{\sum_i C_{ii} \mu_i d_i}
	q^{\sum_{i,j} C_{ij} (\mu_i d_i) (\mu_j d_j)}
	\left( \prod_{i>0}\frac{x_i^{d_i}}{(q^{2\mu_i};q^{2\mu_i})_{d_i}}\right)
	\right]\,.
\end{split}
\ee
Iterating on all remaining nodes, we eventually rewrite the~generalized partition function as $P_Q = :\IP_Q:$, the~normal ordering of a~suitable operator 
\be\label{eq:IP-operator-generalized}
	{
	\IP_Q = \Psi_{q^{\mu_m}}(X_m) \cdot \Psi_{q^{\mu_{m-1}}}(X_{m-1}) \cdots  \Psi_{q^{\mu_1}}(X_1) \,,
	}
\ee
with $\Psi$ defined in (\ref{eq:q-dilog-Psi}), and with variables defined as follows:
\be\label{eq:generalized-X-variables}
	X_{i} = (-1)^{ C_{ii}\mu_i} q^{C_{ii}\mu_i^2-\mu_i} x_i \hat y_i^{C_{ii}\mu_i^2} \prod_{j<i} \hat y_j^{\mu_i\mu_j C_{ij}}
	\,.
\ee

\subsubsection{Including refinement}
Let us now turn to the~operator form of the~refined generalized partition function (\ref{eq:generalized-quiver-P-refined}). 
For simplicity, we shall illustrate this for the~case of a~single $d$-times around basic disk with framing zero (i.e. trivial adjacency matrix), whose partition function is
\be\label{eq:single-disk-refined-generalized}
	P_Q  = \sum_{n} 
	\frac{(a^{\alpha} q^{\beta} t^{\delta} x^{d} )^n}{(q^{2};q^{2})_{d n}} \frac{(-q^2 t^{\pm 1};q^2)_{d n}}{(-q^{2 d} t^{\pm 1};q^{2 d})_{n}} \,.
\ee
The main task will be to rewrite this as a~product of $q$-Pochhammers with suitable variables and steps.
To this end, note that we may rewrite $x^{dn} \, (-q^2 t^{\pm 1};q^2)_{d n}/(-q^{2 d} t^{\pm 1};q^{2 d})_{n}$ as
\begin{align}
	x^{dn} \,&
	(-q^{2} t^{\pm1};q^{2d})_n (-q^{4} t^{\pm1};q^{2d})_n \dots (-q^{2d-2} t^{\pm1};q^{2d})_n
	\nonumber
	\\
	& = 
	x^{dn} \,
	\frac{(-q^{2} t^{\pm1};q^{2d})_\infty}{(-q^{2+2dn} t^{\pm1};q^{2d})_\infty} 
	\frac{(-q^{4} t^{\pm1};q^{2d})_\infty}{(-q^{2+2dn} t^{\pm1};q^{2d})_\infty} \dots 
	\frac{(-q^{2d-2} t^{\pm1};q^{2d})_\infty}{(-q^{2+2dn} t^{\pm1};q^{2d})_\infty}
	\\
	& = 
	\frac{1}{
	(-q^{2} t^{\pm1}\hat y;q^{2d})_\infty
	\dots 
	(-q^{2} t^{\pm1}\hat y;q^{2d})_\infty
	}
	x^{dn} \,
	(-q^{2} t^{\pm1}\hat y;q^{2d})_\infty
	\dots 
	(-q^{2d-2} t^{\pm1} \hat y;q^{2d})_\infty\,,\nonumber
\end{align}
where we made use of $\hat y \,x^{dn} = q^{2dn}\, x^{dn} \, \hat y$ (notice that in this case $\hat y = \hat y_i^{d}$, so this is more generally an~instance of $\hat y_i^{\mu_i} \,x_i^{n} = q^{2\mu_i n}\, x_i^{n} \, \hat y_i^{\mu_i}$).
Similarly, we may also rewrite the~denominator as follows:
\begin{align}
	\frac{x^{dn}}{(q^{2};q^{2})_{d n}}
	& = \frac{x^{dn}}{(q^{2d};q^{2d})_{n}}
	\,
	\frac{1}{(q^2;q^{2d})_{n}\dots (q^{2d-2};q^{2d})_{n}} \nonumber
	\\
	& = \frac{x^{dn}}{(q^{2d};q^{2d})_{n}}
	\,
	\frac{(q^{2+2dn};q^{2d})_{\infty}\dots (q^{2d-2+2dn};q^{2d})_{\infty}}{(q^2;q^{2d})_{\infty}\dots (q^{2d-2};q^{2d})_{\infty}}
	\\
	& =
	(q^{2}\hat y;q^{2d})_{\infty}\dots (q^{2d-2}\hat y;q^{2d})_{\infty}
	\frac{x^{dn}}{(q^{2d};q^{2d})_{n}}
	\,
	\frac{1}{(q^2\hat y;q^{2d})_{\infty}\dots (q^{2d-2}\hat y;q^{2d})_{\infty}}\,.\nonumber
\end{align}
Taken together, these allow to  recast the~contribution of a~$d$-times around disk into operator form:
\be\label{eq:single-disk-refined-generalized-unframed}
\begin{split}
	\IP_Q  & = 
	{
	\CO_{d}(q,t,\hat y)
	\,
	\Psi_{q^d}(a^\alpha q^{\beta-d} t^\delta x^d)
	\,
	(\CO_{d}(q,t,\hat y))^{-1}
	}
\end{split}\,,
\ee
where $\Psi_{q^d}(\xi) = (q^d \xi;q^{2d}_\infty)^{-1}$ as in (\ref{eq:q-dilog-Psi}), and  the~conjugation  factor
\be\label{eq:operator-CO}
	{
	\CO_{d}(q,t,\hat y) : =  
	\frac{
	(-q^{2} t^{\pm1}\hat y;q^{2d})_\infty
	\dots 
	(-q^{2d-2} t^{\pm1} \hat y;q^{2d})_\infty
	}{
	(q^2\hat y;q^{2d})_{\infty}\dots (q^{2d-2}\hat y;q^{2d})_{\infty}
	}
	}
\ee
is a~partition  function  of $\hat y$-disks which disappears in the~unrefined limit $t\to-1$.
Once again, this recovers (\ref{eq:single-disk-refined-generalized}) via normal ordering $P = \, :\IP:$.

Expression (\ref{eq:single-disk-refined-generalized-unframed}) is the~operator form for  the~{refined} contribution of a~single $d$-times around basic disk.
For a~collection of basic disks, each wrapping $\mu_i$-times around, and with mutual (and self-) linking encoded by a~symmetric matrix $C_{ij}$, we have the~more general expression
\be\label{eq:generalized-IP-operator-refined}
	\IP_Q =  \Xi_{q^{\mu_m}}(X_m,\hat y_m) \cdot \Xi_{q^{\mu_{m-1}}}(X_{m-1},\hat y_{m-1}) \cdots \Xi_{q^{\mu_1}}(X_1,\hat y_1)\,,
\ee
where $X_i$ are taken as in (\ref{eq:generalized-X-variables}), with the~understanding that $x_i\sim (-t)^{C_{ii}} x^{\mu_i}$ and we define
\be
	\Xi_{q^{\mu_i}}(X_i,\hat y_i) : = 
	\CO_{\mu_i}(q,t,\hat y_i^{\mu_i})
	\,
	\Psi_{q^{\mu_i}}(X_i)
	\,
	(\CO_{\mu_i}(q,t,\hat y_i^{\mu_i}))^{-1}\,.
\ee

\subsection{\texorpdfstring{$A$}{A}-polynomials for generalized partition function}

Let us next turn to recursion relations obeyed by generalized quiver partition functions.
The advantage of the~operator form of the~partition function is that we can easily read off the~recursion relation.
Notice that
\be\label{eq:single-node-unrefined-quantum-curve}
	(1-\hat y - \hat x^\mu) \Psi_{q^\mu}(x^\mu) = 0\,.
\ee
This may also  be recast as $(1-\hat y_i^\mu - \hat x_i) \Psi_{q^\mu}(x_i) = 0$ in the~language  of single-node variables where $x_i\sim x^\mu, \hat y\sim \hat y_i^\mu$.

Next, let us introduce a~symmetrized version of variables $X_i$ (\ref{eq:generalized-X-variables}) akin to (\ref{eq:X-variables-symmetrized}):
\be\label{eq:X-variables-symmetrized-generalized}
	X'_{i} = (-1)^{ C_{ii}\mu_i} q^{C_{ii}\mu_i^2-\mu_i} x_i \hat y_i^{C_{ii}\mu_i^2} \prod_{j=i}^{m} \hat y_j^{\mu_i\mu_j C_{ij}}\,.
\ee 
These are commuting variables $X'_{i}X'_{j}=X'_{j}X'_{i}$.
Moreover, the~partition function  (\ref{eq:generalized-IP-operator-refined}) may as well be  replaced by an~analogous expression $\IP_Q'$ obtained from $\IP_Q$ by replacing $X_i\to X_i'$. The~two have the~same normal ordering $:\IP_Q': \,=\, :\IP_Q: \, = P_Q$ since the~substitution of $X_i'$ corresponds to right-multiplication by suitable powers of $\hat y_i$, as previously discussed in section \ref{eq:noncomm-variables}.

Then, the~partition function (\ref{eq:generalized-IP-operator-refined}), or more precisely its surrogate $\IP_Q'$, must be  annihilated by operators $\hat  A_{i}$:
\be
	\hat  A_{i} \, \IP_Q' = 0 \qquad \forall i=1,\dots, m\,,
\ee
defined as
\be
	\hat  A_{i} = 
	(\CO_{\mu_i}(q,t,\hat y_i^{\mu_i}))^{-1}
	\,
	\left(1-\hat y_i^{\mu_i} -  X'_i\right)
	\,
	\CO_{\mu_i}(q,t,\hat y_i^{\mu_i})\,.
\ee
This statement is obvious for $i=m$ since it corresponds to the~first factor in (\ref{eq:generalized-IP-operator-refined}). But now recall that the~choice of ordering of nodes in $\IP_Q'$,  the~surrogate of (\ref{eq:generalized-IP-operator-refined}) upon  replacing  $X_i$ by $X_i'$, is arbitrary since variables $X_i'$ commute with each other. 
Therefore, if $\hat A_m$ annihilates the~partition function, so must $\hat A_i$ for any $i=1,\dots,m$.

As they are written, the~operators $\hat  A_{i} $ are not polynomial, but they can be simplified using the~$q$-commutation rules of $\hat x_i$ with $\hat  y_j$:
\be
\begin{split}
	\hat  A_{i} 
	& =
	\frac{
	(q^2\hat y_i^{\mu_i};q^{2\mu_i})_{\infty}\dots (q^{2\mu_i-2}\hat y_i^{\mu_i};q^{2\mu_i})_{\infty}
	}{
	(-q^{2} t^{\pm1}\hat y_i^{\mu_i};q^{2\mu_i})_\infty
	\dots 
	(-q^{2\mu_i-2} t^{\pm1} \hat y_i^{\mu_i};q^{2\mu_i})_\infty
	}
	\cdot
	\left(1-\hat y_i^{\mu_i} -  X'_i\right)
	\\
	&\qquad\qquad\qquad\qquad\times
	\frac{
	(-q^{2} t^{\pm1}\hat y_i^{\mu_i};q^{2\mu_i})_\infty
	\dots 
	(-q^{2\mu_i-2} t^{\pm1} \hat y_i^{\mu_i};q^{2\mu_i})_\infty
	}{
	(q^2\hat y_i^{\mu_i};q^{2\mu_i})_{\infty}\dots (q^{2\mu_i-2}\hat y_i^{\mu_i};q^{2\mu_i})_{\infty}
	}
	\\
	& = 
	1-\hat y_i^{\mu_i} -	\frac{
	(q^2\hat y_i^{\mu_i};q^{2\mu_i})_{\infty}\dots (q^{2\mu_i-2}\hat y_i^{\mu_i};q^{2\mu_i})_{\infty}
	}{
	(q^{2-2\mu_i}\hat y_i^{\mu_i};q^{2\mu_i})_{\infty}\dots (q^{-2}\hat y_i^{\mu_i};q^{2\mu_i})_{\infty}
	}
	\\
	&\qquad\qquad\qquad\qquad\times
	\frac{
	(-q^{2-2\mu_i} t^{\pm1}\hat y_i^{\mu_i};q^{2\mu_i})_\infty
	\dots 
	(-q^{-2} t^{\pm1} \hat y_i^{\mu_i};q^{2\mu_i})_\infty
	}{
	(-q^{2} t^{\pm1}\hat y_i^{\mu_i};q^{2\mu_i})_\infty
	\dots 
	(-q^{2\mu_i-2} t^{\pm1} \hat y_i^{\mu_i};q^{2\mu_i})_\infty
	}
	\cdot
	X'_i
	\\
	& = 
	1-\hat y_i^{\mu_i} 
	-
	\frac{
	(1+q^{2-2\mu_i} t^{\pm1}\hat y_i^{\mu_i})
	\dots 
	(1+q^{-2} t^{\pm1} \hat y_i^{\mu_i})
	}{
	(1-q^{2-2\mu_i}\hat y_i^{\mu_i})\dots (1-q^{-2}\hat y_i^{\mu_i})
	}
	\cdot X'_i\,, 
	\\
\end{split}
\ee
which is now a~rational  function of $\hat y_i, \hat  x_i$.

We may as well get rid of the~overall denominator, and simply redefine properly polynomial recursion  relations as follows:
\be\label{eq:quantum A-polynomial}
	\hat A_i 
	 = 
	(1-q^{2-2\mu_i}\hat y_i^{\mu_i})\dots (1-\hat y_i^{\mu_i})  - 
	(1+q^{2-2\mu_i} t^{\pm1}\hat y_i^{\mu_i})
	\dots 
	(1+q^{-2} t^{\pm1} \hat y_i^{\mu_i})
	\, X'_i \,,
\ee
with  $X_i'$ as defined in (\ref{eq:X-variables-symmetrized-generalized}).
In the~unrefined limit this reduces to the~quantum curve
\be\label{eq:unrefined-quantum-curve}
	\hat A_i = 
	1-\hat y_i^{\mu_i} -  X'_i \,.
\ee
This is consistent with the~fact that the~operator $\CO$ defined in (\ref{eq:operator-CO}) trivializes when $t=-1$, leaving the~unrefined partition function (\ref{eq:IP-operator-generalized}) whose factors are indeed annihilated by these operators, as already  noticed in  (\ref{eq:single-node-unrefined-quantum-curve}).

\subsubsection{Example: a~twice-around  basic disk}
If the~above derivation may appear a~bit formal, we can easily rederive this result for the~special case of a~single $\mu$-times around basic disk with partition function (\ref{eq:generalized-quiver-P-refined}). Taking for simplicity $\mu=2$ and $x_i$ equal to $x^\mu$, we have 
\begin{equation}
P_Q(x,q)=\sum_{d}\frac{x^{2d}}{(q^{2};q^{2})_{2d}}\frac{(-q^{2}t;q^{2})_{2d}}{(-q^{4}t;q^{4})_{d}}=\sum_{d}x^{2d}\frac{(-q^{2}t;q^{4})_{d}}{(q^{2};q^{2})_{2d}}.\label{eq:conjectured t-deformation}
\end{equation}
Now notice that
\be
\begin{split}
	\hat{x}^{2}P_Q(x,q)
	& =\sum_{d}x^{2(d+1)}\frac{(-q^{2}t;q^{4})_{d}}{(q^{2};q^{2})_{2d}}
	\sum_{d}x^{2d}\frac{(-q^{2}t;q^{4})_{d-1}}{(q^{2};q^{2})_{2(d-1)}}
	\\
	& =\sum_{d}x^{2d}\frac{(-q^{2}t;q^{4})_{d}}{(q^{2};q^{2})_{2d}}\frac{(1-q^{4d-2})(1-q^{4d})}{(1+q^{4d-2}t)}=\frac{(1-q^{-2}\hat{y})(1-\hat{y})}{(1+q^{-2}t\hat{y})}P_Q(x,q),
\end{split}
\ee
so
\be\label{eq:quantum A-polynomial-twice-disk}
	{
	\hat{A}(\hat{x},\hat{y},q,t)=(1-q^{-2}\hat{y})(1-\hat{y})- (1+q^{-2}t\hat{y})\hat{x}^{2}\,,
	}
\ee
in agreement with the~specialization of (\ref{eq:quantum A-polynomial}) to $\mu_i=2$.

\subsection{The toric brane property}\label{eq:toric-brane-property}

We now discuss a~new connection between quiver descriptions of generic knot conormals, and the~toric brane corresponding to the~unknot conormal.
This property is best explained in the~formalism of quantum variables introduced in our previous work \cite{Ekholm:2019lmb}, suitably modifided.

\subsubsection{A distinguished basis of quantum variables}

The presentation of the~quiver partition function as the~normal-ordered expression of  $\IP_{Q}$ is clearly not unique. 
It is ambiguous by right-multiplication by any function of the~quantum torus variables $\hat y_i$ that evaluates to $1$ when all $y_i\to 1$.
We will use this freedom to switch to a~new set of noncommuting variables that reveals new structures of the~knot homology encoded by the~quiver. From now we assume that all quiver nodes correspond to basic disks wrapping once-around, \emph{i.e.} such that $\mu_i=1$ for all $i$; the~general case will be discussed later.

As a~first intermediate step, we introduce the~variables
\be\label{eq:X-variables-symmetrized}
	{X'_i = (-q)^{C_{ii}} \hat x_i  \prod_{j=1}^{m} \hat y_j^{C_{ij}}}\,, 
\ee 
which appear in the~quantum quiver $A$-polynomials, see \cite{Ekholm:2019lmb}. Note that these variables are mutually commutative: $X'_iX'_j=X'_jX'_i$.

For the~purpose of studying structures in knot homologies, we  consider a~slightly different set of variables defined as
\be
	Z_i = X'_i \, \prod_{j<i}\hat y_j\,.
\ee
These obey quantum torus relations:
\be\label{eq:Z-commutator}
	Z_j Z_i = q^{2\, \sgn(j-i)} Z_i Z_j \,.
\ee
We will denote the~$Z_i$ the~\emph{toric brane variables}. 

\begin{rmk}
Thinking geometrically of the~boundaries of the~disks as curves very close to the~central $S^1$ in $S^1\times \R^2$, the~variables $X_i'$ correspond to an~unlinked collecton of parallel circles, whereas the~collection $Z_i$ correspond to a~collection of curves obtained from this by introducing linking so that strands $i$ and $j$ link with linking number $i-j$.
\end{rmk}
Note that the~definition of $Z_i$ and their algebra depend on a~choice of ordering of nodes of the~quiver. Later we will discuss a~certain preferred choice.
With any choice of ordering for quiver nodes, we claim that the~operator $\tilde \IP_{Q} = \sum_{r\geq 0} \frac{1}{(q^2;q^2)_r} \tilde \IP_r$ with
\be\label{eq:Pr-sum}
	\tilde \IP_r = \sum_{|\boldsymbol{d}|=r} \left[\begin{array}{c} r \\ \boldsymbol{d} \end{array}\right]_{q^2}  {Z_1}^{d_1} \dots {Z_m}^{d_m}
\ee
has the~property that its normal ordering recovers the~quiver partition function $P_{Q}$:
\be
	:\, \tilde \IP_{Q} \, :  \ = \ P_{Q}(\boldsymbol{x};q)\,.
\ee
This is because monomials of $\tilde \IP_{Q}$ are related to those of $\IP_{Q}$ by right multiplication by $\hat y_i$-variables, which is invisible to normal ordering.

Recalling the~definition of  the~$q$-multinomial,
\be\label{eq:q-multinomial-def}
	 \left[\begin{array}{c} r \\ \boldsymbol{d} \end{array}\right]_{q^2} = \frac{(q^2;q^2)_r}{(q^2;q^2)_{d_1}\dots (q^2;q^2)_{d_m}}\,,
\ee
and the~algebraic property (\ref{eq:Z-commutator}), we find that $\tilde \IP_r$ takes a~very simple form:
\be\label{eq:Pr-power}
	{\tilde\IP_r = (Z_1 + \dots + Z_n)^r} \,.
\ee
A proof of this formula is given in appendix \ref{app:q-multinomial-proof}.
In the~limit  $q\to 1 $, this statement reduces to the~multinomial theorem.

\subsubsection{The toric brane recursion relation}

Expression (\ref{eq:Pr-power}) shows that there is a~universal recursion relation among knot homologies (at least for those knots which admit a~quiver description with once-around generators):
\be\label{eq:Z-recursion}
	{\tilde \IP_r = (Z_1 + \dots + Z_n) \, \tilde \IP_{r-1}}\,.
\ee

The recursion relation (\ref{eq:Z-recursion}) is  reminiscent of that satisfied by the~partition function of a~toric brane in $\IC^3$ (which corresponds to the~unknot conormal partition function, in reduced normalization). In fact the~quiver partition function written in terms of $Z_i$ takes the~same exact form:
\be
\begin{split}
	\tilde\IP_{Q}  
	& = \sum_{r\geq 0} \frac{1}{(q^2;q^2)_{r}} (Z_1+\dots +Z_r)^r
	= \prod_{n\geq 0} (1 - q^{2n} Z)^{-1}
	= (Z; q^2)_{\infty}^{-1}
\end{split}
\ee
in terms of the~variable 
\be\label{eq:Z-toric-brane}
	Z = Z_1+\dots +Z_m\,.
\ee
In other words, the~quiver partition function takes the~form of a~toric brane partition function in this coordinate, after normal ordering
\be
	P_{Q}(\boldsymbol{x};q) \, = \, 
	:\, (Z; q^2)_{\infty}^{-1} \, :   
\ee
This is the~remarkable \emph{toric brane property} of quivers (so far we have assumed that all basic nodes wrap exactly once-around the~conormal, but soon we will extend this to generalized quivers).

Recalling that $\hat y = \prod_i \hat y_i$ is the~quantum meridian holonomy on $\partial L_K$, it follows that $Z$ satisfies
\be
	\hat y\, Z = q^2 \, Z\, \hat y \,.
\ee
Using this relation, it is easy to see that
\be\label{eq:quantum-A-poly-simple}
	{(1-\hat y-Z)\,  P_{Q}(\boldsymbol{x};q) = 0}\,.
\ee
This is exactly the~quantum curve of a~toric brane in $\IC^3$.
The change of variables $x_i\to Z_i$ then shows that the~HOMFLY-PT partition function of \emph{any}  knot (admitting a~quiver description with once-around generators) obeys a~new relation, which takes the~form of an~unknot quantum curve (the partition function  of a~toric brane in $\IC^3$ corresponds to the~unknot conormal in the~conifold in reduced normalization).
This has consequences for the~structure of knot homologies, which we will analyze in section \ref{sec:Structures in HOMLFY-PT homology}.

Finally we relax the~assumption that all quiver nodes are basic disks with $\mu_i=1$, and consider the~more general case 
where $\mu_i>1$ for some nodes.
If longitudes of basic disks are given by $x_{i}\sim x^{\mu_{i}}$, then meridians contribute in proportional amounts to the~overall meridian at $\partial L_K$, namely  $\hat{y}=\hat{y}_{1}^{\mu_{1}}\hat{y}_{2}^{\mu_{2}}\ldots \hat{y}_{m}^{\mu_{m}}$.
Ordering nodes so that $\mu_{1}\geq \mu_{2}\geq\ldots\geq \mu_{m}$, we introduce $Z^{(n)}=\sum_{i\in I_{n}}Z_{i}$,
where $I_{n}$ is a~set of indices for which $\mu_{i}\geq n$. $I_{0}=I_{1}$
contains all indices and $Z^{(0)}=Z^{(1)}=Z$. $I_{2}$ contains indices
for which $\mu_{i}\geq2$ (for all thin knots it is empty), and so on. Then  the~generalization of the~toric brane  property is encoded by the~following quantum curve:
\begin{align}
\hat{y}P_{Q}(\boldsymbol{x};q)= & \hat{y}_{1}^{\mu_{1}}\hat{y}_{2}^{\mu_{2}}\ldots \hat{y}_{m}^{\mu_{m}}\textrm{Exp}\left(\frac{Z_{1}+Z_{2}+\ldots+Z_{m}}{1-q^{2}}\right)\\
= & \textrm{Exp}\left(\frac{(q^{2}-1)q^{2n_{max}}Z^{(n_{max})}+\ldots+(q^{2}-1)q^{2}Z^{(2)}+(q^{2}-1)Z^{(1)}+Z^{(0)}}{1-q^{2}}\right)\nonumber\\
= & \left(1-q^{2n_{max}}Z^{(n_{max})}\right)\ldots\left(1-q^{2}Z^{(2)}\right)\left(1-Z^{(1)}\right)P_{Q}(\boldsymbol{x};q)\,.\nonumber
\label{eq:Recursion general updated derivation}
\end{align}

\begin{rmk}
The definition (\ref{eq:Z-toric-brane}) of the~toric brane variables shares tantalizing similarities with partition functions of framed BPS states. 
In \cite{Gaiotto:2010be} it was observed that the~expectation values of a~`UV' line operator of a~rank-$N$ theory, engineered by a~stack of M5-branes wrapping a~complex curve $C$, admits a~decomposition as a~linear combination of expectation values of `IR' line operators of a~low-energy $U(1)$ theory, arising from a~single M5 wrapping a~covering $\Sigma\to C$.
Both UV and IR line operators are engineered by M2-branes attached to the~M5-system.
In our setup the~$Z_i$ correspond to expectation values of $U(1)$ Wilson lines on the~Lagrangian $L$ wrapped by a~single M5-brane. The~expansion of $Z$ as a~sum with integer coefficient in the~$Z_i$ then suggests an interpretation of $Z$ as the~expectation value of a~line operator in a~`UV' theory, obtained by wrapping a~stack of M5 on another three-manifold $L'$. The~quantum curve (\ref{eq:quantum-A-poly-simple}) identifies a~natural candidate: $L'$ should be the~toric brane, or unknot conormal, on which a~stack of M5 is wrapped. 
The line operator on $L'$ is described by (\ref{eq:quantum-A-poly-simple}), and arises from an M2-brane ending on $L'$ along a~curve.
In the~infrared the~multiple M5-branes on $L'$ merge and branch, resolving into a~single brane wrapping a~covering of $L'$, corresponding to $L$. The~M2-curve gets lifted to the~various basic disks on $L$, linking as described by the~quiver. This interpretation of $Z$ as a~line operator in the~toric brane is therefore consistent with viewing $L$ as a~covering of $L'$, in line with the~general assumption, underlying the~geometric origin of quivers, that boundaries of basic holomorphic curves in $L$ all lie very close to the~zero-section $S^1\times \text{point}$. 
\end{rmk}

\section{Homology structure encoded in the quiver}\label{sec:Structures in HOMLFY-PT homology}

The~possibility of describing knot invariants using quivers enables translating quiver properties to HOMFLY-PT homology. In this section we will study how redundant pairs and toric brane variables are reflected in structures of HOMFLY-PT homology. For simplicity, we will assume that all nodes of the~quiver have multiplicity $\mu_i=1$.

\subsection{Redundant pairs and \texorpdfstring{$\mathfrak{sl}_1$}{sl1} pairs}\label{sec:Redundant pairs}

In section \ref{sec:pathcategorification} we described the~degree $\mu$ redundant pair of nodes. Now we want to study its relationship with the~$d_N$ differential of \cite{GS1112}. We consider the~simplest possible case in which we have a~quiver consisting of a~single degree~1 redundant pair of nodes without any other vertices and arrows. This means that the~quiver adjacency matrix is given by
\begin{equation}\label{eq:redundant pair quiver}
C = \left(
\begin{array}{cc}
	0 & 0 \\
	0 & 1
\end{array}
\right)
\end{equation}
and we demand $x_1 = qx_2$.

Now we consider the~$t$-deformation of the~degree 1 redundant pair of nodes, following the~rule $t_i=C_{ii}$. This means that
\begin{equation}
\label{eq:redundant pair x}
	x_2 = x_1  q^{-1}(-t)\,.
\end{equation}
Note that after $t$-deformation nodes 1 and 2 are no longer redundant and the~quiver partition function is given by
\begin{equation}
\label{eq:redundant pair P_Q}
    P_Q(x_1,x_2;q)=1+\frac{x_1 + (-q) x_2}{1-q^2}+\ldots 
	= 1+ \frac{1 + t}{1-q^2} x_1 +\ldots \,.
\end{equation}
(As a~cross-check, we can see that $t=-1$ leads to $P_Q=1$.)

We observe that relation (\ref{eq:redundant pair x}) is  reminiscent of  
the shift in $(a,q,t)$-degrees involved in the~definition of 
$d_N$ differentials in~\cite{GS1112}.
Indeed, deforming \eqref{eq:redundant pair x} into
\begin{equation}\label{eq:dN-pair}
x_2 = x_1 a^2 q^{-2N-1}(-t)
\end{equation}
corresponds to the~statement that homology generators corresponding  to nodes 1 and 2 are connected by a~$d_N$ differential which shifts $(a,q,t)$-degrees by $(-2,2N,-1)$ respectively, which matches \cite{GS1112} for \(N>0\) (after adapting conventions). Note that when $a=q^N$, (\ref{eq:dN-pair})~reduces back to  (\ref{eq:redundant pair x}).

Keeping $a$ generic and taking $N=1$ leads to the~most important building block of the~uncolored homology: the~canceling differential $d_1$. Since $(\mathcal{H}_r,d_1)$  is isomorphic to one-dimensional $\mathcal{H}_1^{\mathfrak{sl}_1}$, all quiver nodes but one are grouped into pairs of the~form
\begin{equation}
\label{eq:sl1 pair}
\left(x_i ,    x_j = x_i a^2 q^{-3}(-t)  \right)\,.
\end{equation}
We call such $(x_i,x_j)$ $\mathfrak{sl}_1$ pairs. Note that elements of $\mathfrak{sl}_1$ pairs cancel each other at any level (any representation $S^r$) in the~$d_1$-cohomology. This is in line with the~expectation  that $\mathfrak{sl}_1$ knot homology is the~same  for all knots, with only one  generator surviving for each  symmetric coloring (the one corresponding to the~unknot). In particular, the~unique surviving generator in the~first homology  $\mathcal{H}_1^{\mathfrak{sl}_1}$ must correspond to a~node of the~quiver, which we call the~\emph{spectator} node.

\subsubsection{Relation to prequivers}

Nodes in $\mathfrak{sl}_1$ pairs are closely related to a~particular prequiver introduced in \cite{JKLNS}. 
The~basic idea is that all nodes of the~quiver come in pairs, except for the~spectator node. 
Furthermore, within each pair the~two variables are related as in (\ref{eq:sl1 pair}), suggesting that node $x_j$ can be viewed as a~bound state of node $x_i$ with an~additional $x$-independent node with variable $x_{*}=a^2 q^{-3} (-t)$. Indeed, it was observed in \cite{Ekholm:2019lmb} that introducing such an~extra node with a~single unit  of linking to $x_i$ would result in a~bound state like the~node $x_j$, via a~procedure of unlinking. 
In \cite{JKLNS} this idea was developed in greater detail, suggesting the~following:
\begin{con}\label{conj:prequivers}
For each knot $K$ with refined partition function that has a~quiver description as in the~original knots-quivers correspondence (i.e. with $x_i\sim x$) there exists a~unique quiver $Q$ corresponding to $K$ with adjacency  matrix of the~form
\begin{equation}
C=\left(\begin{array}{ccccc}
0 & F_{1} & F_{2} & \ldots & F_{k}\\
F_{1}^{T} & D_{1} & U_{12} & \ldots & U_{1k}\\
F_{2}^{T} & U_{12}^{T} & D_{2} & \ldots & U_{2k}\\
\vdots & \vdots & \vdots & \ddots & \vdots\\
F_{k}^{T} & U_{1k}^{T} & U_{2k}^{T} & \ldots & D_{k}
\end{array}\right),\label{eq:preferred form of Q}
\end{equation}
where
\begin{equation}
\label{eq:blocks in the preferred form of Q}
F_{i}=\left[\begin{array}{cc}
\check{C}_{0i} & \check{C}_{0i}\end{array}\right],
\quad 
D_{i}=\left[\begin{array}{cc}
\check{C}_{ii} & \check{C}_{ii}\\
\check{C}_{ii} & \check{C}_{ii}+1
\end{array}\right],
\quad 
U_{ij}=\left[\begin{array}{cc}
\check{C}_{ij} & \check{C}_{ij}\\
\check{C}_{ij}+1 & \check{C}_{ij}+1
\end{array}\right],
\quad
\check{C}_{mn}\in \mathbb{Z}.
\end{equation}
\end{con}
Here the~diagonal block $D_{i}$ corresponds to the~self-interaction of  the~$i$-th
$\mathfrak{sl}_{1}$ pair (note that $D_{i}$ is a~framed version of \eqref{eq:redundant pair quiver}), the~block $U_{ij}$ corresponds to the~interaction between
$i$-th and $j$-th $\mathfrak{sl}_{1}$ pairs, the~zero in the~top-left
corner of $C$ to the~spectator node, and $F_{i}$~corresponds to the~interaction
between the~spectator and $i$-th $\mathfrak{sl}_{1}$ pair.

Using equations (\ref{eq:preferred form of Q})-(\ref{eq:blocks in the preferred form of Q}), we can define a~\emph{prequiver}  $\check{Q}$  \cite{JKLNS} with an~adjacency matrix given by
\begin{equation}
\check{C}=\left(\begin{array}{ccccc}
0 & \check{C}_{01} & \check{C}_{02} & \ldots & \check{C}_{0k}\\
\check{C}_{01}^{T} & \check{C}_{11} & \check{C}_{12} & \ldots & \check{C}_{1k}\\
\check{C}_{02}^{T} & \check{C}_{12}^{T} & \check{C}_{22} & \ldots & \check{C}_{2k}\\
\vdots & \vdots & \vdots & \ddots & \vdots\\
\check{C}_{0k}^{T} & \check{C}_{1k}^{T} & \check{C}_{2k}^{T} & \ldots & \check{C}_{kk}
\end{array}\right).\label{eq:prequiver}
\end{equation}
Nodes of the~prequiver correspond to the~spectator node and $\mathfrak{sl}_{1}$
pairs, while arrows correspond to interactions among them. Note that using equations (\ref{eq:preferred form of Q})-(\ref{eq:prequiver}) one can easily reconstruct $C$ from $\check{C}$. Moreover,
since the~change of variables for the~$\mathfrak{sl}_{1}$ pair ($x_i$,$x_j$) satisfies~(\ref{eq:sl1 pair}), the~changes of variables for $Q$ and $\check{Q}$ read
\begin{equation}
    \boldsymbol{x} = 
    \left(\begin{array}{c}
    \check{x}_0\\
    \check{x}_1\\
    \check{x}_1a^{2}q^{-3}(-t)\\
    \check{x}_2\\
    \check{x}_2a^{2}q^{-3}(-t)\\
    \vdots\\
    \check{x}_k\\
    \check{x}_k a^{2}q^{-3}(-t)\\
    \end{array}\right)\,,
    \qquad
    \check{\boldsymbol{x}} = 
    \left(\begin{array}{c}
    \check{x}_0\\
    \check{x}_1\\
    \check{x}_2\\
    \vdots\\
    \check{x}_k\\
    \end{array}\right)\,,
\end{equation}
and we have
\begin{equation}
\label{eq:Q and preQ}
\sum_{\boldsymbol{d}}(-q)^{\boldsymbol{d}\cdot C\cdot\boldsymbol{d}}\frac{\boldsymbol{x}^{\boldsymbol{d}}}{(q^{2};q^{2})_{\boldsymbol{d}}}=\sum_{\check{\boldsymbol{d}}}(-a^{2}q^{-2}t;q^{2})_{\check{d}_{1}+\ldots+\check{d}_{k}}(-q)^{\check{\boldsymbol{d}}\cdot \check{C}\cdot\check{\boldsymbol{d}}}\frac{\check{\boldsymbol{x}}^{\check{\boldsymbol{d}}}}{(q^{2};q^{2})_{\check{\boldsymbol{d}}}}.
\end{equation}

Let us see how it works on the~example of the~trefoil. In that case the~quiver and the~change of variables are given by \cite{Kucharski:2017poe,Kucharski:2017ogk}
\begin{equation}\label{eq:3_1 quiver}
    C
    =
    \left(\begin{array}{cc}
    0 & F_1\\
    F^T_1 & D_1
    \end{array}\right)
    =
    \left(\begin{array}{c:cc}
    0 & 1 & 1\\
    \hdashline
    1 & 2 & 2\\
    1 & 2 & 3
    \end{array}\right), 
    \qquad  \qquad
    \boldsymbol{x}=
    \left(\begin{array}{c}
    xa^2q^{-2}\\
    xa^2 (-t)^2\\
    xa^4 q^{-3} (-t)^3
    \end{array}\right),
\end{equation}
whereas the~prequiver and corresponding change of variables read \cite{JKLNS}
\begin{equation}
\label{eq:3_1 prequiver}
    \check{C}=\left(\begin{array}{c:c}
    0 & 1\\
    \hdashline
    1 & 2
    \end{array}\right), \phantom{\qquad} \phantom{\qquad} \boldsymbol{\check{x}} = \left(\begin{array}{c}
    x a^2 q^{-2}\\
    x a^2 (-t)^2
    \end{array}\right)\,.
\end{equation}
One can also check that equation \eqref{eq:Q and preQ} is satisfied.

As we mentioned earlier, the~transition between \eqref{eq:3_1 prequiver} and \eqref{eq:3_1 quiver}, called splitting in~\cite{JKLNS}, can be interpreted in terms of the~unlinking procedure described in \cite{Ekholm:2019lmb}. We can consider a~quiver node corresponding to a~curve that does not wind around $L_K$ and links node number 2:
\begin{equation}\label{eq:3_1 prequiver linked}
    C
    =
    \left(\begin{array}{c:c:c}
    0 & 1 & 0\\
    \hdashline
    1 & 2 & 1\\
     \hdashline
    0 & 1 & 0
    \end{array}\right), 
    \qquad  \qquad
    \boldsymbol{x}=
    \left(\begin{array}{c}
    xa^2q^{-2}\\
    xa^2 (-t)^2\\
    a^2 q^{-3} (-t)
    \end{array}\right)
    \,.
\end{equation}
Then unlinking node number 2 and the~new node leads to
\begin{equation}\label{eq:3_1 prequiver unlinked}
    C
    =
    \left(\begin{array}{c:cc:c}
    0 & 1 & 1 & 0\\
    \hdashline
    1 & 2 & 2 & 0\\
    1 & 2 & 3 & 0\\
    \hdashline
    0 & 0 & 0 & 0
    \end{array}\right), 
    \qquad  \qquad
    \boldsymbol{x}=
    \left(\begin{array}{c}
    xa^2q^{-2}\\
    xa^2 (-t)^2\\
    xa^4 q^{-3} (-t)^3\\
    a^2 q^{-3} (-t)
    \end{array}\right).
\end{equation}
After unlinking, the~last ($x$-independent) node is completely detached from the~rest of the~quiver. 
Its contribution to the~partition function   is an~overall multiplicative factor $(a^2 q^{-3} (-t);q^2)^{-1}$. We view it as part of the~closed sector and when discarding this node, we recover the~trefoil quiver and its partition  function \eqref{eq:3_1 quiver}.

We conjecture that the~refined partition function of every knot corresponding to a~quiver with $x_i\sim x$ can be recovered (up to a~$x$-independent prefactor) from a~prequiver linked to an~extra node which does not wind around $L_K$. This is true for all examples analyzed in \cite{JKLNS}.

We expect conjecture \ref{conj:prequivers} to have a~natural extension to include multiply-wrapped basic disks, where some of the~multiply-wrapped disks can be viewed as products of unlinking between a~`primitive' multiply-wrapped disk, and a~distinguished disk which does not wrap around the~longitude of $L_K$ and that links one of the~$\mu_i$ strands. It is not clear from currently available calculations how such pairs are organized. Direct computations for colored HOMFLY-PT polynomials for knots that require multiply-wrapped generators would help understanding this.

\subsection{Knot homologies in toric brane variables: a~new grading}

In this subsection and the~next one we use the~toric brane variables to illustrate a~property of colored knot homologies.
We start here by defining a~certain grading that will later be used to describe the~structure of a~spectral sequence involving knot homologies with different colorings.

Recall the~ambiguity in the~definition of  toric brane variables due to a~choice of ordering of the~nodes of the~quiver $Q$. 
Recall that each node corresponds to a~basic disk, with interactions among basic disks determined by their mutual linking in $L_K$ corresponding to quiver arrows, and the~spectrum of bound states is encoded in $P_{Q}$ \cite{Ekholm:2018eee}.

While everything that follows holds independently of the~specific choice made (\emph{mutatis mutandis}), it will be convenient to fix this choice at least partially.
For this purpose, we recall that for any knot its homology in the~fundamental representation and \emph{reduced} normalization must have an~odd number of generators. There is a~distinguished spectator disk, while the~remaining disks arrange in $\sl_1$ pairs, as discussed in section \ref{sec:Redundant pairs}.

We adopt an~ordering where the~spectator node comes first and $\sl_1$ pairs are labeled by $2i, 2i+1$ for $i\geq 1$.
Let us  recall that, in the~language of quivers, an~$\sl_1$ pair of disks is characterized by 
\be
	C_{2i+1,2i+1}=C_{2i,2i}+1
	\,,
	\qquad
	x_{2i+1} = x_{2i} a^2 q^{-3} (-t) \,. 
\ee
We do not specify how $\sl_1$ pairs should be ordered among themselves, but simply assign $Z_1$ to the~distinguished spectator node and $Z_{2i}, Z_{2i+1} $ to the~pairs.
Note  that
\be
	Z_{2i+1} = Z_{2i} \, a^2 q^{-2}  t \, \hat y_{2i}\, \prod_{j} \hat y_{j} ^{C_{2i+1,j} - C_{2i,j}}\,.
\ee
In particular, this means that in (\ref{eq:Z-recursion}), when $Z_{2i}, Z_{2i+1}$ act on the~\emph{same} term $\sim \boldsymbol{x}^{\boldsymbol{d}}\subset \tilde \IP_{r-1}$ 
 with $|\boldsymbol{d}| = r-1$, they produce two terms in $\tilde \IP_r$ with well-defined shift of degrees:
\be\label{eq:Q-diff-shifts}
	:\, Z_{2i+1} \boldsymbol{x}^{\boldsymbol{d}} \, : = 
	 a^2 t\, q^{-2+2 d_{2i} + 2\sum_{j=1}^{m}  d_j(C_{2i+1,j}  - C_{2i,j}) }\, :\, Z_{2i} \boldsymbol{x}^{\boldsymbol{d}} \, :\,.
\ee
This relation will be instrumental for defining a~new differential on the~$S^r$ homology.

If a~knot $K$ has a~corresponding quiver $Q$, one may consider a~grading on the~$S^r$-colored HOMFLY-PT homology $\CH_r$ induced by the~dimension vector $\boldsymbol{d}$.
The $\boldsymbol{d}$-grading subsumes $t$-grading and $a$-grading, but not the~$q$-grading -- this is clear from the~form of the~quiver partition function with $x_i=x a^{a_i} q^{l_i} (-t)^{C_{ii}}$ (we have $l_i=q_{i}-C_{ii}$ and assume that $\mu_i=1$ for all quiver nodes):
\be\label{eq:naive-grading}
\begin{split}
	P_{Q} 
	& = \sum_{\boldsymbol{d}} (-q)^{\sum_{i,j=1}^{m} C_{ij} d_i d_j} \prod_{i=1}^{m} \frac{(x a^{a_i} q^{l_i} (-t)^{C_{ii}})^{d_i}  }{(q^2;q^2)_{d_i}} \\
	& = \sum_r \frac{1}{(q^2;q^2)_r} \sum_{|\boldsymbol{d}|=r}  {\left[\begin{array}{c} r \\ \boldsymbol{d} \end{array}\right]_{q^2}}(-q)^{\sum_{i,j=1}^{m} C_{ij} d_i d_j} \prod_{i=1}^{m} (x a^{a_i} q^{l_i} (-t)^{C_{ii}})^{d_i}  \,,
\end{split}
\ee
In general, the~grading by $(\boldsymbol{d}, q)$ is more refined than (or at least as refined as) the~triple grading (\ref{eq:HOMFLY as Euler}). As the~latter can always be recovered, we will henceforth focus on $(\boldsymbol{d},q)$ gradings.

Recall that we are working under the~assumption that $x_i\sim x$ for all quiver nodes. Then the~partition function of the~quiver can be organized by $x$-degree, corresponding to $S^r$-coloring of the~HOMFLY-PT polynomial, as in the~second  line of (\ref{eq:naive-grading}).
The $S^r$-colored superpolynomial is then a~sum of terms with distinct $\boldsymbol{d}$-degrees such that $|\boldsymbol{d}|=r$. 
In particular, terms of degree $\boldsymbol{d}$ take the~form of the~$q$-binomial {\small${\left[\begin{array}{c} r \\ \boldsymbol{d} \end{array}\right]_{q^2}}$} times a~monomial.
Let $\kappa_{n}(\boldsymbol{d})$ be the~coefficients of {\small$\left[\begin{array}{c} r \\ \boldsymbol{d} \end{array}\right]_{q^2}$} $= \sum_{n\geq 0} \kappa_n(\boldsymbol{d}) \, q^{2n}$, which are positive  integers.\footnote{To see positivity, one may note that it holds for coefficients of the~$q$-binomial \cite[Theorem 6.1]{kac2001quantum}. Then one may recursively express the~$q$-multinomial in terms of nested $q$-binomials, as we did in appendix \ref{app:q-multinomial-proof}.}
The partition function may be then expressed as
\be
	P_{Q} 
	= \sum_r \frac{1}{(q^2;q^2)_r} \sum_{|\boldsymbol{d}|=r} \sum_{n\geq 0} \kappa_n(\boldsymbol{d}) \, q^{2n}  (-q)^{\sum_{i,j=1}^{m} C_{ij} d_i d_j} \prod_{i=1}^{m} (x a^{a_i} q^{l_i} (-t)^{C_{ii}})^{d_i}  \,.
\ee
Using the~$\boldsymbol{d}$-degree we may introduce a~nonlinear shift of the~$q$-grading, to just retain the~`$n$-grading' induced by the~$q$-multinomial
\be
	n_\text{degree} = \frac{1}{2} \left(  q_\text{degree} - \sum_{i,j=1}^{m} C_{ij} d_i d_j - \sum_i l_i d_i \right) \,.
\ee

This way of grading states in $\CH_r$ is precisely what is accomplished by switching to variables $Z_i$. For example, at level $r=2$ in (\ref{eq:Pr-sum}) and (\ref{eq:Pr-power}) one has terms such as
\be
	\left[\begin{array}{c} 2 \\ \boldsymbol{d}=(1,1,0,\dots) \end{array}\right]_{q^2} Z_1 Z_2 = (\underbrace{1}_{m=0} + \underbrace{q^2}_{m=1}) \cdot Z_1 Z_2  = Z_1 Z_2 + Z_2 Z_1 \quad\subset \quad (Z_1+Z_2+\dots)^2 \,.
\ee
These two terms both have $\boldsymbol{d} = (1,1,0,\dots)$ but are distinguished by $n=0$ and $n=1$.

Adopting the~$(n,\boldsymbol{d})$-grading of $\CH_r$, we deduce its universal structure from~(\ref{eq:Pr-power}):
\be
	{\CH_r = \bigoplus_{|\boldsymbol{d}|=r, \, n\geq 0} \IC^{\kappa_n(\boldsymbol{d})} \otimes \CH_{\boldsymbol{d}, n}}\,,
\ee
where  $\CH_{\boldsymbol{d}, n} \simeq \IC$ is one-dimensional.\footnote{The choice of ordering of basic disks does  not influence this structure. Different choices are related by elements of ${\rm End} \,\IC^{\kappa_n(\boldsymbol{d})}$.}

\subsection{Traces of a~spectral sequence of symmetrically colored knot homologies}

Having introduced the~necessary notions of quiver grading on knot homologies together with the~toric brane property, we are now in a~position to lay out some of their most interesting consequences.
We present a~simple structure that appears on the~knot homologies, which is  reminiscent  of a spectral sequence. 

Pages of the~spectral sequence correspond to HOMFLY-PT homologies with different colorings by symmetric representations. The~nontrivial data that we will provide corresponds to a~definition for the~differentials of the~spectral sequence.
(These should \emph{not} be confused with the~differentials that {define} the~colored HOMFLY-PT homologies.
The~structure we are about to discuss is more similar to, although not quite the~same as, the~one conjectured in \cite{GGS1304}.)
We should stress that our considerations are motivated by structures observed at the~level of partition  functions, and we will provide only partial information about differentials in the~spectral  sequence. 

We introduce linear maps 
\be
	\dd_r : \CH_r\to \CH_r\,,
\ee
which map vectors (more properly, one-dimensional subspaces) with degrees
\be
	\left\{
	\begin{array}{l}
	\boldsymbol{d} = (d_1, \dots, d_{2i}, d_{2i+1},\dots) \\
	n
	\end{array}\right.
	\quad\mapsto \quad
	\left\{
	\begin{array}{l}
	\boldsymbol{d}' = (d_1, \dots, d_{2i}+1 , d_{2i+1}-1,\dots)\\
	n-2 d_{2i}\,.
	\end{array}\right.
\ee
Of course, this does not fully specify the~map, because different subspaces of $\CH_{\boldsymbol{d}, n}$ may be mapped to different $\CH_{\boldsymbol{d}', n'}$, depending on the~possible ways (labeled by $i$) of trading a~unit of $d_{2i+1}$ for a~unit $d_{2i}$.

Nevertheless, we claim that the~map exists, as a~consequence of formula (\ref{eq:Z-recursion}).
It follows from the~same formula that the~map can be defined in such a~way that it is nilpotent:
\be
	\dd_r^2 = 0 \,.
\ee

One way to describe the~maps  $\dd_r$ in some detail is to consider the~free group $F_{m}$ generated by $S=\{a_1,\dots a_{m}\}$, where $m$ is the number of nodes in $Q$.
We then consider an~element of the~group ring $\IZ[F_{m}]$ defined by
\be
	g_r = (a_1+\dots +a_{m})^r = \sum_{(i_1, \dots, i_r)}\sum_{\sigma \in S_r} a_{\sigma(i_1)}\dots a_{\sigma(i_{m})}\,,
\ee
where the~first sum runs over all distinct partially ordered sets $(i_1,\dots, i_r)$ with $i_1\leq i_2\leq \ldots \leq i_r$ with $1\leq i_{k}\leq m$. 
By construction, all monomials are distinct and there are $m^r = \dim \CH_r$ of them. 
In particular, monomials of $g_r$ are in 1-1 correspondence with generators of $\CH_r$ in a~suitable basis:
\be
	\CH_r \simeq  \sum_{(i_1, \dots, i_r)}\sum_{\sigma \in S_r} a_{\sigma(i_1)}\dots a_{\sigma(i_{m})} \cdot \IC \,.
\ee
We do not specify the~details of this map. The~only constraint is that the~choice of map must be compatible with the~one obtained by replacing $a_i\to Z_i$, and with a~subsequent identification of the~normally ordered expression with the~polynomial of degree $x^r$ in $P_{Q}$.\footnote{Each monomial of $g_r$ maps to a~certain $(\boldsymbol{d},n)$ degree. There are
$\kappa_n(\boldsymbol{d})$ mapping to a~given $(\boldsymbol{d},n)$. On the~other hand $\CH_{\boldsymbol{d},n}$ is  $\kappa_n(\boldsymbol{d})$ -dimensional. So there are $\kappa_n(\boldsymbol{d})!$ ways to map, any choice is admissible.}

The linear map $\dd_r$ can now be described as follows: its action depends only on the~first letter $a_j$ in a~given monomial $a_j\dots \subset g_r$, labeling the~corresponding one-dimensional subspace of $\CH_r$:
\be\label{eq:dr-details}
	\dd_r : \left\{
	\begin{array}{l}
	(a_{1}\dots) \cdot \IC \, \mapsto 0\\
	(a_{2i}\dots) \cdot \IC \, \mapsto 0\\
	(a_{2i+1}\dots) \cdot \IC \, \mapsto (a_{2i}\dots)\cdot \IC\,.
	\end{array}\right. 
\ee
Ellipses denote any word of $r-1$ letters, which is furthermore understood to be identical on the~\emph{lhs} and \emph{rhs} of the~last line.
Up to normalization of the~basis, this defines $\dd_r$ explicitly.
Nilpotency follows directly.

An interesting property of the~maps  $\dd_r$ is that they endow the~$S^r$-colored HOMFLY-PT  homologies of a~knot with the~structure  of a~spectral sequence, in  the~sense that
\be\label{eq:dr-cohomology}
	{H^*(\CH_{r}, \dd_r) \simeq \CH_{r-1}}\,,
\ee
where $\simeq$ involves a~{regrading}, which will be detailed shortly.
The $\CH_r$ are pages of a~spectral sequence, where the~$r$-ordering is decreasing.

The claim (\ref{eq:dr-cohomology}) can be established easily with the~aid of $\tilde \IP_r$ written in terms of variables $Z_i$, as in (\ref{eq:Z-recursion}). In fact, the~specific form of $\dd_r$ given in (\ref{eq:dr-details}) corresponds to pairing up all terms of the~type $(Z_i + Z_{i+1}) \, \prod_{k=1}^{r-1} q^{(\dots)} Z_{i_1}\dots Z_{i_k}$, where $ \prod_{k=1}^{r-1} q^{(\dots)} Z_{i_1}\dots Z_{i_k}$ is any monomial in $\tilde \IP_{r-1}$.
What remains after pairwise cancellations of these terms is precisely a~regraded copy of $\tilde\IP_{r-1}$:
\be\label{eq:Z-homology-terms}
	\tilde \IP_r \supset Z_1  \cdot \tilde\IP_{r-1}   = \left[ x a^\Sigma q^{-\Sigma} \, (-q)^{C_{11}} \, \prod_{j} \hat y_j^{C_{1j}} \right] \cdot \tilde  \IP_{r-1}\,,
\ee
where the~shift of various degrees can be summarized as follows :
\be\label{eq:Q-diff-regrading}
\begin{split}
	&x^{r-1} \to x^r \,,\qquad
	a^{\#}\to a^{\#+\Sigma}\,,\\
	&t^\#\to t^\#\,,\qquad
	q^\# \to q^{\# - \Sigma + C_{11} + 2\sum_{j=1}^{m} d_j\,C_{1j}}\,.
\end{split}
\ee

\subsubsection{Example: trefoil}

For the~trefoil knot in reduced normalization, the~generators of $\CH_1$  and $\CH_2$ have respectively the~following degrees:
\be
\begin{array}{c|ccc}
	\CH_1 & a~& q & t \\
	 \hline
	Z_1 & 2& -2  & 0 \\
	Z_2 & 2& 2 & 2\\ 
	Z_3 & 4& 0 & 3\\
	\end{array}
\qquad\qquad
	\begin{array}{c|ccc}
	\CH_2 & a~& q & t \\
	 \hline
	Z_1^2 & 4& -4  & 0 \\
	Z_2^2 & 4& 8 & 4\\ 
	Z_3^2 & 8& 6 & 6\\
	Z_1 Z_2 & 4& 2  & 2 \\
	Z_2 Z_1 & 4& 4 & 2\\ 
	Z_2 Z_3 & 6& 6 & 5\\
	Z_3 Z_2 & 6&  8 &5 \\
	Z_1 Z_3 & 6& 0 & 3\\ 
	Z_3 Z_1 & 6 & 2  & 3 \\
	\end{array}
\ee 
Here the~spectator node is $Z_1$, while the~$\mathfrak{sl}_1$ pair is $(Z_2, Z_3)$.
Therefore, in $\CH_2$ the~differential $\dd_r$ pairs up terms of types $(Z_2 Z_i, Z_3 Z_i)$ for $i=1,2,3$ (cf. (\ref{eq:Q-diff-shifts})):
\be
\begin{split}
	& 
	Z_2 Z_1  : (4,4,2)  \quad\text{with}\quad
	Z_3 Z_1 : (6,2,3) \,,
	\\
	& 
	Z_2 Z_2  : (4,8,4)  \quad\text{with}\quad
	Z_3 Z_2 : (6,8,5) \,,
	\\
	& 
	Z_2 Z_3  : (6,6,5)  \quad\text{with}\quad
	Z_3 Z_3 : (8,6,6) \,,
	\\
\end{split}
\ee
leaving three terms corresponding to a~regrading of $\IP''_1$ (cf. (\ref{eq:Q-diff-regrading})):
\be
	Z_1^2 : (4,-4,0)
	\qquad
	Z_1 Z_2 : (4,2,2)
	\qquad
	Z_1 Z_3 : (6,0,3) \,.
\ee

\subsubsection{Sketch of the~differential from $\CH_{r}$ to $\CH_{r-l}$}

Once we have a~differential relating $\CH_{r}$ to $\CH_{r-1}$, it is natural to consider its composition, decreasing color by multiple steps $S^{r}\rightarrow S^{r-l}$ for any $l\leq r$.
Let us begin by rewriting
\begin{equation}
P_{r}=:(Z_{1}+Z_{\mathfrak{sl}_{1}})^{r}:\,,
\end{equation}
where $Z_{\mathfrak{sl}_{1}}=Z-Z_{1}$ contains all $\mathfrak{sl}_{1}$
pairs (and excludes the~spectator):
\begin{equation}
Z_{\mathfrak{sl}_{1}}=\sum_{i=1}^{k}Z_{2i}+Z_{2i+1}.
\end{equation}
In the~simplest case when $l=1$, we have
\begin{equation}
P_{r}=:(Z_{1}+Z_{\mathfrak{sl}_{1}})P_{r-1}:
\end{equation}
 and we can see that $Z_{1}P_{r-1}$ is just a~regrading of $P_{r-1}$.
We can cancel $Z_{\mathfrak{sl}_{1}}P_{r-1}$ by defining a~differential:
\begin{equation}
\begin{split}Z_{1}P_{r-1}\overset{\mathbf{d}}{\longmapsto} & 0\,,\\
Z_{2i}P_{r-1}\overset{\mathbf{d}}{\longmapsto} & Z_{2i+1}P_{r-1}\,,\\
Z_{2i+1}P_{r-1}\overset{\mathbf{d}}{\longmapsto} & 0\,,
\end{split}
\end{equation}
whose homology corresponds (by construction) to $Z_{1}P_{r-1}$, a~regrading of $\CH_{r-1}$.

For $l=2$ we have
\begin{equation}
P_{r}=:(Z_{1}^{2}+Z_{1}Z_{\mathfrak{sl}_{1}}+Z_{\mathfrak{sl}_{1}}Z_{1}+Z_{\mathfrak{sl}_{1}})P_{r-2}:\,.
\end{equation}
Then we define
\begin{equation}
\begin{split}Z_{1}^{2}P_{r-2}\overset{\mathbf{d}}{\longmapsto} & 0\,,\\
Z_{1}Z_{2i}P_{r-2}\overset{\mathbf{d}}{\longmapsto} & Z_{1}Z_{2i+1}P_{r-2}\,,\\
Z_{1}Z_{2i+1}P_{r-2}\overset{\mathbf{d}}{\longmapsto} & 0\,,\\
Z_{2i}Z_{j}P_{r-2}\overset{\mathbf{d}}{\longmapsto} & Z_{2i+1}Z_{j}P_{r-2}\,,\\
Z_{2i+1}Z_{j}P_{r-2}\overset{\mathbf{d}}{\longmapsto} & 0\,,
\end{split}
\end{equation}
whose homology corresponds, by construction, to $Z_{1}P_{r-2}$, a~regrading of $\CH_{r-2}$.

For general $l$ we write
\begin{equation}
\begin{split}Z_{1}^{l}P_{r-l}\overset{\mathbf{d}}{\longmapsto} & 0\,,\\
Z_{1}^{l-1}Z_{2i}P_{r-l}\overset{\mathbf{d}}{\longmapsto} & Z_{1}^{l-1}Z_{2i+1}P_{r-l}\,,\\
Z_{1}^{l-1}Z_{2i+1}P_{r-l}\overset{\mathbf{d}}{\longmapsto} & 0\,,\\
\vdots\\
Z_{1}^{l-n}Z_{2i}Z_{j_{1}}Z_{j_{2}}\ldots Z_{j_{m-1}}P_{r-l}\overset{\mathbf{d}}{\longmapsto} & Z_{1}^{l-n}Z_{2i+1}Z_{j_{1}}Z_{j_{2}}\ldots Z_{j_{m-1}}P_{r-l}\,,\\
Z_{1}^{l-n}Z_{2i+1}Z_{j_{1}}Z_{j_{2}}\ldots Z_{j_{m-1}}P_{r-l}\overset{\mathbf{d}}{\longmapsto} & 0\,,
\end{split}
\end{equation}
whose homology corresponds (by construction) to $Z_{1}P_{r-l}$, a~suitable regrading  of $\CH_{r-l}$.

\section{Examples with multiply-wrapped basic disks}
\label{sec:Examples}

In this section we discuss concrete examples of geometries where multiply-wrapped basic disks appear naturally, and cross-check proposals discussed in previous sections. We point out that the~available data to which we compare is limited and sometimes conjectural.

\subsection{The line in real projective 3-space}\label{sec:orbifold example}
In this section we study the~example of the~conormal of the~real projective line in real projective space, where a~multiply wrapped disk appears in a~smooth, non-orbifold geometry.

Consider the~curve
\be\label{eq:conifold-orbifold}
    1-y+a^2 x^2 y - x^2+\gamma = 0\,.
\ee
This curve was derived in \cite{Ekholm:2018iso} as the~zero set of the~augmentation polynomial of the~knot contact homology of the~projective line $\R\IP^1$ in $\IR\IP^3$. Viewing the~latter as a~$(\IZ/2\IZ)$-orbifold of $S^3$, in the~large $N$ limit this is also the~moduli space of a~toric brane sitting at the~orbifold point in a~$(\IZ/2\IZ)$-orbifold of the~resolved conifold. 
To see this directly, we observe that this is a~genus-one curve with six asymptotic regions.
There are horizontal asymptotic legs at $x\to 0$ with $y\to \gamma$, and at $x\to\infty$ with $y\to 1/a^2$.
There are also pairs of vertical asymptotic legs, located at  $x\to \pm a^{-2}$ with  $y\to \infty$, and at $x\to \pm \sqrt{1+\gamma}$ with $y\to 0$.

Let us focus on the~limit $\gamma=0$, in which the~curve is doubly covered by the~unknot mirror curve $1-y-x+a^2 xy=0$ through the~change of variable $x\to \pm x^{1/2}$. 
The vertical external legs of the~conifold geometry now sit at $x=\pm a^{-1}$ and $x=\pm 1$.
In particular, each pair of legs is separated by $\log (-1)  = i\pi$, precisely as the~external legs of figure \ref{fig:C3-Z2-diagram-flipped} in the~limit $Q\to -1$.
Indeed, taking also $a\to 0$ and switching orientation $y\to y^{-1}$ turns (\ref{eq:conifold-orbifold}) into (\ref{eq:C2-orbifold-curve}). This confirms that $(\ref{eq:C2-orbifold-curve})$ describes a~toric brane in the~$(\IZ/2\IZ)$ orbifold of $\IC^3$, namely the~half-geometry of the~$(\IZ/2\IZ)$ orbifold of the~resolved conifold obtained by sending $a^2\to 0$.

Since the~brane described by (\ref{eq:conifold-orbifold}) at $\gamma=0$ sits at the~orbifold point, the~upshot of this analysis is that any holomorphic curve ending on it must have boundary that wraps around the~longitude \emph{twice}. This is so because otherwise such a~curve in the~brane would not bound any holomorphic curve away from the~orbifold point.
This is the~large $N$ dual of the~statement that such a~brane was engineered as the~conormal of the~line in $\IR\IP^3$, which is likewise non-bounding: only its double cover can bound curves in $T^*\IR\IP^3$.
This observation provides an~independent confirmation from knot contact homology (a rigorous check, in fact), that curve counting in our orbifold toy model (\ref{eq:orbifold-classical-curve}) really features a~basic twice-around curves, as claimed in (\ref{eq:orbifold-partition-function}).
In particular, this dispels any potential ambiguities concerning reinterpretations of (\ref{eq:orbifold-partition-function}) as two once-around disks in the~vein of (\ref{eq:2-to-1-disks}), since the~boundaries in class $x$ cannot bound any curves in the~orbifold, where the~brane is stuck at the~orbifold locus.

A similar analysis of the~conormal of the~shortest closed geodesic in lens space $L(\mu,1)$ shows how degree $\mu$ basic disk generators appears.   

\subsection{The knot \texorpdfstring{$\bf{9_{42}}$}{942}}

Our next example of multiply-wrapped basic holomorphic curves is the~knot conormal of the~knot $9_{42}$.
The superpolynomial in the~fundamental representation is \cite{DGR0505}
\be\label{eq:942-P1}
	P_1(a,q,t) = a^{-2}\left(\frac{1}{q^2 t^2}+q^2 \right)+\left(q^4 t^3+\frac{1}{q^4 t}+2 t+1\right)
	+ a^2 \left(q^2 t^4+\frac{t^2}{q^2}\right)\,.
\ee
In a~standard quiver  picture, each  generator of the~first homology $\CH_1$ would correspond to a~node whose change of  variables (\ref{eq:KQ-corr-refined}) would involve the~following $(a,q,t)$ degrees:
\be\label{eq:once-around-disks-942}
\begin{array}{c|c|c}
a_i  & q_i-C_{ii} & t_i=C_{ii} \\
\hline
 2 & -2 & 4 \\
 2 & -4 & 2 \\
 0 & 1 & 3 \\
 0 & -1 & 1 \\
 0 & -1 & 1 \\
 0 & 0 & 0 \\
 0 & -3 & -1 \\
 -2 & 2 & 0 \\
 -2 & 0 & -2 \\
\end{array}
\ee
Now let us temporarily assume that all higher BPS states arise from bound states of the~basic once-around curves corresponding to the~generators (\ref{eq:once-around-disks-942}), as in the~original  formulation of  the~knots-quivers correspondence of \cite{Kucharski:2017ogk,Kucharski:2017poe}.
Then the~quiver partition function would predict the~following contributions at level $x^2$:
\begin{align}
	P_{Q}\big|_{x^2} 
	& = 
	a^{-2}
	\Big(
	\left(q^2+1\right) t^3 q^{2 C_{3,8}+6}+\frac{\left(q^2+1\right) q^{2 C_{6,9}-2}}{t^2}+\frac{\left(q^2+1\right) q^{2 C_{7,9}-6}}{t^3}
	\nonumber\\
	& 
	+\left(q^2+1\right) t q^{2 C_{3,9}+2} +\left(q^2+1\right) t q^{2 C_{4,8}+2}+\frac{\left(q^2+1\right) q^{2 C_{4,9}-2}}{t}
	\nonumber\\
	& 
	+\left(q^2+1\right) t q^{2 C_{5,8}+2}+\frac{\left(q^2+1\right) q^{2 C_{5,9}-2}}{t}+\frac{\left(q^2+1\right) q^{2 C_{7,8}-2}}{t}
	+\left(q^2+1\right) q^{2 C_{6,8}+2}
	\Big)
	\nonumber\\
	&
	+
	a^2 \Big(
	\left(q^2+1\right) t^7 q^{2 C_{1,3}+6}+\left(q^2+1\right) t^5 q^{2 C_{1,4}+2}+\left(q^2+1\right) t^5 q^{2 C_{1,5}+2}
    \\
	& 
	+\left(q^2+1\right) t^4 q^{2 C_{1,6}+2}+\left(q^2+1\right) t^3 q^{2 C_{1,7}-2}+\left(q^2+1\right) t^5 q^{2 C_{2,3}+2}
	\nonumber\\
	& 
	+\left(q^2+1\right) t^3 q^{2 C_{2,4}-2}+\left(q^2+1\right) t^3 q^{2 C_{2,5}-2}+\left(q^2+1\right) t^2 q^{2 C_{2,6}-2}+\left(q^2+1\right) t q^{2 C_{2,7}-6}
	\Big)
	\nonumber\\
	&
	+
	a^4 \left(\left(q^2+1\right) t^6 q^{2 C_{1,2}}+q^{12} t^8+t^4\right)
	+a^{-4} \Big(\frac{\left(q^2+1\right) q^{2 C_{8,9}}}{t^2}+\frac{1}{q^8 t^4}+q^4\Big)
	\nonumber\\
	& 
	+\left(q^2+1\right) t^4 q^{2 C_{1,8}+4}+\left(q^2+1\right) t^2 q^{2 C_{1,9}}+\left(q^2+1\right) t^2 q^{2 C_{2,8}}+\left(q^2+1\right) t^4 q^{2 C_{3,4}+4}
	\nonumber\\
	& 
	+\left(q^2+1\right) t^4 q^{2 C_{3,5}+4}+\left(q^2+1\right) t^3 q^{2 C_{3,6}+4}+\left(q^2+1\right) t^2 q^{2 C_{3,7}}+\left(q^2+1\right) t^2 q^{2 C_{4,5}}
	\nonumber\\
	& 
	+\left(q^2+1\right) t q^{2 C_{4,6}}+\left(q^2+1\right) t q^{2 C_{5,6}}+\frac{\left(q^2+1\right) q^{2 C_{6,7}-4}}{t}+\left(q^2+1\right) q^{2 C_{2,9}-4}
	\nonumber\\
	& 
	+\left(q^2+1\right) q^{2 C_{4,7}-4}+\left(q^2+1\right) q^{2 C_{5,7}-4}+q^{14} t^6+2 q^2 t^2+\frac{1}{q^{10} t^2}+1 \nonumber
\end{align}
in terms of unknown linking numbers $C_{ij}$.

This prediction should be compared to the~superpolynomial for the~second  symmetric representation \cite{GS1112}:
\be\label{eq:942-P2}
\begin{split}
	P_{2}(a,q,t) 
	& = 1+ 
	\left(q^2+1\right) \left(\frac{1}{a^2 q^4 t^3}+1\right) \left(\frac{q^2}{a^2 t}+1\right) \left(a^2 q^2 t^4+\frac{a^2 t^2}{q^2}\right)
	\\
	& + 
	\left(q^2+1\right) \left(\frac{1}{a^2 q^6 t^3}+1\right) \left(\frac{1}{a^2 q^4 t^3}+1\right) \left(\frac{q^2}{a^2 t}+1\right)
	\\
	&\qquad\qquad\times 
	\left(a^2 q^6 t^6+a^2 q^6 t^5+a^4 q^4 t^6+2 a^2 q^2 t^5+a^2 t^4+q^2 t^4+t^3\right)
	\\
	& + 
	\left(q^2+1\right) \left(\frac{1}{t}+1\right) \left(\frac{1}{a^2 q^6 t^3}+1\right) \left(\frac{1}{a^2 q^4 t^3}+1\right) \left(\frac{q^2}{a^2 t}+1\right) 
	\\
	&\qquad\qquad\times 
	\left(a^4 q^8 t^8+a^2 q^8 t^7+a^2 q^4 t^5+a^4 q^2 t^6+a^2 t^4+q^6 t^5+t^3\right)
	\\
	& +
	\left(\frac{1}{a^2 t}+1\right) \left(\frac{1}{a^2 q^6 t^3}+1\right) \left(\frac{1}{a^2 q^4 t^3}+1\right) \left(\frac{q^2}{a^2 t}+1\right) \left(a^4 q^{12} t^8+a^4 t^4\right)\,.
\end{split}
\ee
In  fact, one can readily deduce that the~two expressions cannot quite agree 
\be
	P_2\neq P_{Q} \big|_{x^2}\,,
\ee  
because of the~terms of degree $a^{-6}$ present in  $P_2$.

Although knowing the~first and  second colored superpolynomials is not quite enough to  fix all the~linking numbers $C_{ij}$,  it is nevertheless instructive to make an~ansatz for the~latter so as to reproduce at  least part  of the~second-degree superpolynomial.
Upon making such a~(non-unique) ansatz for the~linking  numbers $C_{ij}$ (this affects only terms of $a$-degree $\geq -4$), we may  then compute the~difference between the~two, which  turns out to  be
\be\label{eq:942-level-2-difference}
    P_2 - P_{Q}|_{x^2} = a^{-6} \tilde p_{-6} + a^{-4}\tilde p_{-4} + a^{-2}\tilde p_{-2} + a^{0}\tilde p_{0} + a^{2}\tilde p_{2} + a^{4}\tilde p_{4}\,,
\ee
where
\begin{align*}
	\tilde p_{-6}
	& =  
	\left(\frac{1}{q^6 t^4}+\frac{1}{q^8 t^4}+\frac{1}{q^8 t^5}+\frac{1}{q^2 t^3}\right) \left(q^2 t+1\right)
	+
	\left(\frac{1}{q^8 t^4}+\frac{1}{q^2 t^2}\right) \left(\frac{q^2}{t}+1\right)\,,
	\nonumber
	\\
	\tilde p_{-4} & =
	\left(\frac{1}{q^2 t^2}+\frac{2}{q^4 t^2}+\frac{1}{q^4 t^3}+\frac{1}{q^6 t^3}+\frac{2}{q^8 t^3}+\frac{1}{q^{10} t^4}+\frac{2}{q^2 t}+q^4+q^2+\frac{2}{t}\right) \left(q^2 t+1\right)
	\nonumber	\\
	& +
	\left(\frac{q^2}{t}+1\right) \left(\frac{1}{q^4 t^2}+\frac{1}{q^8 t^2}+\frac{1}{q^8 t^3}+\frac{2}{q^{10} t^3}+q^4 t+q^2 t+\frac{2}{q^2 t}+\frac{3}{q^4 t}+\frac{1}{q^2}+1\right)\,,
	\nonumber	\\
	\tilde p_{-2} & =
	\left(q^2 t+1\right) \Big(q^8 t^3+q^8 t^2+q^6 t^2+2 q^4 t^2+\frac{2}{q^6 t^2}+\frac{1}{q^8 t^2}
	\nonumber\\
	& \qquad\qquad\qquad + \frac{1}{q^{10} t^2}+\frac{1}{q^{10} t^3}+3 q^2 t+\frac{3}{q^4 t}+\frac{2}{q^6 t}+\frac{1}{q^8 t}+\frac{4}{q^2}+\frac{2}{q^4}+2 t+3\Big)
	\nonumber\\
	& + 
	\left(\frac{q^2}{t}+1\right) \Big(q^8 t^4+2 q^4 t^3+q^4 t^2+3 q^2 t^2
	\\
	& \qquad\qquad\qquad +\frac{1}{q^{10} t^2}+\frac{3 t}{q^2}+\frac{t}{q^4}+\frac{2}{q^6 t}+\frac{3}{q^4}+\frac{1}{q^6}+t^2+3 t\Big)\,,
	\end{align*}
    \begin{align*}
	\tilde p_{0} & =
	\left(q^2 t+1\right) \Big(q^{10} t^5+q^8 t^4+4 q^6 t^3+5 q^4 t^3+q^4 t^2
	\nonumber\\
	& \qquad\qquad\qquad +3 q^2 t^3+2 q^2 t^2 + q^2 t+\frac{4 t}{q^2}+\frac{2 t}{q^4}+\frac{1}{q^8 t}+\frac{2}{q^6}+2 t^2\Big)
	\nonumber\\
	& +
	\left(\frac{q^2}{t}+1\right) \left(q^{10} t^6+2 q^8 t^5+q^6 t^5+2 q^6 t^4+q^2 t^3+\frac{5 t^2}{q^2}+\frac{2 t^2}{q^4}+\frac{2 t}{q^6}+\frac{1}{q^8}+2 t^3\right)\,,
	\\
	\tilde p_{2} & =
	\left(q^2 t+1\right) \left(q^{12} t^6+q^{10} t^6+q^8 t^6+q^6 t^5+2 q^6 t^4+2 q^4 t^4+2 q^2 t^4+\frac{t^2}{q^2}+\frac{t^2}{q^4}+t^4+t^3+t^2\right)
	\\
	& +
	\left(\frac{q^2}{t}+1\right) \left(q^8 t^7+q^6 t^6+q^4 t^5+2 q^2 t^5+\frac{t^3}{q^4}+t^4\right)\,,
	\\
	\tilde p_{4} & =
	\left(q^2 t+1\right) \left(q^8 t^7+q^2 t^5\right)
	+
	\left(\frac{q^2}{t}+1\right) \left(q^8 t^8+q^2 t^6\right)\,.
\end{align*}

The number of generators at each level is
\be
\begin{array}{c|cccccccc}
	 & a^{-6} & a^{-4} & a^{-2} &  a^{0} & a^2 & a^4 \\
	 \hline
	P_2 & 12 & 60 & 124 & 129 & 64 &  12\\
	P_{Q}|_{x^2} & 0 & 4 & 20 & 33 & 20 & 4\\ 
\end{array}
\ee
The difference, divided by two to account for the~correction to the~denominator $1-q^4$ of~(\ref{eq:generalized-quiver-P}) vs $(1-q^2)(1-q^4)$ of knot homology, gives the~number of new disks wrapping around $L_K$ twice.

As far  as we are aware, $9_{42}$ is  the~simplest knot where multiply-wrapped basic disks appear. The~known expressions for the~first and  second colored superpolynomials provide robust evidence for deviations from the~standard form of the~knots-quivers correspondence.

In  addition, we observe how the~difference (\ref{eq:942-level-2-difference}),  corresponding precisely to the~contributions of  the~new twice-wrapped basic disks, organizes into sums of terms with overall factors of the~form  $(1+t^{\pm1}q^2)$, level by level in powers of  $a$.
This fact is nontrivial evidence supporting our proposal for the~refined version of the~generalized partition function (\ref{eq:generalized-quiver-P-refined}), noting that setting $d_i=1,\,\mu_i=2$ gives
\be
	\frac{1}{(q^{2};q^{2})_{\mu_id_i}} \frac{(-q^2 t^{\pm 1};q^2)_{d_i \mu_i}}{(-q^{2\mu_i} t^{\pm 1};q^{2\mu_i})_{d_i}}
	=
	\frac{1}{(q^{2};q^{2})_{2}} \frac{(-q^2 t^{\pm 1};q^2)_{2}}{(-q^{4} t^{\pm 1};q^{4})_{1}}
	=
	\frac{1+q^2 t^{\pm 1}}{(1-q^{2})(1-q^4)} \,.
\ee
The denominator matches with the~one accompanying the~superpolynomial for $\CH_2$, while the~numerator is exactly the~type of binomial that organizes (\ref{eq:942-level-2-difference}).
It should be noted that the~expression (\ref{eq:942-P2}) is in fact conjectural.
In this regard, the~evidence supporting our proposal for the~refined generalized partition function (\ref{eq:generalized-quiver-P-refined}) is not conclusive.

\subsection{The knot \texorpdfstring{$\bf{10_{132}}$}{10132}}

Another example is provided by the~knot  conormal of the~knot $10_{132}$. In this section we discuss it following the~steps from the~analysis of $9_{42}$, but since some formulas are very long, they are presented in full form in appendix \ref{sec:10132}.

The superpolynomial in the~fundamental representation is given by \cite{DGR0505}
\[
	P_1(a,q,t) = 
	a^2 \left(q^2 \left(t^3+t^2\right)+\frac{t+1}{q^2}\right)
	+a^4 \left(q^4 t^6+\frac{t^2}{q^4}+\left(2 t^4+t^3\right)\right)
	+a^6 \left(q^2 t^7+\frac{t^5}{q^2}\right)
\]
This corresponds to quiver generators with $(a,q,t)$ degrees
\be\label{eq:once-around-disks-10132}
\begin{array}{c|c|c}
a_i  & q_i-C_{ii} & t_i=C_{ii} \\
\hline
 6 & -5 & 7 \\
 6 & -7 & 5 \\
 4 & -2 & 6 \\
 4 & -4 & 4 \\
 4 & -4 & 4 \\
 4 & -3 & 3 \\
 4 & -6 & 2 \\
 2 & -1 & 3 \\
 2 & 0 & 2 \\
 2 & -3 & 1 \\
 2 & -2 & 0 \\
\end{array}
\ee
Now let us again temporarily assume that all higher BPS states arise from bound states of the~basic once-around curves corresponding to the~generators (\ref{eq:once-around-disks-10132}), as in the~original  formulation of  the~knots-quivers correspondence of \cite{Kucharski:2017ogk,Kucharski:2017poe}.
Then, the~quiver partition function would predict a~certain contribution to the~partition function for the~coefficient of $x^2$, reported in (\ref{eq:10132-PQ-x2}).
This prediction can be compared to the~superpolynomial for the~second  symmetric representation given in (\ref{eq:10132-P2}).
As for $9_{42}$, the~main point is that the~two expressions cannot agree:
\be
	P_2\neq P_{Q} \big|_{x^2}\,,
\ee  
this time because of the~terms of degree $a^{0}$ and $a^2$ present in  $P_2$.

The difference is computed in (\ref{eq:10132-level-2-difference}) and it turns out to have the~general form
\be\label{eq:10132-level-2-difference-short}
	P_2 - P_{Q}|_{x^2}
	 =  
	\left(q^2 t+1\right) \Big[\dots\Big]
	+
	\left(\frac{q^2}{t}+1\right) \Big[\dots\Big]\,.
\ee
The number of generators at each level is
\be
\begin{array}{c|cccccccc}
	 & a^{0} & a^{2} & a^{4} &  a^{6} & a^8 & a^{10} & a^{12} \\
	 \hline
	P_2 & 36 & 172 & 352 & 384 & 237 & 80 & 12 \\
	P_{Q}|_{x^2} & 0 & 0 & 16 & 40 & 41 & 20 &4\\ 
\end{array}
\ee
The difference, divided by two to account for the~correction to the~denominator $1-q^4$ of (\ref{eq:generalized-quiver-P}) vs $(1-q^2)(1-q^4)$ of knot homology, gives the~number of new disks wrapping around~$L_K$ twice.

The known expressions for the~first and  second colored superpolynomials provide further robust evidence for deviations from the~standard form of the~knots-quivers correspondence, adding to  the~case of $9_{42}$.
In  addition, we observe how the~difference (\ref{eq:10132-level-2-difference-short}),  corresponding precisely to the~contributions of  the~new twice-wrapped basic disks, organizes into sums of terms with overall factors of the~form  $(1+t^{\pm1}q^2)$, level by level in powers of  $a$.

As for $9_{42}$, it should be noted that the~expression (\ref{eq:10132-P2}) for the~second symmetrically colored HOMFLY-PT is conjectural.
In this regard, the~evidence supporting our proposal for the~refined generalized partition function (\ref{eq:generalized-quiver-P-refined}) is not conclusive.

\section*{Acknowledgements}

We thank Marko Sto\v{s}i\'c and Paul Wedrich for insightful discussions and for sharing results.
T.E. is supported by the~Knut and Alice Wallenberg Foundation as a~Wallenberg scholar KAW2020.0307 and by the~Swedish Research Council VR2020-04535.
P.K. was supported by the~Polish Ministry of Education and Science through its program Mobility Plus (decision no.~1667/MOB/V/2017/0).
P.L. is supported by NCCR SwissMAP, funded by the~Swiss National Science Foundation.

\newpage
\appendix

\section{Inductive derivation of the~\texorpdfstring{$q$}{q}-multinomial identity}\label{app:q-multinomial-proof}

We would like to prove that
\be
	\sum_{|\boldsymbol{d}|=r} \left[\begin{array}{c} r \\ \boldsymbol{d} \end{array}\right]_{q^2}  {Z_1}^{d_1} \dots {Z_m}^{d_m} = 
	(Z_1+\dots+Z_m)^r
\ee
with $Z_i$ obeying the~algebra (\ref{eq:Z-commutator}) and with the~$q$-multinomial coefficient a~finite polynomial in $q$ with positive integer coefficients, defined as in (\ref{eq:q-multinomial-def}).

For $m=1$, the~identity is trivial. Now suppose it holds for a~given $m$, and consider adding one variable
\be
	\sum_{|\boldsymbol{d}|=r} \left[\begin{array}{c} r \\ \boldsymbol{d} \end{array}\right]_{q^2}  {Z_1}^{d_1} \dots {Z_m}^{d_m} {Z_{m+1}}^{d_{m+1}} = 
	(Z_1+\dots+Z_m+Z_{m+1})^r\,.
\ee
The \emph{lhs} can be recast as follows:
\be
	\sum_{k=0}^{r} \left(\sum_{|\boldsymbol{d}'|=k} \left[\begin{array}{c} r-k \\ \boldsymbol{d}' \end{array}\right]_{q^2}  {Z_1}^{d_1} \dots {Z_m}^{d_m}\right) \left[\begin{array}{c} r \\ k \end{array}\right]_{q^2} {Z_{m+1}}^{k}\,, 
\ee
where $k=d_{m+1}$, $\boldsymbol{d}'$ is a~dimension vector consisting of the~first $m$ entries in $\boldsymbol{d} = (\boldsymbol{d}', k)$, and we used
\be
	\left[\begin{array}{c} r \\ \boldsymbol{d} \end{array}\right]_{q^2} = 
	\left[\begin{array}{c} r-k \\ \boldsymbol{d}' \end{array}\right]_{q^2} \, \left[\begin{array}{c} r \\ k \end{array}\right]_{q^2}\,,
\ee
which follows from the~definition of the~$q$-multinomial.
Using the~identity on $m$ variables, the~\emph{lhs} can be further simplified:
\be
	\sum_{k=0}^{r}  \left[\begin{array}{c} r \\ k \end{array}\right]_{q^2} Z^{r-k} Z_{m+1}^k\,,
\ee
where $Z=Z_1+\dots Z_m$. Notice that $Z_{m+1} \cdot Z = q^2 \ Z \cdot Z_{m+1}$.
Therefore, we only need to prove the~following binomial identity with noncommutative variables:
\be
	\sum_{k=0}^{r}  \left[\begin{array}{c} r \\ k \end{array}\right]_{q^2} Z^{r-k} Z_{m+1}^k = (Z+Z_{m+1})^r\,.
\ee
In fact, this equation is a~consequence of Gauss' $q$-binomial theorem, see \cite[Theorem 5.1]{kac2001quantum}.

\section{Details on \texorpdfstring{$\bf{10_{132}}$}{10132}}\label{sec:10132}
This appendix collects some details on computations for $10_{132}$.
Assuming that all higher BPS states arise from bound states of the~basic once-around curves corresponding to the~generators (\ref{eq:once-around-disks-10132}), as in the~original  formulation of  the~knots-quivers correspondence of \cite{Kucharski:2017ogk,Kucharski:2017poe},
 the~quiver partition function would predict the~following contributions at level~$x^2$:
\begin{align}
	P_{Q}\big|_{x^2} 
	& = 
	\Big(
	\left(q^2+1\right) t^{12} q^{2 c_{1,2}}+t^{14} q^{18}+t^{10} q^6\Big) a^{12}
	\nonumber\\
	&
	+\Big(\left(q^2+1\right) t^{13} q^{2 c_{1,3}+6}+\left(q^2+1\right) t^{11} q^{2 c_{1,4}+2}+\left(q^2+1\right) t^{11} q^{2 c_{1,5}+2}
	\nonumber\\
	&
	+\left(q^2+1\right) t^{10} q^{2 c_{1,6}+2}+\left(q^2+1\right) t^9 q^{2 c_{1,7}-2}+\left(q^2+1\right) t^{11} q^{2 c_{2,3}+2}
	\nonumber\\
	&
	+\left(q^2+1\right) t^9 q^{2 c_{2,4}-2}+\left(q^2+1\right) t^9 q^{2 c_{2,5}-2}+\left(q^2+1\right) t^8 q^{2 c_{2,6}-2}+\left(q^2+1\right) t^7 q^{2 c_{2,7}-6}\Big) a^{10}
	\nonumber\\
	&
	+\Big(
	\left(q^2+1\right) t^{10} q^{2 c_{1,8}+4}+\left(q^2+1\right) t^9 q^{2 c_{1,9}+4}+\left(q^2+1\right) t^8 q^{2 c_{1,10}}+\left(q^2+1\right) t^7 q^{2 c_{1,11}}
	\nonumber\\
	&
	+\left(q^2+1\right) t^8 q^{2 c_{2,8}}+\left(q^2+1\right) t^7 q^{2 c_{2,9}}+\left(q^2+1\right) t^6 q^{2 c_{2,10}-4}+\left(q^2+1\right) t^5 q^{2 c_{2,11}-4}
	\nonumber\\
	&
	+\left(q^2+1\right) t^{10} q^{2 c_{3,4}+4}+\left(q^2+1\right) t^{10} q^{2 c_{3,5}+4}+\left(q^2+1\right) t^9 q^{2 c_{3,6}+4}+\left(q^2+1\right) t^8 q^{2 c_{3,7}}
	\nonumber\\
	&
	+\left(q^2+1\right) t^8 q^{2 c_{4,5}}+\left(q^2+1\right) t^7 q^{2 c_{4,6}}+\left(q^2+1\right) t^6 q^{2 c_{4,7}-4}+\left(q^2+1\right) t^7 q^{2 c_{5,6}}
	\nonumber\\
	&
	+\left(q^2+1\right) t^6 q^{2 c_{5,7}-4}+\left(q^2+1\right) t^5 q^{2 c_{6,7}-4}+t^{12} q^{20}+2 t^8 q^8+t^6 q^6+\frac{t^4}{q^4}\Big) a^8
	\label{eq:10132-PQ-x2}
	\\
	&
	+\Big(
	\left(q^2+1\right) t^9 q^{2 c_{3,8}+6}+\left(q^2+1\right) t^8 q^{2 c_{3,9}+6}+\left(q^2+1\right) t^7 q^{2 c_{3,10}+2}
	\nonumber\\
	&
	+\left(q^2+1\right) t^6 q^{2 c_{3,11}+2}+\left(q^2+1\right) t^7 q^{2 c_{4,8}+2}+\left(q^2+1\right) t^6 q^{2 c_{4,9}+2}
	\nonumber\\
	&
	+\left(q^2+1\right) t^5 q^{2 c_{4,10}-2}+\left(q^2+1\right) t^4 q^{2 c_{4,11}-2}+\left(q^2+1\right) t^7 q^{2 c_{5,8}+2}
	\nonumber\\
	&
	+\left(q^2+1\right) t^6 q^{2 c_{5,9}+2}+\left(q^2+1\right) t^5 q^{2 c_{5,10}-2}+\left(q^2+1\right) t^4 q^{2 c_{5,11}-2}
	\nonumber\\
	&
	+\left(q^2+1\right) t^6 q^{2 c_{6,8}+2}+\left(q^2+1\right) t^5 q^{2 c_{6,9}+2}+\left(q^2+1\right) t^4 q^{2 c_{6,10}-2}
	\nonumber\\
	&
	+\left(q^2+1\right) t^3 q^{2 c_{6,11}-2}+\left(q^2+1\right) t^5 q^{2 c_{7,8}-2}+\left(q^2+1\right) t^4 q^{2 c_{7,9}-2}
\nonumber	\\
	&
	+\left(q^2+1\right) t^3 q^{2 c_{7,10}-6}+\left(q^2+1\right) t^2 q^{2 c_{7,11}-6}\Big) a^6
\nonumber	\\
	&
	+\Big(\left(q^2+1\right) t^5 q^{2 c_{8,9}+4}+\left(q^2+1\right) t^4 q^{2 c_{8,10}}+\left(q^2+1\right) t^3 q^{2 c_{8,11}}+\left(q^2+1\right) t^3 q^{2 c_{9,10}}
\nonumber	\\
	&
	+\left(q^2+1\right) t^2 q^{2 c_{9,11}}+\left(q^2+1\right) t q^{2 c_{10,11}-4}
	+t^6 q^{10}+t^4 q^8+\frac{t^2}{q^2}+\frac{1}{q^4}\Big) a^4\,,
	\nonumber
\end{align}
in terms of unknown linking numbers $C_{ij}$.

This prediction should be compared to the~superpolynomial for the~second  symmetric representation \cite{Stosic}
\be\label{eq:10132-P2}
P_{2}(a,q,t) = a^{0}p_{0} + a^{2}p_{2} + a^{4}p_{4} + a^{6}p_{6} + a^{8}p_{8} + a^{10}p_{10} + a^{12}p_{12}\,,
\ee
where
\begin{align*}
    p_{0} & = \frac{t^{2 {n_1}-8}}{q^6}+\frac{t^{2 {n_1}-8}}{q^8}+\frac{t^{2 {n_2}-7}}{q^4}+\frac{t^{2 {n_2}-7}}{q^6}+\frac{t^{2 {n_3}-8}}{q^2}+t^{2 {n_3}-8}
	+q^2 t^{2 {n_4}-7}+t^{2 {n_4}-7}
	\\
	&	+\frac{t^{2 {n_5}-8}}{q^2}+\frac{t^{2 {n_5}-8}}{q^4}+\frac{t^{2 {n_6}-7}}{q^2}+\frac{t^{2 {n_6}-7}}{q^4}+q^6 t^3+q^4 t^3+q^6 t^2+q^4 t^2+q^2 t^2+t^2+\frac{3 t}{q^2}+3 t
	\\
	& 	+\frac{2}{q^2}+\frac{1}{q^4}+\frac{1}{q^6}+\frac{2}{q^6 t}+\frac{2}{q^8 t}+\frac{1}{q^6 t^2}+\frac{1}{q^8 t^2}+2\,,
\end{align*}

\begin{align*}
    p_{2} & =\frac{t^{2 {n_1}-7}}{q^6}+\frac{2 t^{2 {n_1}-7}}{q^8}
	+\frac{t^{2 {n_1}-7}}{q^{10}}+\frac{2 t^{2 {n_1}-5}}{q^2}+\frac{t^{2 {n_1}-5}}{q^4}+t^{2 {n_1}-5}+\frac{t^{2 {n_2}-6}}{q^4}
	\\
	&
	+\frac{2 t^{2 {n_2}-6}}{q^6}+\frac{t^{2 {n_2}-6}}{q^8}+q^2 t^{2 {n_2}-4}+\frac{t^{2 {n_2}-4}}{q^2}+2 t^{2 {n_2}-4}+\frac{2 t^{2 {n_3}-7}}{q^2}
	\\
	&
	+\frac{t^{2 {n_3}-7}}{q^4}+t^{2 {n_3}-7}+q^6 t^{2 {n_3}-5}+2 q^4 t^{2 {n_3}-5}+q^2 t^{2 {n_3}-5}+q^2 t^{2 {n_4}-6}+\frac{t^{2 {n_4}-6}}{q^2}
	\\
	&
	+2 t^{2 {n_4}-6}+q^8 t^{2 {n_4}-4}+2 q^6 t^{2 {n_4}-4}+q^4 t^{2 {n_4}-4}+\frac{t^{2 {n_5}-7}}{q^2}+\frac{2 t^{2 {n_5}-7}}{q^4}+\frac{t^{2 {n_5}-7}}{q^6}
	\\
	&
	+q^4 t^{2 {n_5}-5}+2 q^2 t^{2 {n_5}-5}+t^{2 {n_5}-5}+\frac{t^{2 {n_6}-6}}{q^2}+\frac{2 t^{2 {n_6}-6}}{q^4}+\frac{t^{2 {n_6}-6}}{q^6}
	\\
	&
	+q^4 t^{2 {n_6}-4}+2 q^2 t^{2 {n_6}-4}+t^{2 {n_6}-4}+q^{12} t^6+2 q^{10} t^6+q^8 t^6+q^{12} t^5+2 q^{10} t^5+3 q^8 t^5+3 q^6 t^5
	\\
	&
	+q^4 t^5+q^8 t^4+6 q^6 t^4+9 q^4 t^4+4 q^2 t^4+4 q^6 t^3+7 q^4 t^3+8 q^2 t^3+\frac{2 t^3}{q^2}+7 t^3+q^4 t^2+2 q^2 t^2
	\\
	&
	+\frac{12 t^2}{q^2}+\frac{5 t^2}{q^4}+8 t^2+\frac{7 t}{q^2}+\frac{5 t}{q^4}+\frac{3 t}{q^6}+\frac{t}{q^8}+4 t+\frac{1}{q^2}+\frac{1}{q^4}
	\\
	&
	+\frac{2}{q^6}+\frac{4}{q^8}+\frac{2}{q^{10}}+\frac{1}{q^6 t}+\frac{2}{q^8 t}+\frac{1}{q^{10} t}\,,
	\\
	\\
    p_{4} & = \frac{t^{2 {n_1}-6}}{q^8}+\frac{t^{2 {n_1}-6}}{q^{10}}+\frac{3 t^{2 {n_1}-4}}{q^2}+\frac{3 t^{2 {n_1}-4}}{q^4}+\frac{t^{2 {n_1}-4}}{q^6}
	\\
	&
	+t^{2 {n_1}-4}+q^4 t^{2 {n_1}-2}+q^2 t^{2 {n_1}-2}+\frac{t^{2 {n_2}-5}}{q^6}+\frac{t^{2 {n_2}-5}}{q^8}+q^2 t^{2 {n_2}-3}+\frac{3 t^{2 {n_2}-3}}{q^2}
	\\
	&
	+\frac{t^{2 {n_2}-3}}{q^4}+3 t^{2 {n_2}-3}+q^6 t^{2 {n_2}-1}+q^4 t^{2 {n_2}-1}+\frac{t^{2 {n_3}-6}}{q^2}+\frac{t^{2 {n_3}-6}}{q^4}+q^6 t^{2 {n_3}-4}
	\\
	&
	+3 q^4 t^{2 {n_3}-4}+3 q^2 t^{2 {n_3}-4}+t^{2 {n_3}-4}+q^{10} t^{2 {n_3}-2}+q^8 t^{2 {n_3}-2}+\frac{t^{2 {n_4}-5}}{q^2}+t^{2 {n_4}-5}+q^8 t^{2 {n_4}-3}
	\\
	&
	+3 q^6 t^{2 {n_4}-3}+3 q^4 t^{2 {n_4}-3}+q^2 t^{2 {n_4}-3}+q^{12} t^{2 {n_4}-1}+q^{10} t^{2 {n_4}-1}+\frac{t^{2 {n_5}-6}}{q^4}+\frac{t^{2 {n_5}-6}}{q^6}
	\\
	&
	+q^4 t^{2 {n_5}-4}+3 q^2 t^{2 {n_5}-4}+\frac{t^{2 {n_5}-4}}{q^2}+3 t^{2 {n_5}-4}+q^8 t^{2 {n_5}-2}+q^6 t^{2 {n_5}-2}+\frac{t^{2 {n_6}-5}}{q^4}
	\\
	&
	+\frac{t^{2 {n_6}-5}}{q^6}+q^4 t^{2 {n_6}-3}+3 q^2 t^{2 {n_6}-3}+\frac{t^{2 {n_6}-3}}{q^2}+3 t^{2 {n_6}-3}+q^8 t^{2 {n_6}-1}+q^6 t^{2 {n_6}-1}
	\\
	&
	+q^{16} t^9+q^{14} t^9+q^{16} t^8+2 q^{14} t^8+3 q^{12} t^8+2 q^{10} t^8+q^{14} t^7+4 q^{12} t^7+9 q^{10} t^7+7 q^8 t^7
	\\
	&
	+q^6 t^7+2 q^{12} t^6+8 q^{10} t^6+12 q^8 t^6+14 q^6 t^6+8 q^4 t^6+q^2 t^6+q^{10} t^5+6 q^8 t^5+13 q^6 t^5
	\\
	&
	+20 q^4 t^5+15 q^2 t^5+3 t^5+q^8 t^4+4 q^6 t^4+13 q^4 t^4+17 q^2 t^4+\frac{8 t^4}{q^2}+\frac{t^4}{q^4}+15 t^4+3 q^4 t^3
	\\
	&
	+8 q^2 t^3+\frac{14 t^3}{q^2}+\frac{11 t^3}{q^4}+\frac{2 t^3}{q^6}+10 t^3+q^4 t^2+q^2 t^2+\frac{8 t^2}{q^2}+\frac{7 t^2}{q^4}+\frac{4 t^2}{q^6}+\frac{2 t^2}{q^8}+2 t^2
	\\
	&
	+\frac{t}{q^2}+\frac{3 t}{q^4}+\frac{2 t}{q^6}+\frac{2 t}{q^8}+\frac{2 t}{q^{10}}+\frac{1}{q^4}+\frac{1}{q^8}+\frac{1}{q^{10}}\,,
\end{align*}

\begin{align*}
    p_{6} & =\frac{t^{2 {n_1}-3}}{q^2}+\frac{2 t^{2 {n_1}-3}}{q^4}+\frac{t^{2 {n_1}-3}}{q^6}+q^4 t^{2 {n_1}-1}+2 q^2 t^{2 {n_1}-1}
	\\
	&
	+t^{2 {n_1}-1}+q^6 t^{2 {n_2}}+2 q^4 t^{2 {n_2}}+q^2 t^{2 {n_2}}+\frac{2 t^{2 {n_2}-2}}{q^2}+\frac{t^{2 {n_2}-2}}{q^4}+t^{2 {n_2}-2}
	\\
	&
	+q^4 t^{2 {n_3}-3}+2 q^2 t^{2 {n_3}-3}+t^{2 {n_3}-3}+q^{10} t^{2 {n_3}-1}+2 q^8 t^{2 {n_3}-1}+q^6 t^{2 {n_3}-1}+q^{12} t^{2 {n_4}}
	\\
	&
	+2 q^{10} t^{2 {n_4}}+q^8 t^{2 {n_4}}+q^6 t^{2 {n_4}-2}+2 q^4 t^{2 {n_4}-2}+q^2 t^{2 {n_4}-2}+q^2 t^{2 {n_5}-3}+\frac{t^{2 {n_5}-3}}{q^2}
	\\
	&
	+2 t^{2 {n_5}-3}+q^8 t^{2 {n_5}-1}+2 q^6 t^{2 {n_5}-1}+q^4 t^{2 {n_5}-1}+q^8 t^{2 {n_6}}+2 q^6 t^{2 {n_6}}+q^4 t^{2 {n_6}}+q^2 t^{2 {n_6}-2}
	\\
	&
	+\frac{t^{2 {n_6}-2}}{q^2}+2 t^{2 {n_6}-2}+q^{18} t^{11}+q^{16} t^{11}+q^{18} t^{10}+3 q^{16} t^{10}+3 q^{14} t^{10}+q^{12} t^{10}+3 q^{16} t^9 
	\\
	&
	+6 q^{14} t^9+10 q^{12} t^9+9 q^{10} t^9+2 q^8 t^9+3 q^{14} t^8+9 q^{12} t^8+15 q^{10} t^8+14 q^8 t^8+5 q^6 t^8+q^{12} t^7
	\\
	&
	+9 q^{10} t^7+18 q^8 t^7+22 q^6 t^7+17 q^4 t^7+5 q^2 t^7+q^{10} t^6+9 q^8 t^6+18 q^6 t^6+21 q^4 t^6+18 q^2 t^6
	\\
	&
	+7 t^6+q^8 t^5+2 q^6 t^5+9 q^4 t^5+15 q^2 t^5+\frac{9 t^5}{q^2}+\frac{2 t^5}{q^4}+14 t^5+q^4 t^4+7 q^2 t^4+\frac{9 t^4}{q^2}+\frac{6 t^4}{q^4}
	\\
	&
	+\frac{2 t^4}{q^6}+11 t^4+q^2 t^3+\frac{3 t^3}{q^2}+\frac{4 t^3}{q^4}+\frac{2 t^3}{q^6}+\frac{t^3}{q^8}+t^3+\frac{t^2}{q^4}+\frac{2 t^2}{q^6}+\frac{t^2}{q^8}\,,
\\
\\
    p_8 & = q^2 t^{2 {n_1}}+t^{2 {n_1}}+q^4 t^{2 {n_2}+1}+q^2 t^{2 {n_2}+1}+q^8 t^{2 {n_3}}+q^6 t^{2 {n_3}}+q^{10} t^{2 {n_4}+1}
	\\
	&
	+q^8 t^{2 {n_4}+1}+q^6 t^{2 {n_5}}+q^4 t^{2 {n_5}}+q^6 t^{2 {n_6}+1}+q^4 t^{2 {n_6}+1}+q^{20} t^{12}+2 q^{18} t^{12}+3 q^{16} t^{12}
	\\
	&
	+q^{14} t^{12}+2 q^{18} t^{11}+4 q^{16} t^{11}+4 q^{14} t^{11}+2 q^{12} t^{11}+2 q^{16} t^{10}+7 q^{14} t^{10}+12 q^{12} t^{10}
	\\
	&
	+13 q^{10} t^{10}+6 q^8 t^{10}+2 q^{14} t^9+9 q^{12} t^9+14 q^{10} t^9+13 q^8 t^9+6 q^6 t^9+3 q^{10} t^8+12 q^8 t^8
	\\
	&
	+15 q^6 t^8+14 q^4 t^8+6 q^2 t^8+3 q^8 t^7+12 q^6 t^7+14 q^4 t^7+9 q^2 t^7+4 t^7+q^6 t^6+2 q^4 t^6
	\\
	&
	+6 q^2 t^6+\frac{3 t^6}{q^2}+\frac{t^6}{q^4}+6 t^6+q^2 t^5+\frac{4 t^5}{q^2}+\frac{t^5}{q^4}+4 t^5+\frac{t^4}{q^4}\,,
\\
\\
    p_{10} & = t^{13} q^{20}+2 t^{13} q^{18}+t^{12} q^{18}
	+3 t^{13} q^{16}+3 t^{12} q^{16}+2 t^{13} q^{14}+3 t^{12} q^{14}+3 t^{11} q^{14}+t^{12} q^{12}
	\\
	&
	+6 t^{11} q^{12}+3 t^{10} q^{12}+8 t^{11} q^{10}+7 t^{10} q^{10}+5 t^{11} q^8+6 t^{10} q^8+3 t^9 q^8+2 t^{10} q^6+5 t^9 q^6
	\\
	&
	+2 t^8 q^6+4 t^9 q^4+4 t^8 q^4+2 t^9 q^2+2 t^8 q^2+t^7 q^2+t^7\,,
    \\
    \\
    p_{12} & = t^{14} q^{18}+t^{14} q^{16}+t^{13} q^{16}+t^{14} q^{14}+t^{13} q^{14}
	\\
	&
	+t^{12} q^{12}+2 t^{12} q^{10}+t^{11} q^{10}+t^{12} q^8+t^{11} q^8+t^{10} q^6\,,
\end{align*}
and $n_1,\ldots,n_6$ are some (unknown) integer numbers.

As for $9_{42}$, the~main point is that the~two expressions cannot agree:
\be
	P_2\neq P_{Q} \big|_{x^2}\,,
\ee  
this time because of the~terms of degree $a^{0}$ and $a^2$ present in  $P_2$.

In fact, we may compute the~difference, upon making a~(non-unique) ansatz for the~linking  numbers $C_{ij}$ (this affects only terms of degree $\geq a^{4}$)
\be\label{eq:10132-level-2-difference}
P_2 - P_{Q}|_{x^2}=a^{0}\tilde p_{0} + a^{2}\tilde p_{2} + a^{4}\tilde p_{4} + a^{6}\tilde p_{6} + a^{8}\tilde p_{8} + a^{10}\tilde p_{10}  + a^{12}\tilde p_{12}\,,
\ee
where
\begin{align*}
    \tilde p_{0} & =  
	\left(q^2 t+1\right) \left(q^4 t^2+\frac{1}{q^8 t^2}+\frac{1}{q^2 t^8}+\frac{1}{q^4 t^8}+\frac{1}{q^6 t^8}+\frac{1}{q^8 t^8}+\frac{t}{q^2}+\frac{1}{q^6 t}+\frac{1}{q^8 t}+\frac{2}{q^2}+\frac{1}{t^8}+t\right)
	\\
	& +
	\left(\frac{q^2}{t}+1\right) \left(q^4 t^3+\frac{1}{q^4 t^7}+\frac{2 t}{q^2}+\frac{1}{q^8 t}\right)\,,
\\
\\
    \tilde p_{2} & = 
	\left(q^2 t+1\right) \Big(q^{10} t^5+q^8 t^5+2 q^6 t^4+\frac{q^6}{t^5}+2 q^4 t^4+3 q^4 t^3+\frac{2 q^4}{t^5}+3 q^2 t^3+\frac{2 q^2}{t^5}
	\\
	&
	+\frac{5 t^2}{q^2}+\frac{2}{q^2 t^5}+\frac{1}{q^4 t^5}+\frac{2}{q^2 t^7}+\frac{2}{q^4 t^7}+\frac{2}{q^6 t^7}+\frac{2}{q^8 t^7}+\frac{1}{q^{10} t^7}
	\\
	&
	+\frac{2 t}{q^2}+\frac{3 t}{q^4}+\frac{t}{q^6}+\frac{1}{q^{10} t}+\frac{1}{q^4}+\frac{1}{q^6}+\frac{2}{q^8}+4 t^2+\frac{2}{t^5}+\frac{1}{t^7}\Big)
	\\
	& +
	\left(\frac{q^2}{t}+1\right) \Big(q^{10} t^6+q^8 t^6+q^6 t^5+q^4 t^5+4 q^4 t^4+4 q^2 t^4+q^2 t^3+\frac{q^2}{t^4}+\frac{2 t^3}{q^2}
	\\
	&
	+\frac{4 t^2}{q^2}+\frac{4 t^2}{q^4}+\frac{1}{q^4 t^6}+\frac{1}{q^6 t^6}+\frac{t}{q^4}+\frac{t}{q^8}+\frac{1}{q^8}+\frac{2}{q^{10}}+2 t^3+\frac{1}{t^4}\Big)	\,,
	\\
	\\
	\tilde p_{4} & =
	\left(q^2 t+1\right) \Big(
	q^{14} t^8+q^{12} t^7+2 q^{10} t^7+2 q^{10} t^6+\frac{q^{10}}{t^2}+q^8 t^7+5 q^8 t^6+3 q^8 t^5+\frac{q^8}{t^2}
	\\
	&
	+4 q^6 t^6+5 q^6 t^5+q^6 t^4+\frac{q^6}{t^2}+\frac{q^6}{t^4}+8 q^4 t^5+q^4 t^3+\frac{q^4}{t^2}+\frac{3 q^4}{t^4}+q^2 t^5
	\\
	&
	+9 q^2 t^4+\frac{q^2}{t^2}+\frac{4 q^2}{t^4}+\frac{8 t^3}{q^2}+\frac{6 t^3}{q^4}+\frac{3 t^2}{q^2}+\frac{4 t^2}{q^4}+\frac{2 t^2}{q^6}
	\\
	&
	+\frac{t^2}{q^8}+\frac{4}{q^2 t^4}+\frac{3}{q^4 t^4}+\frac{1}{q^6 t^4}+\frac{1}{q^2 t^6}+\frac{1}{q^4 t^6}+\frac{1}{q^6 t^6}+\frac{1}{q^8 t^6}
	\\
	&
	+\frac{1}{q^{10} t^6}+\frac{t}{q^4}+\frac{2 t}{q^6}+\frac{t}{q^8}+\frac{t}{q^{10}}+\frac{1}{q^{10}}+t^5+t^4+3 t^3+\frac{4}{t^4}\Big)
	\\
	& + 
	\left(\frac{q^2}{t}+1\right) \Big(q^{14} t^9+q^{12} t^8+q^{10} t^8+2 q^{10} t^7+2 q^8 t^7+q^8 t^6+q^6 t^7+2 q^6 t^6+7 q^4 t^6
	\\
	&
	+2 q^4 t^5+13 q^2 t^5+2 q^2 t^4+q^2 t^3+\frac{q^2}{t^3}+\frac{2 t^4}{q^2}+\frac{t^4}{q^4}+\frac{t^3}{q^2}+\frac{3 t^3}{q^4}
	\\
	&
	+\frac{t^3}{q^6}+\frac{t^2}{q^6}+\frac{1}{q^2 t^3}+\frac{1}{q^6 t^5}+\frac{q^6}{t}+\frac{t}{q^{10}}+2 t^5+5 t^4+\frac{2}{t^3}\Big)\,,
\end{align*}

\begin{align*}
    \tilde p_{6} & =
	\left(q^2 t+1\right) \Big(q^{16} t^{10}+q^{16} t^9+q^{14} t^{10}+2 q^{14} t^9+q^{14} t^8+q^{12} t^9+3 q^{12} t^8+q^{12} t^7
	\\
	&
	+7 q^{10} t^8+4 q^{10} t^7+6 q^8 t^8+6 q^8 t^7+4 q^8 t^6+q^6 t^8+6 q^6 t^7+8 q^6 t^6+2 q^6 t^5+q^4 t^7
	\\
	&
	+11 q^4 t^6+4 q^4 t^5+\frac{q^4}{t^3}+10 q^2 t^6+5 q^2 t^5+3 q^2 t^4+\frac{2 q^2}{t^3}+\frac{3 t^4}{q^2}+\frac{5 t^4}{q^4}+\frac{t^3}{q^2}+\frac{t^2}{q^8}+\frac{2}{q^2 t^3}
	\\
	&
	+\frac{2}{q^4 t^3}+\frac{1}{q^6 t^3}+\frac{q^{10}}{t}+\frac{2 q^8}{t}+\frac{2 q^6}{t}+\frac{2 q^4}{t}+\frac{2 q^2}{t}+3 t^6+5 t^5+4 t^4+t^3+\frac{2}{t^3}+\frac{1}{t}\Big)
	\\
	& +
	\left(\frac{q^2}{t}+1\right) \Big(q^{14} t^{10}+q^{12} t^{10}+q^{12} t^9+2 q^{10} t^9+q^8 t^9+q^8 t^8+q^8 t^7+3 q^6 t^8+2 q^6 t^7
	\\
	&
	+q^6 t^6+3 q^4 t^7+2 q^2 t^7+q^2 t^6+q^2 t^5+\frac{t^5}{q^2}+\frac{2 t^5}{q^4}+\frac{t^4}{q^4}+\frac{2 t^4}{q^6}+\frac{t^3}{q^8}
	\\
	&
	+\frac{1}{q^2 t^2}+q^6+q^4+4 t^6+3 t^5+t^4+\frac{1}{t^2}\Big)\,,
	\\
	\\
	\tilde p_{8} & =
	\left(q^2 t+1\right) \Big(2 q^{16} t^{11}+q^{14} t^{11}+q^{14} t^{10}+q^{12} t^{10}+4 q^{12} t^9+9 q^{10} t^9+5 q^8 t^9+2 q^8 t^8
	\\
	&
	+3 q^6 t^8+4 q^6 t^7+9 q^4 t^7+2 q^2 t^7+3 q^2 t^6+\frac{2 t^5}{q^2}+q^8+q^6+q^4+q^2+t^6+3 t^5+1\Big)
	\\
	&+
	\left(\frac{q^2}{t}+1\right) \Big(2 q^{16} t^{12}+q^{14} t^{12}+2 q^{14} t^{11}+2 q^{12} t^{11}+2 q^{12} t^{10}
	\\
	&
	+5 q^{10} t^{10}+3 q^8 t^{10}+3 q^8 t^9+4 q^6 t^9+3 q^6 t^8+8 q^4 t^8+2 q^2 t^8+2 q^2 t^7+q^4 t+t^6\Big)\,,
	\\
	\\
	\tilde p_{10} & = 
	\left(q^2 t+1\right) \Big(q^{18} t^{12}+2 q^{16} t^{12}+2 q^{14} t^{12}+q^{12} t^{11}+2 q^{12} t^{10}
	\\
	&
	+5 q^{10} t^{10}+4 q^8 t^{10}+q^6 t^9+2 q^6 t^8+3 q^4 t^8+q^2 t^8\Big)
	\\
	&+
	\left(\frac{q^2}{t}+1\right) \left(q^{14} t^{13}+q^{12} t^{12}+q^{10} t^{11}+2 q^8 t^{11}+q^6 t^{10}\right)\,,
	\\
	\\
	\tilde p_{12} & = 
	\left(q^2 t+1\right) \left(q^{14} t^{13}+q^{10} t^{11}+q^8 t^{11}\right)
	+
	 \left(\frac{q^2}{t}+1\right)q^{14} t^{14} \,.
\end{align*}

\section{Comparing \texorpdfstring{$T[L_K]$}{T[LK]} and \texorpdfstring{$T[Q]$}{T[Q]} in view of knot homology}\label{app:TLK-vs-TQ}
We collect here some comments on how the~theory $T[L_K]$ is more closely related to knot homology than its dual description $T[Q]$.

At level  $x^2$,  the~quiver partition  function has terms proportional  to  $x_i x_j / [(q^2;q^2)_1]^2$.
The partition function of $T[L_K]$ is obtained by substitution of quiver variables $x_i=x_i(x,a,q)$. Such terms appear as $(1+q^2) x_i x_j / (q^2;q^2)_2$, i.e., with a~different denominator: 
\be
    \frac{1}{(1-q^2)^2} \to \frac{1+q^2}{(1-q^2)(1-q^4)}\,.
\ee
This is somewhat unnatural from the~viewpoint of  $T[Q]$ since it `doubles' the~contribution of certain generators of the~Hilbert space of BPS vortices.
On the~other hand, from the~viewpoint of $T[L_K]$, this change of denominator, and similar changes at higher orders in~$x$, arise naturally and therefore the~Hilbert space of level $r$ BPS vortices in $T[L_K]$ is more closely related to HOMFLY-PT homologies of  knots.
Physically, the~difference between the~denominator $(q^2;q^2)_2$ in $T[L_K]$ and $((q^2;q^2)_1)^2$ in $T[Q]$ comes from  vieweing a~vortex with vorticity $2$ as a~bound state of two \emph{distinguishable} vortices (each with vorticity~$1$) in~$T[Q]$ versus viewing it as a~bound state of two \emph{indistinguishable} vortices with vorticity~$1$ in~$T[L_K]$.

Let us discuss another point concerning denominators in the~partition function. 
Consider a~$d$-times around node which is \emph{not} a~bound state of once-around disks. 
It will contribute to the~partition function $P_K(x,a,q)$ by a~term of the~form
\be
 \frac{ a^\alpha q^\beta  x^2}{1-q^{2d}}\,.
\ee
The contribution of this term to the~HOMFLY-PT polynomial with $S^d$-coloring will then be obtained  by correcting the~denominator by multiplication and division by $(1-q^2)\dots(1-q^{2(d-1)})$:
\be
\frac{(1-q^2)\dots(1-q^{2(d-1)}) a^\alpha q^\beta}{(q^2;q^2)_d} \,.
\ee
From the~viewpoint of a~quiver-like description it is somewhat unnatural to compensate the~denominator by this extra factor: it multiples the~contribution of a~single fundamental curve (a $d$ times around basic  disk) by $2^{d-1}$ and it produces terms with  opposite signs in the~numerator. See definition \ref{def:genpartfunction} for a~framework where the~corrected denominator appears naturally and where there is a~refinement that retains positivity.

\newpage

\bibliography{main}{}
\bibliographystyle{JHEP}

\end{document}